\documentclass[preprint]{aastex}

\usepackage{graphicx,amssymb,multirow,gensymb,lscape}
\usepackage{natbib}
\newcommand{\Jybeam}{\mbox{Jy beam$^{-1}$}}% mJy / beam
\newcommand{\mJybeam}{\mbox{mJy beam$^{-1}$}}
\newcommand{\vol}{\mbox{cm$^{-3}$}} %cgs volume (cm^-3)
\newcommand{\cden}{\mbox{cm$^{-2}$}} %column density
\newcommand{\kms}{\mbox{km s$^{-1}$}}% km/s
\newcommand{\um}{\mbox{$\mu$m}}% microns
\newcommand{\Msun}{\mbox{M$_{\odot}$}}
\newcommand{\cmg}{\mbox{cm$^2$ g$^{-1}$}}

\newcommand{\NHH}{\mbox{N(H$_2$)}}

\newcommand{\Kkms}{\mbox{K km s$^{-1}$}}

\newcommand{\kappanu}{\mbox{$\kappa_{\nu,d}$}}

\shorttitle{Constraining Dust Properties in the Perseus B1 Clump}
\shortauthors{Sadavoy et al.}

\title{The Herschel and JCMT Gould Belt Surveys: Constraining Dust Properties in the Perseus B1 Clump with PACS, SPIRE, and SCUBA-2}
\author{S. I. Sadavoy\altaffilmark{1,2},
	J. Di Francesco\altaffilmark{1,2},
	D. Johnstone\altaffilmark{1,2,3},
	M. J. Currie\altaffilmark{3},
	E. Drabek\altaffilmark{4},
	J. Hatchell\altaffilmark{4},
	D. Nutter\altaffilmark{5},
	Ph. Andr\'{e}\altaffilmark{6},
	D. Arzoumanian\altaffilmark{7},
	M. Benedettini\altaffilmark{8},
	J.-P. Bernard\altaffilmark{9,10},
	A. Duarte-Cabral\altaffilmark{11,12},
	C. Fallscheer\altaffilmark{1,2},
	R. Friesen\altaffilmark{13},
	J. Greaves\altaffilmark{14},
	M. Hennemann\altaffilmark{6},
	T. Hill\altaffilmark{6},
	T. Jenness\altaffilmark{3},
	V. K\"{o}nyves\altaffilmark{6,7},
	B. Matthews\altaffilmark{1,2},
	J. C. Mottram\altaffilmark{15},
	S. Pezzuto\altaffilmark{8},
	A. Roy\altaffilmark{6},
	K. Rygl\altaffilmark{8},
	N. Schneider-Bontemps\altaffilmark{11,12},
	L. Spinoglio\altaffilmark{8},
	L. Testi\altaffilmark{16},
	N. Tothill\altaffilmark{17},
	D. Ward-Thompson\altaffilmark{18},
	G. White\altaffilmark{19}
	}
\author{and the JCMT and \emph{Herschel} Gould Belt Survey teams}

\altaffiltext{1}{Department of Physics \& Astronomy, University of Victoria, PO Box 355, STN CSC, Victoria BC, V8W 3P6, Canada}
\altaffiltext{2}{National Research Council Canada, 5071 West Saanich Road, Victoria BC Canada, V9E 2E7}
\altaffiltext{3}{Joint Astronomy Centre, 660 N. A$^{\prime}$oh\={o}k\={u} Place, University Park, Hilo, Hawaii, 96720, USA}
\altaffiltext{4}{School of Physics, University of Exeter, Stocker Road, Exeter EX4 4QL, U.K.}
\altaffiltext{5}{School of Physics and Astronomy, Cardiff University, Queen's Buildings, The Parade, Cardiff CF24 3AA, UK}
\altaffiltext{6}{Laboratoire AIM, CEA/DSM-CNRS-Universit\'{e} Paris Diderot, IRFU/Service dÕAstrophysique, Saclay, 91191 Gif-sur-Yvette, France}
\altaffiltext{7}{IAS, CNRS (UMR 8617), Universit\'{e} Paris-Sud 11, B\^{a}timent 121, 91400 Orsay, France}
\altaffiltext{8}{Istituto di Astrofisica e Planetologia Spaziali, via Fosso del Cavaliere 100, 00133, Rome, Italy}
\altaffiltext{9}{CNRS, IRAP, 9 Av. colonel Roche, BP 44346, 31028 Toulouse Cedex 4, France}
\altaffiltext{10}{Universit\'{e} de Toulouse, UPS-OMP, IRAP, 31028 Toulouse Cedex 4, France}
\altaffiltext{11}{Universit\'{e} de Bordeaux, LAB, UMR 5804, F-33270, Floirac, France}
\altaffiltext{12}{CNRS, LAB, UMR 5804, F-33270, Floirac, France}
\altaffiltext{13}{Dunlap Institute, University of Toronto, Toronto, ON, M5S 3H8, Canada}
\altaffiltext{14}{School of Physics and Astronomy, University of St. Andrews, North Haugh, St. Andrews, Fife KY16 9SS}
\altaffiltext{15}{Leiden University, 506 Huygens Laboratory, PO Box 9513, 2300 RA, Leiden, Netherlands}
\altaffiltext{16}{ESO, Karl Schwarzschild-Strasse 2, 85748 Garching bei M\"{u}nchen, Germany}
\altaffiltext{17}{University of Western Sydney, Locked Bag 1797, Penrith NSW 2751, Australia}
\altaffiltext{18}{Jeremiah Horrocks Institute, University of Central Lancashire, Preston, Lancashire, PR1 2HE, UK }
\altaffiltext{19}{The Open University, Department of Physics and Astronomy, Milton Keynes MK7 6AA, UK}
\begin{document}

%%%%%%%%%%%   ABSTRACT  %%%%%%%%%%%%%%%%
\begin{abstract}
 
We present \emph{Herschel} observations from the \emph{Herschel} Gould Belt Survey and SCUBA-2 science verification observations from the JCMT Gould Belt Survey of the B1 clump in the Perseus molecular cloud. We determined the dust emissivity index using four different techniques to combine the \emph{Herschel} PACS+SPIRE data at $160 - 500$ \um\ with the SCUBA-2 data at 450 \um\ and 850 \um. Of our four techniques, we found the most robust method was to filter-out the large-scale emission in the \emph{Herschel} bands to match the spatial scales recovered by the SCUBA-2 reduction pipeline. Using this method, we find $\beta \approx 2$ towards the filament region and moderately dense material and lower $\beta$ values ($\beta \gtrsim 1.6$) towards the dense protostellar cores, possibly due to dust grain growth. We find that $\beta$ and temperature are more robust with the inclusion of the SCUBA-2 data, improving estimates from \emph{Herschel} data alone by factors of $\sim 2$ for $\beta$ and by $\sim 40$\%\ for temperature. Furthermore, we find core mass differences of $\lesssim 30$\%\ compared to \emph{Herschel}-only estimates with an adopted $\beta=2$, highlighting the necessity of long wavelength submillimeter data for deriving accurate masses of prestellar and protostellar cores.

\end{abstract}

%\keywords{keywords go here}

%%%%%%%%%%%%%%%%%%%%%%%%%%%%%%%%%%%%%
%%%%%%%%%%%   INTRODUCTION  %%%%%%%%%%%%%%%%
%%%%%%%%%%%%%%%%%%%%%%%%%%%%%%%%%%%%%
\section{Introduction\label{Intro}}

Molecular clouds are composed of dust and molecular gas ranging from $\sim 10^2$ \vol\ for the large-scale, low-density cloud to $> 10^4$ \vol\ for the small-scale, dense star-forming ``cores'' (\citealt{BerginTafalla07}). These molecular clouds are often traced by their cold dust emission, which peak at far-infrared or submillimeter wavelengths, and many studies have used optically thin dust emission to measure masses and column densities of cores and filaments (e.g., \citealt{Kirk06}; \citealt{Enoch09}; \citealt{Andre10}; \citealt{Arzoumanian11}). A key parameter to convert dust emission into mass (or column density) is the dust opacity, which corresponds to the ability of dust grains to absorb radiation. Unfortunately, the dust opacity is one of the most difficult observational measurements, and uncertainties in the dust opacity can result in significant mass uncertainties (\citealt{Henning95}; \citealt{Shirley11}; \citealt{Martin12}). Since dust populations in molecular clouds are expected to change due to dust coagulation and the formation of icy mantles in the densest environments (\citealt{DraineLee84}; \citealt{Ossenkopf94}), it is important to determine the dust opacity for both the large-scale cloud and the small-scale cores. 

Generally, submillimeter dust opacities assume a power-law form such that $\kappa_{\nu} \sim \nu^{\beta}$, where $\beta$ is the dust emissivity index. Since the spectral energy distribution (SED) of cold, dusty star forming regions are best represented by greybody emission, $\beta$ is critical to the SED fit. Several studies have attempted to determine $\beta$ towards specific prestellar or protostellar cores using long-wavelength submillimeter emission (e.g., \citealt{Friesen05}, \citealt{Schnee10}, \citealt{Shirley11}). These studies, however, were limited in wavelength coverage and poor map sensitivities. Conversely, recent observations with the \emph{Herschel} Space Observatory (\citealt{Pilbratt10}) extend our ability to measure cold dust emission to larger-scales over multiple wavelengths. In particular, the Photodetector Array Camera and Spectrometer (PACS; \citealt{Poglitsch10}) and the Spectral and Photometric Imaging Receiver (SPIRE; \citealt{Griffin10}) have provided unprecedented sensitivity for far-infrared and submillimeter wavelengths at 70 - 500 \um, which cover the SED peak from cold dust at $T \sim 10 - 40$ K. Both $\beta$ and temperature, however, affect the curvature of the black body function, and thus, are degenerate towards the SED peak. To break this degeneracy, $\beta$ can be constrained from the Rayleigh-Jeans tail of the SED (\citealt{Doty94}; \citealt{Shetty09}), i.e., $\lambda \gg 300$ \um\ for cold star-forming regions at $T \lesssim 30$ K. Therefore, additional long-wavelength data at good resolution and sensitivity are necessary to constrain the dust emissivity. 

These requirements make the Submillimetre Common-User Bolometer Array-2 (SCUBA-2; \citealt{Holland13}) at the James Clerk Maxwell Telescope (JCMT) an excellent complement to PACS and SPIRE data. SCUBA-2 can simultaneously observe at 450 \um\ and 850 \um, both at higher resolutions than the three SPIRE bands (the JCMT primary mirror is five times larger than \emph{Herschel}'s). Therefore, SCUBA-2 observations provide necessary data along the Rayleigh-Jeans tail and excellent spatial resolution at longer wavelengths. 

To explore possible variations of $\beta$ towards cold cores, we have combined PACS+SPIRE data with SCUBA-2 data from the JCMT Gould Belt Survey (GBS) for the first time. For this study, we have selected the B1 clump of the Perseus molecular cloud. B1 was previously observed with \emph{Herschel} as part of the \emph{Herschel} GBS (\citealt{Sadavoy12}; \citealt{Pezzuto12}), and was the first field of the JCMT GBS (\citealt{Ward-T07GBS}) that nominally reached the SCUBA-2 survey depth. B1 is a nearby star-forming region at 235 pc (\citealt{Hirota08}) with a total clump mass of $\sim 1200$ \Msun\ and $\sim 10$ dense submillimeter cores grouped into a central cluster (\citealt{Kirk06}). B1 is also relatively young (compared with the nearby NGC 1333 and IC 348 clumps), containing a high percentage ($> 40$ \%) of Class I young stellar objects (YSOs), of which several drive outflows (\citealt{Jorgensen08}; \citealt{Bally08}). Furthermore, a recent study by \citet{Pezzuto12} suggests that two dense cores in B1 are excellent candidates for first hydrostatic cores based on \emph{Herschel} detections at $\gtrsim 100$ \um\ with no corresponding Spitzer detections at 24 \um. Additionally,  \emph{Herschel} observations of B1 revealed several pillar-like structures (\citealt{Sadavoy12}), suggesting that a strong ionization front may be influencing the clump. This front likely originates from 40~Per, a high mass main sequence star in the Perseus OB2 region (see \citealt{Bally08}). 

This paper is organized as follows: In Section \ref{data}, we describe our \emph{Herschel} and SCUBA-2 datasets, as well as CO ($3-2$) observations of B1.  In Section \ref{results}, we show our results, first from our SCUBA-2 continuum and CO ($3-2$) observations, and then from fitting SEDs to the \emph{Herschel}-only bands. In Section \ref{combo}, we determine $\beta$ using the PACS+SPIRE and SCUBA-2 bands following four different techniques to combine these data. In Section \ref{discussion}, we use our most robust technique to probe the variations in dust opacity and core mass, and we discuss the implications of our analyses for future studies that will combine SCUBA-2 and PACS+SPIRE data. In Section \ref{conc}, we summarize our results.

%%%%%%%%%%%%%%%%%%%%%%%%%%%%%%%%%%%%%%%%
%%%%%%%%%%%%%%  The DATA section  %%%%%%%%%%%%%%%%
%%%%%%%%%%%%%%%%%%%%%%%%%%%%%%%%%%%%%%%%
\section{Data}\label{data}

\subsection{Herschel Observations}

The western half of Perseus, including the B1 clump, was observed with \emph{Herschel} in February of 2010 as part of the the \emph{Herschel} GBS. These data consist of parallel photometric observations with the PACS and SPIRE instruments at 70 \um, 160 \um, 250 \um, 350 \um, and 500 \um\ at a 60 arcsec s$^{-1}$ scan rate. The PACS and SPIRE raw data were reduced with version 7.0 of HIPE using modified reduction scripts by M. Sauvage (PACS) and P. Panuzzo (SPIRE) and updated calibration corrections; we used the PACS Calibration Set v26 and the SPIRE Calibration Tree 6.1 for our PACS and SPIRE reductions, respectively. The final maps were produced using version 11 of the \emph{scanamorphos} routine (\citealt{Roussel12}). For more information, see \citet{Sadavoy12}. For more information about the observations of Perseus and an alternative map making routine, see \citet{Pezzuto12}. The \emph{Herschel} observation have resolutions of $\sim 8 - 36$\arcsec.

\subsection{SCUBA-2 Observations}\label{scuba2}

B1 was observed at 450 \um\ and 850 \um\ with SCUBA-2 as part of Science Verification (hereafter, S2SV) observations for the JCMT GBS in October 2011. SCUBA-2 is a bolometer detector with $\sim 10 000$ pixels spread over eight science arrays for a field of view of $\sim 45$ arcmin$^2$ (\citealt{Holland13}).  For comparison, the predecessor bolometer detector, SCUBA, contained only 128 pixels over two science arrays for a field of view of $\sim 5$ arcmin$^2$ (\citealt{Holland99}). For S2SV, all eight science grade arrays (four arrays for each band) were available. 

The B1 clump was observed six times on October 18 in very dry (Grade 1; $\tau_{225} < 0.05$) weather and three more times on October 19 in poorer conditions (Grade 3; $\tau_{225} < 0.1$). Each observation used a PONG1800 observing pattern (\citealt{Dempsey12}; \citealt{Holland13}), which involved mapping a $38\arcmin \times 38$\arcmin\ square box centered at $3^h33^m12^s.7$, $+31\degree06\arcmin48\arcsec$ (J2000) with a mapping speed of 480 arcsec s$^{-1}$ and scan spacing of 30\arcsec. The box was mapped 5 times, rotating the sides by 18$\degree$ with respect to the sky between each coverage to produce a well-sampled circular map of $\sim$ 30\arcmin\ diameter. This pattern was repeated to increase the sensitivity of the map. Two of the six observations on October 18 were made at high elevations ($> 65\degree$), which caused erratic telescope tracking motions during the PONG-mode mapping and introduced corrugated artifacts into the respective images. 

The S2SV observations were reduced using the \emph{Starlink} SMURF version 1.4.0 package and the \emph{makemap} routine (\citealt{Jenness11}; \citealt{Chapin13}), which  generates a map of the sky by iteratively fitting a number of different signals in the data. These signals included the common mode signal (caused by atmospheric emission and oscillations in the refrigerator system), the gain of each bolometer, the atmospheric extinction, and the astronomical signal. The mapmaker iterated until a convergence parameter is reached\footnote{The convergence parameter was defined as $\bar{C_i} = \frac{1}{N_i}\sum{|F_{i,p} - F_{i-1,p}|/\sqrt{\sigma_{i,p}}}$, where $i$ is the $i^{th}$ iteration, $p$ corresponds to each pixel within the current mask, $N_i$ is the total number of pixels within the mask, $F_{i,p}$ is the flux in each pixel for the i$^{th}$ iteration of the map, and $\sigma_{i,p}$ is the variance in each pixel for the i$^{th}$ iteration. The convergence parameter was measured as an average over all pixels. The i$^{th}$ iteration converged when $\bar{C_i} \le 0.001$.}.  Each observation was reduced independently in this manner and then an initial coadded map was produced from the independent observations. We combined the four good observations in Grade 1 weather, rejecting the three Grade 3 observations and two corrugated Grade 1 observations. 

\begin{figure}[h!]
\includegraphics[scale=0.48]{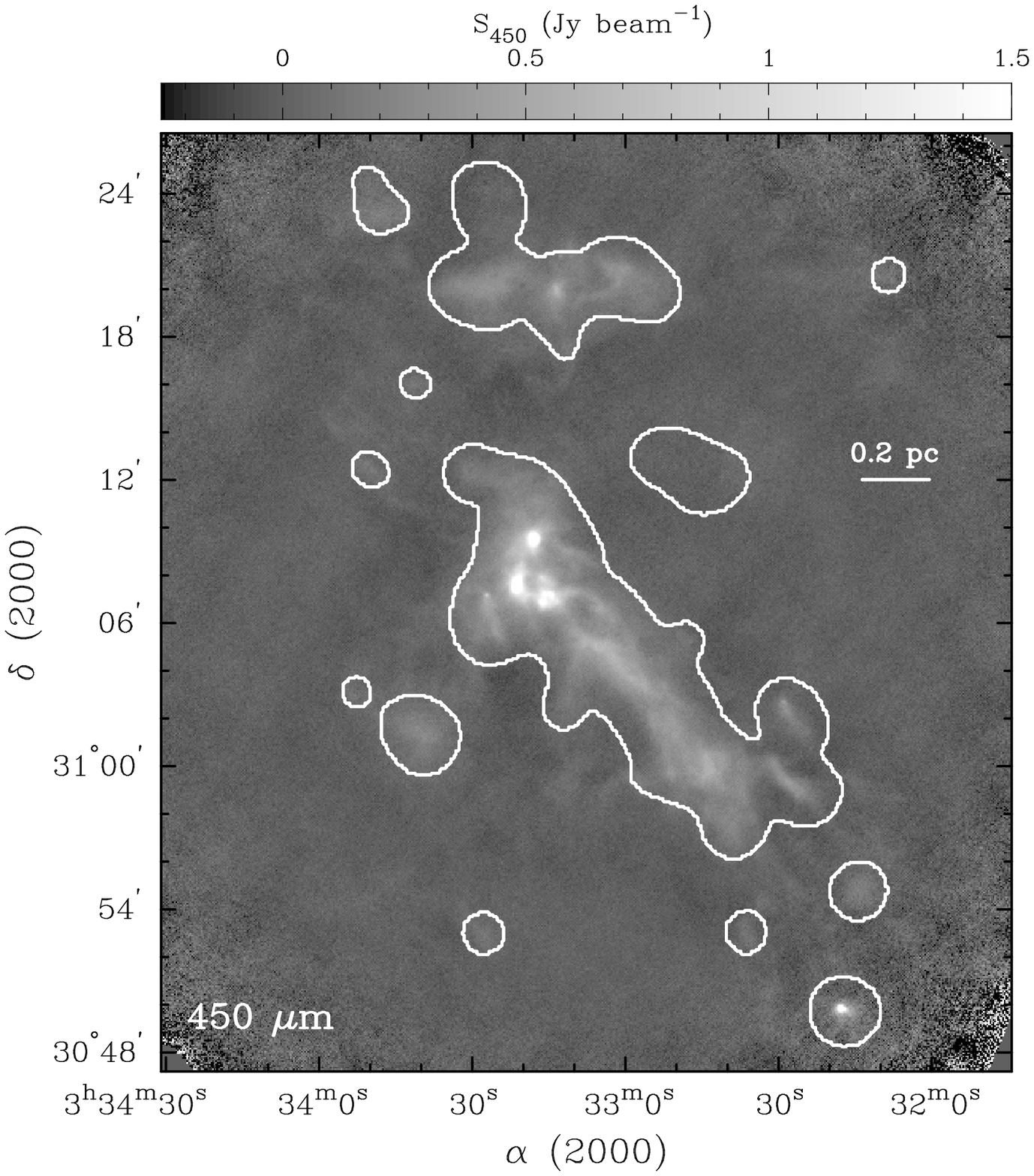}
\includegraphics[scale=0.48]{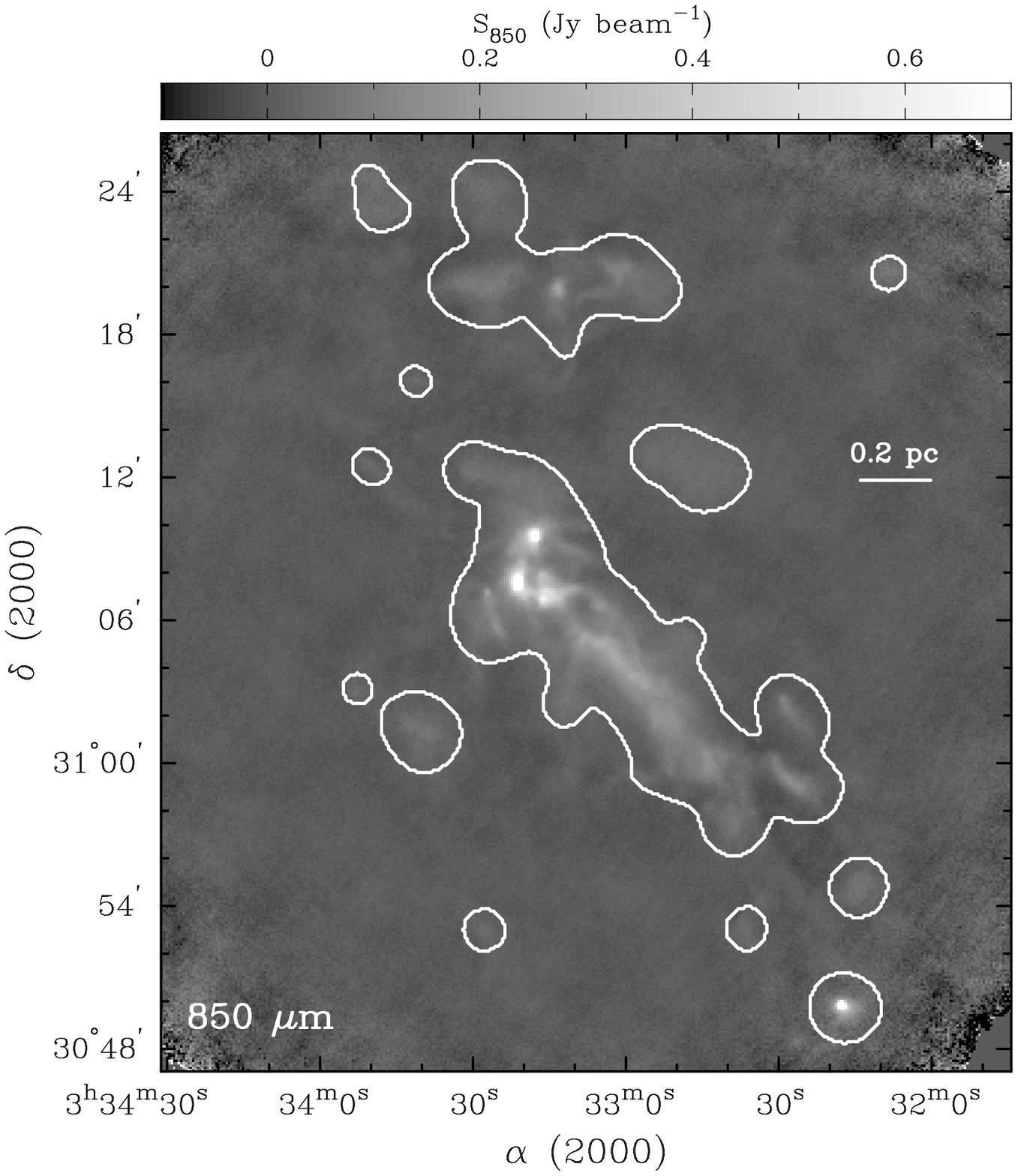}
\caption{SCUBA-2 observations of the B1 clump at 450 \um\ (left) and 850 \um\ (right). Contours show the masked regions used in the reductions. Areas outside of the mask have uncertain fluxes. \label{S2Maps}}
\end{figure}

An initial coadded map was used to generate a mask that encompassed all parts of the map that contain significant levels of flux. This mask was used to re-reduce all of the data in the manner described above, except the astronomical signal is only retained between each iteration if it lies within the masked area.  The purpose of using the mask in this manner was to prevent spurious artifacts in the low sensitivity regions of the map. For simplicity, the mask was generated corresponding to those regions with signal-to-noise ratios $\gtrsim 5$ at 850 \um.  Regions outside the mask, however, were not used in any analysis, as the data there did not iterate to convergence. We tested the robustness of the maps by inserting artificial sources into the raw data and recovering reliable fluxes in the final map. Figure \ref{S2Maps} shows the final reduced maps, highlighting the masked regions. The same mask was used for the 450 \um\ reductions. 

We used flux conversion factor (FCF) values of 491 \Jybeam\ pW$^{-1}$ and 556 \Jybeam\ pW$^{-1}$ for the 450 \um\ and 850 \um\ maps, respectively, based on observations of standard calibration sources (\citealt{Dempsey12}, \citealt{Dempsey13}). These FCF values corresponded to the version of \emph{Starlink} used in our reduction. We adopted a calibration uncertainty of 10\%\ across the SCUBA-2 850 \um\ map (\citealt{Dempsey12}). For the 450 \um\ band, however, atmospheric variability becomes more significant, even in Grade 1 weather. Therefore, we adopted a 50\%\ calibration uncertainty at 450 \um, which was the typical flux error at 450 \um\ adopted for the SCUBA Legacy Survey (\citealt{difran08}). We acknowledge that such a large calibration error is conservative and with further investigation, will be improved. The final $1\ \sigma$ rms sensitivity of the observed SCUBA-2 map was $\sim 7$ \mJybeam\ at 850 \um\ over a 6\arcsec\ pixel grid. The effective $1\ \sigma$ point source sensitivity integrated over the beam area was $\sim$ 3 mJy, the target set by the SCUBA-2 Gould Belt Legacy Survey (\citealt{Ward-T07GBS}). For the 450 \um\ map, the final $1\ \sigma$ rms sensitivity was $\sim 28$ \mJybeam\ over 4$\arcsec$ pixel grid for an equivalent point source sensitivity of $\sim 11$ mJy over the beam area. 

The SCUBA-2 beam contains two components, the primary beam and an error beam (e.g., \citealt{difran08}). The 450 \um\ and 850 \um\ primary beams are $\sim 8\arcsec$ and $\sim 13\arcsec$ and the error beams are $\sim 25\arcsec$ and $\sim 48\arcsec$, respectively. At 450 \um, the error beam accounted for $\sim 40$\%\ of the volume, whereas at 850 \um, the error beam accounted for $\sim 25$\%\ of the volume (\citealt{Dempsey13}; see also Section \ref{s2+harp}). For this paper, we adopted effective Gaussian beams with FWHM values of 9.3\arcsec\ and 14.2\arcsec\ for 450 \um\ and 850 \um, respectively.

\subsection{HARP Observations}

We mapped the B1 clump in CO ($3-2$) line emission with the Heterodyne Array Receiver Program (HARP; see \citealt{Buckle09} and references therein) on June 29, 2012 in good Grade 2 weather ($\tau_{225} < 0.08$) over 3 hours. In brief, the HARP instrument consists of sixteen heterodyne detectors arranged in a $4\times4$ pixel grid and covering a frequency range of 325 - 375 GHz. Observations from HARP are processed by the Auto-Correlation Spectral Imaging System (ACSIS; \citealt{Jenness08}). For these observations, only fourteen receptors were functional. We made two position-switch raster maps (in orthogonal directions) over a $30\arcmin \times 30\arcmin$ region coincident with the S2SV-mapped region. The backend was configured to 1 GHz bandwidth over 2048 channels for a velocity resolution of $\sim 0.42$ \kms. 

The data were reduced using the {ORAC-DR} (\citealt{Cavanagh08}) pipeline for ACSIS, and the reduction recipe outlined below can be found in the latest \emph{Starlink} release.  Heterodyne detectors are notorious for noisy baselines due to external interferences. Rather than reject the entire spectra from these detectors, we performed an iterative reduction to remove suspect spectra from the time series while keeping the good spectra, thereby improving the signal-to-noise and the coverage of the final spectral map.  The reduction pipeline can be broadly divided into two parts:

First, we performed an initial processing of each observation, which includes chronological sorting of the raw time-series cube and a quality-assurance stage.  Of particular relevance to the B1 data was the removal of spectra affected by high and low frequency interferences.  To identify these interferences, we trimmed the excessively noisy ends of the spectra and excluded the spectral region where the astronomical signal was present.  High-frequency noise (3 to 4 spectral elements) was detected using a one-dimensional Laplacian filter along the spectral axis for each receptor, and the affected blocks of spectra were identified by a threshold rms profile along the time axis. Low-frequency oscillations (or distortions in the baseline) were found by smoothing features smaller than 100 pixels, and we summed the rms deviations from the best-linear baseline fit to measure the non-linearity for each spectrum. Again, we used a threshold to identify spectra that contained low frequency artifacts. Spectra or whole receptors that failed quality assurance were rejected.

Second, we applied an iterative procedure, which refined the baseline subtraction at each cycle.  We combined and regridded all the raw time series cubes into a group spatial cube.  Then, we applied a baseline subtraction and smoothing so that the astronomical emission was located within the group cube using clump finding.  The emission defines a further mask, which is converted back to a time series and is applied to the raw time-series cube for the next iteration.  In practice one iteration was sufficient. 

We converted the final reduced data from $T_A^*$ to $T_{MB}$ using a main beam efficiency of $\eta_{MB} = 0.61$. The final reduced data were smoothed to 0.85 \kms, resulting in a line sensitivity of $\sim 0.3$ K. Since the SCUBA-2 and HARP data were observed at the same facility, both data were taken at very similar resolutions ($\sim$ 14\arcsec). Having both the continuum and line data at the same resolution is very advantageous. \citet{Sun06} made a more complete CO ($3-2$) survey of the entire Perseus cloud with the KOSMA 3m telescope. These data, however, have much lower spatial resolution ($\sim 82\arcsec$), and therefore, comparisons to the HARP or SCUBA-2 data at $\sim 14\arcsec$ are difficult.

%%%%%%%%%%%%%%%%%%%%%%%%%%%%%%%%%%%%%%%%
%%%%%%%%%%%%%%   Results %%%%%%%%%%%%%%%%%%%%%
%%%%%%%%%%%%%%%%%%%%%%%%%%%%%%%%%%%%%%%%

\section{Results}\label{results}

\begin{figure}[h!]
\includegraphics[scale=0.82]{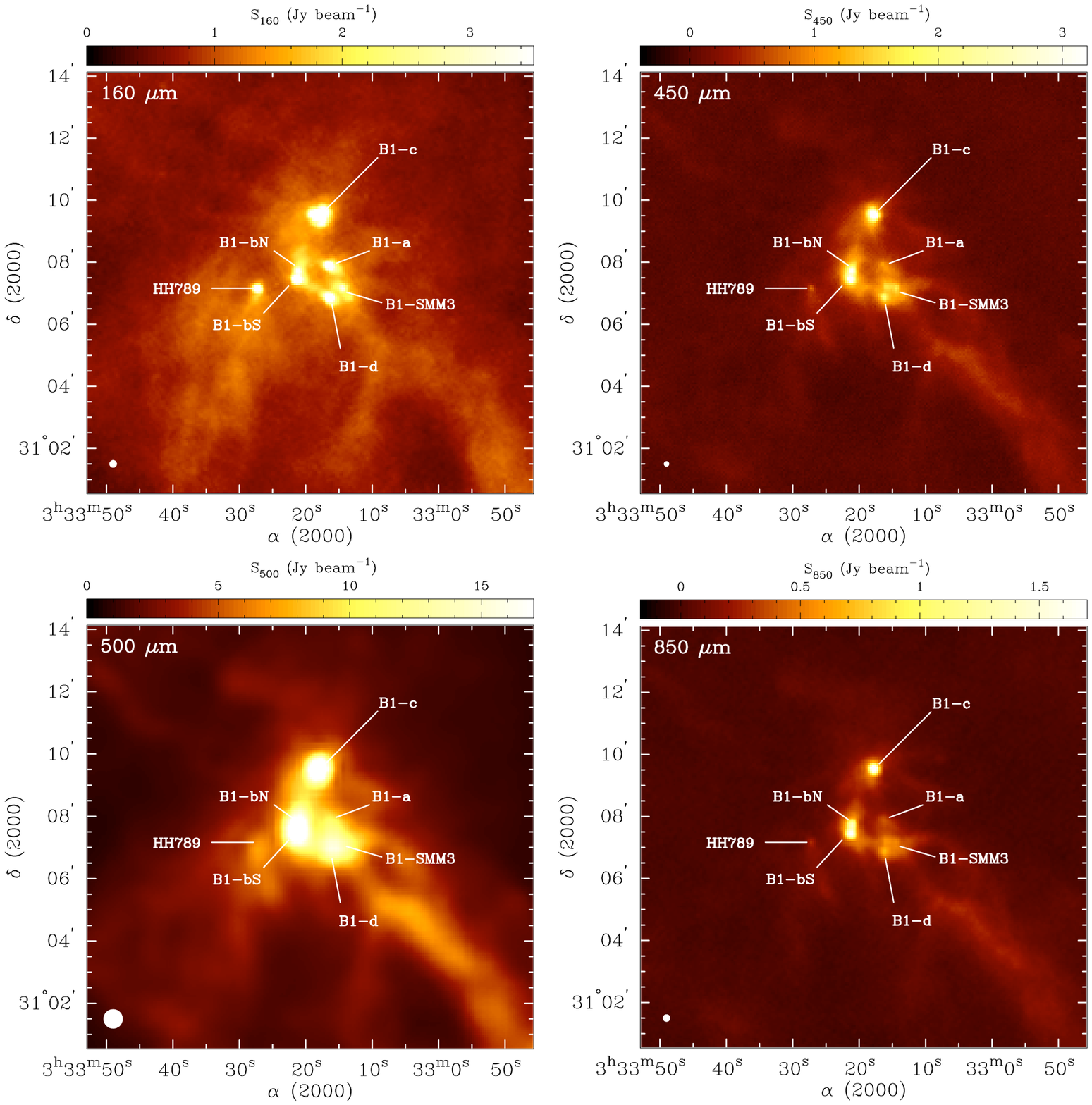}
\caption{Observed continuum maps from PACS at 160 \um\ (upper left), SCUBA-2 at 450 \um\ (upper right), SPIRE at 500 \um\ (lower left), and  SCUBA-2 at 850 \um\ (lower right). Note that several compact objects with PACS and SCUBA-2 are blended at 500 \um. The observed resolutions are $\sim 13\arcsec$ at 160 \um, $\sim 9\arcsec$ at 450 \um, $\sim 36\arcsec$ at 500 \um, and $\sim 14\arcsec$ at 850 \um. These maps reveal several known sources in the central B1 region (see \citealt{Jorgensen07}; \citealt{Hatchell07}). \label{obsMaps}}
\end{figure}

Figure \ref{obsMaps} shows the observations at 160 \um, 450 \um, 500 \um, and 850 \um\ for the central B1 region. Ground-based submillimeter continuum detectors like SCUBA-2  contend with rapidly varying atmospheric signals that greatly affect the detection of radiation. To remove the atmospheric emission, we effectively filter-out all large-scale structure from the maps, including the large scale astronomical signals. For our SCUBA-2 data, this effective spatial filtering corresponds to all angular scales of $\gtrsim 2.5\arcmin$ based on tests of the mapmaker to recover Gaussians of various sizes (see Appendix \ref{filterAppendix} for more details). Therefore, we believe that structures of $\lesssim 2.5\arcmin$ are recovered fully by SCUBA-2. Conversely, space-based instruments like those on-board \emph{Herschel} are free from these limitations and can recover the diffuse large-scale emission. For our \emph{Herschel} observations, we can recover spatial scales within half a scan length ($\sim 1.5\degree$). 

Another key difference between the SCUBA-2 and \emph{Herschel} observations is resolution. The \emph{Herschel} resolutions are $\sim 8.4\arcsec, 13.5\arcsec, 18.2\arcsec, 24.9\arcsec$, and $36.3\arcsec$ at 70 \um, 160 \um, 250 \um, 350 \um, and 500 \um, respectively for parallel mode observations at 60 arcsec s$^{-1}$ scan rates. Comparatively, the SCUBA-2 resolutions are $\sim 9\arcsec$ and $14\arcsec$ for the 450 \um\ and 850 \um\ bands, respectively. For clustered environments, the gain in resolution brought by SCUBA-2 at these long wavelengths is important for measuring the properties of compact sources. For instance, some of the compact objects within the central B1 region are blended at the longer \emph{Herschel} wavelengths but separated out with the SCUBA-2 observations (e.g., see the B1-bN and B1-bS cores in Figure \ref{obsMaps}).

\subsection{SCUBA-2 and HARP Analyses}\label{s2+harp}

Molecular line contamination is significant if the line flux measured over the continuum band boosts the continuum signal over the expected value from dust emission alone (\citealt{White03}). In cold star-forming regions, line contamination is most prominent from rotational transitions of CO, the second most abundant molecule in molecular clouds. While the \emph{Herschel} data are largely free of line contamination\footnote{The 350 \um\ band can be contaminated by the high excitation transition CO ($7-6$) at 806.652 GHz, though we believe such contamination should be negligible towards B1. See Appendix \ref{COAppendix}.}, the SCUBA-2 850 \um\ band is susceptible to contamination from the CO ($3-2$) line transition at 345.796 GHz, which lies in the middle of the band and is commonly detected in star-forming regions. Additionally, CO ($6-5$) line emission at 691.473 GHz can contaminate the 450 \um\ band; see Appendix \ref{COAppendix}. For a recent summary of the effects of line contamination on observed submillimeter fluxes, see \citet{Drabek12}.

Since the 850 \um\ continuum bandpass is very wide ($\sim 35$ GHz), molecular lines need to be both bright and broad to make significant contributions to the observed continuum emission (\citealt{Johnstone03}). We determined the contribution of CO ($3-2$) line emission to the continuum using the same technique as \citet{Drabek12}. For more details, see Appendix \ref{COAppendix}. Unlike SCUBA-2, the HARP CO ($3-2$) line data recover large-scale structures since they were obtained using a reference position free of line emission at a much larger angular distance from the target positions. Therefore, we ran the integrated intensity CO ($3-2$) map through the SCUBA-2 reduction pipeline to produce a filtered CO flux map that inlcudes only emission on the same spatial scales as the 850 \um\ data (see Appendix \ref{filterAppendix}). The filtered CO ($3-2$) line emission contributed $\lesssim 70$ \mJybeam\ ($\lesssim 100$ \Kkms) to the 850 \um\ continuum emission with the most significant contributions towards the outflows associated with the protostellar cores B1-c and B1-d. 

\begin{figure}[h!]
\includegraphics[scale=0.48]{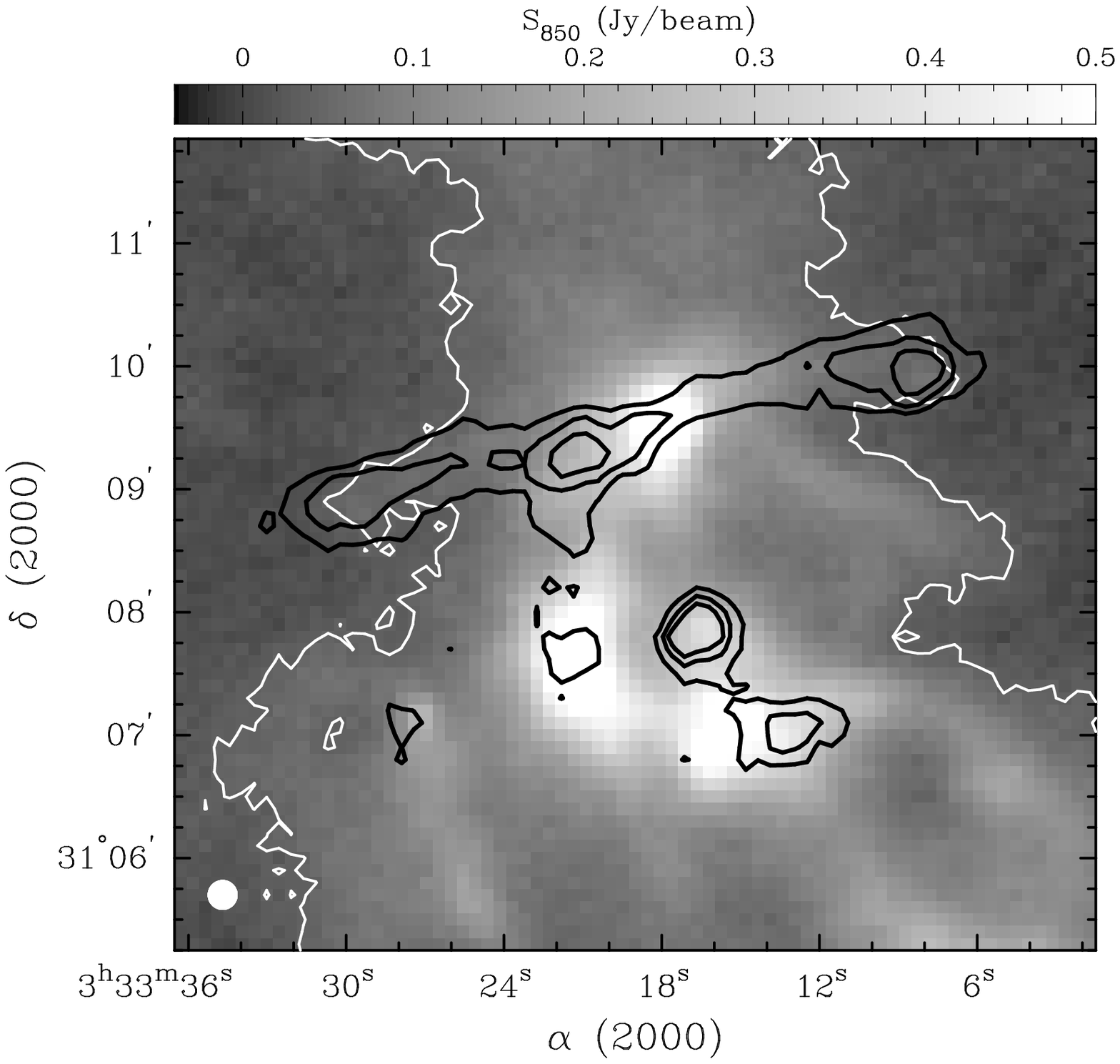}
\includegraphics[scale=0.48]{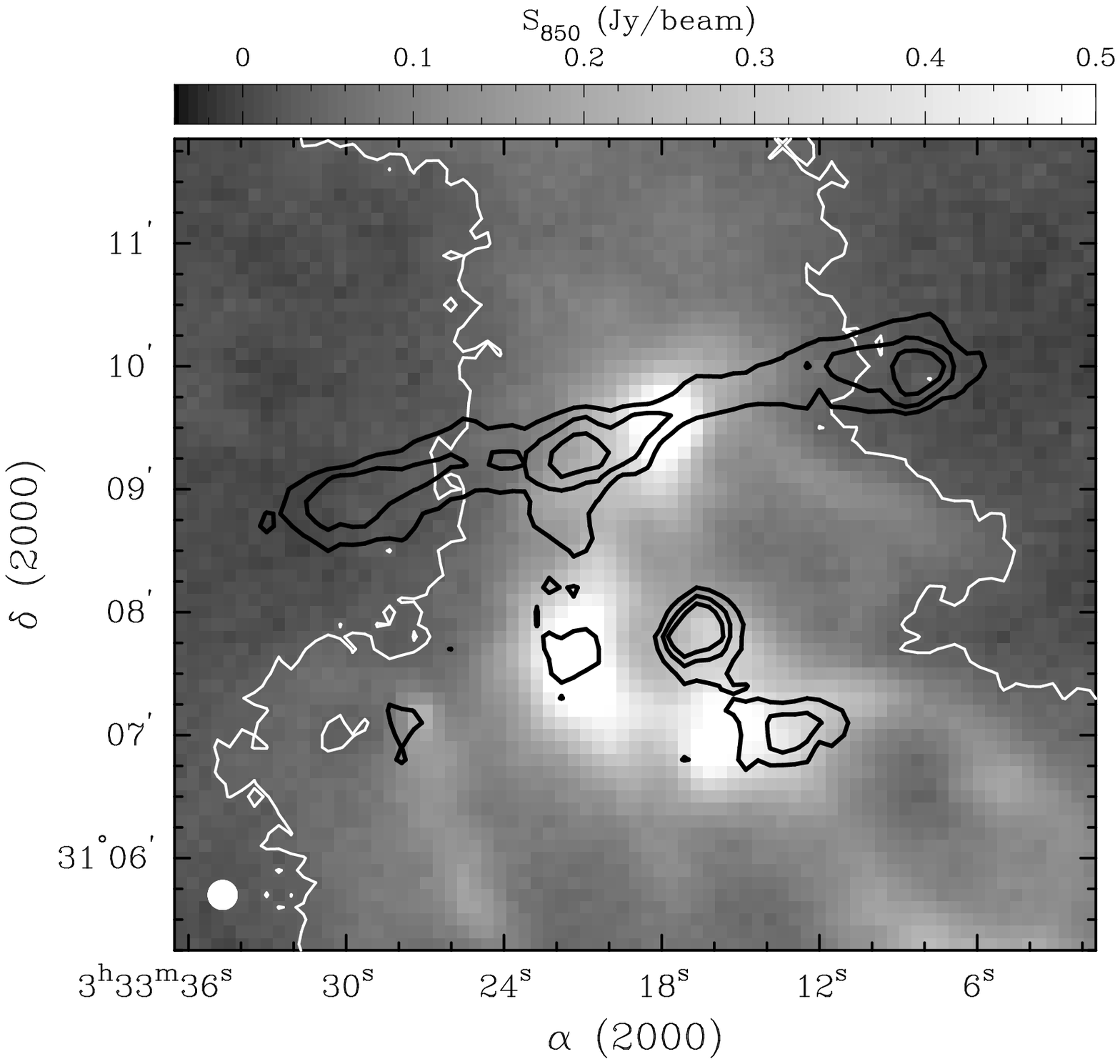}
\caption{Comparisons between 850 \um\ continuum emission and CO ($3-2$) line emission. \emph{Left:} Background image shows the uncorrected 850 \um\ continuum map. Black contours corresponds to CO ($3-2$) line emission at levels of 10 \mJybeam, 20 \mJybeam, and 30 \mJybeam. The CO data were filtered using the SCUBA-2 pipeline (see Appendix \ref{filterAppendix}). The white contours show 850 \um\ continuum emission at 30 \mJybeam. \emph{Right:} Background image shows the 850 \um\ continuum map with the filtered CO line emission subtracted out. The white contours show corrected 850 \um\ emission at 30 \mJybeam. The SCUBA-2 and HARP data have a common resolution of $\sim 14.2\arcsec$, shown as a white circle. \label{compareB1Cluster}}
\end{figure}

Figure \ref{compareB1Cluster} compares the 850 \um\ map before and after the CO ($3-2$) line contamination was removed. In the left panel, the two lobes of CO ($3-2$) emission along the northern outflow clearly dominate the continuum flux, contributing up to $\sim 90\%$ of the total 850 \um\ flux at the positions of the outflow. Aside from these outflows, the corrected 850 \um\ map (right panel) shows little difference with the original map (left). The CO ($3-2$) line emission contributes $\lesssim 15\%$ of the total 850 \um\ flux for the central region of B1 and $\lesssim 1\%$ for the rest of the B1 clump. 

Hereafter, the observed 850 \um\ data have been corrected for CO ($3-2$) line contamination. For comparisons with \emph{Herschel} data, we convolved both the CO-corrected 850 \um\ map and the observed 450 \um\ map to the 500 \um\ resolution of 36.3\arcsec. For the 450 \um\ map, both the main beam and error beam were considered in the convolution\footnote{For the 450 \um\ map, the error beam has a FWHM of $\sim$ 25\arcsec, which is smaller than the 500 \um\ beam.  We convolved a two-component Gaussian beam with a main component of 7.9\arcsec\ and an error component of 25\arcsec\ to  $34.25$\arcsec\ for an effective resolution of 36.3\arcsec. Without considering the error beam, the 450 \um\ fluxes would be overestimated by $\lesssim 10\%$.}, whereas for the 850 \um\ map, we found that the relative power in the error beam was not as significant.

%%%%%%%%%%%
%Herschel-ONLY
%%%%%%%%%%%

\subsection{Herschel-Only Results}\label{sedH}

We used the PACS 160 \um\ and the SPIRE 250 \um, 350 \um, and 500 \um\ data to estimate the dust properties in B1. Since the PACS 70 \um\ band can trace non-equilibrium emission from very small dust grains (\citealt{Martin12}) and warm dust emission from protostellar sources (as opposed to the cool envelope traced by the longer wavelengths), we do not include those data in our analysis. First, we corrected the arbitrary zero-point flux offset in each band using Planck HFI and IRAS data following the method proposed in \citet{Bernard10}. Second, we convolved each map to the same resolution (36.3\arcsec, or the 500 \um\ beam FWHM) and projected all maps onto a common grid of 14\arcsec\ pixels. 

For the PACS and SPIRE bands, minor colour correction factors ($\lesssim 3 \%$) are necessary (for more details, see \citealt{Pezzuto12} and Appendix \ref{ColourCorrectionAppendix}). Table \ref{colourCorr} lists our adopted colour correction factors and uncertainties, assuming $\beta \approx 1.5 - 2.5$ and T $\approx 10\ \mbox{K} - 15$ K. Additionally, we estimated the pixel-to-pixel rms sensitivity by selecting several areas relatively free of diffuse emission in the Perseus West map (see Figure 1 in \citealt{Sadavoy12}). Table \ref{colourCorr} lists the approximate 1 $\sigma$ rms uncertainties at the native resolution of each band and for the convolved maps (36.3\arcsec\ resolution). Since the selected regions were not entirely free of emission, these sensitivity measurements can be considered upper limits to the true map noise. These map sensitivities also depend greatly on the processing, such as the mapmaker. The observed 1 $\sigma$ rms noise estimates are within a factor of $\sim$ 2 of the expected 1 $\sigma$ rms noise errors according to the \emph{Herschel} observation planning tool (HSpot)\footnote{HSpot is a joint development by the Herschel Science Ground Segment Consortium, consisting of ESA, the NASA Herschel Science Center, and the HIFI, PACS and SPIRE consortia.}. 

\begin{table}[h!]
\caption{Adopted Colour Corrections and Flux Uncertainties}\label{colourCorr}
\begin{tabular}{lcccc}
\hline\hline
Band & 160 \um\ & 250 \um\ & 350 \um\ & 500 \um \\
\hline
Colour Correction\tablenotemark{a} & 1.01 & 1.02 & 1.01 & 1.03 \\
Colour Uncertainty\tablenotemark{b} &  5\% & 0.8\% & 1\% & 2\% \\
rms$_{\nu}$ (\mJybeam)\tablenotemark{c} & 20 & 18 & 15 & 20 \\
rms$_{36.3}$ (\mJybeam)\tablenotemark{d} & 80 & 60 & 30 & 20 \\
\hline
\end{tabular}
\tablenotetext{a}{\ Average colour corrections for each band, where $S_{corrected} = CC \times\ S_{obs}$. These values assume $T \approx 10 - 15$ K and  $\beta \approx 1.5 - 2.5$. See Appendix \ref{ColourCorrectionAppendix}.}
\tablenotetext{b}{\ Uncertainties refer to the fractional uncertainty in the average colour correction based on the range of accepted colour corrections. These uncertainties do not include the calibration uncertainties.}
\tablenotetext{c}{\ Approximate 1$\sigma$ rms noise in the \emph{Herschel} maps for relatively blank regions at the native resolution of each band.}
\tablenotetext{d}{\ Approximate 1$\sigma$ rms noise in the convolved \emph{Herschel} maps for relatively blank regions at 36.3\arcsec\ resolution.}
\end{table}

Table 1 excludes additional errors from flux calibration.  Flux calibration depends on a number of factors, including the calibration files used in the reduction and the mapmaker.  Based on point sources, the flux calibrations uncertainties are roughly 5\% for PACS\footnote{PACS flux calibration can be found in the ``Assessment analysis of the extended emission calibration for the PACS red channel'', http://herschel.esac.esa.int/twiki/bin/view/Public/PacsCalibrationWeb.} and 7\% for SPIRE\footnote{SPIRE flux calibration can be found in the SPIRE ObserversÕ Manual (2011), HERSCHEL-DOC-0798, v2.4, http://herschel.esac.esa.int/Docs/SPIRE/pdf/spire\_om.pdf}.  Since we are interested in extended emission and our observations were taken with the fast (60 arcsec s$^{-1}$) scan rate resulting in elongated beams, we adopted 10\% flux calibration errors for both PACS and SPIRE.  To simulate these uncertainties, we generated 1000 random correction factors following a Gaussian with a mean of 1.0 and a HWHM of 0.1. Since the flux calibrations are correlated for each instrument (M. Griffin, private communication), we assumed the same flux calibration corrections for the three SPIRE bands and a separate randomly generated set of corrections for the PACS 160 \um\ band. We applied the randomly-selected PACS and SPIRE calibration correction factors appropriately across the maps.  For each pair of correction factors, we fit the resulting SEDs pixel-by-pixel to identify the broad distribution of best-fit SED parameters within the calibration uncertainties.

We fitted the SEDs of individual pixels using the IDL program \emph{mpfitfun}, which performs an iterative least-squares comparison between a set of data and a model function until a best fit is achieved (\citealt{Markwardt09}). We fit each SED with the modified black body function,
\begin{equation}
I_{\nu} = \kappa_{\nu_0}(\nu/\nu_0)^{\beta}B_{\nu}(T)\Sigma \label{modBB}
\end{equation}
where $\kappa_{\nu_0}$ is a reference dust opacity per unit gas and dust mass at reference frequency $\nu_0$, $\beta$ is the dust emissivity power law index, $B_{\nu}$ is the black body function, $T$ is the dust temperature, and $\Sigma= \mu m_H \NHH$ is the gas mass column density of material with mean molecular mass $\mu m_H$ and gas column density \NHH. For consistency with other papers from the \emph{Herschel} Gould Belt Survey (e.g., \citealt{Andre10}), we assumed that $\kappa_{\nu_0} = 0.1$ \cmg\  at $\nu_0 = 1000$ GHz (\citealt{Hildebrand83}). We adopt a mean molecular weight per unit hydrogen mass of $\mu = 2.8$, for a cloud of 71\%\ molecular hydrogen gas, 27\%\ helium, and 2\%\ metals (e.g., \citealt{Kauffmann08}). 

\begin{figure}[h!]
\includegraphics[scale=0.49]{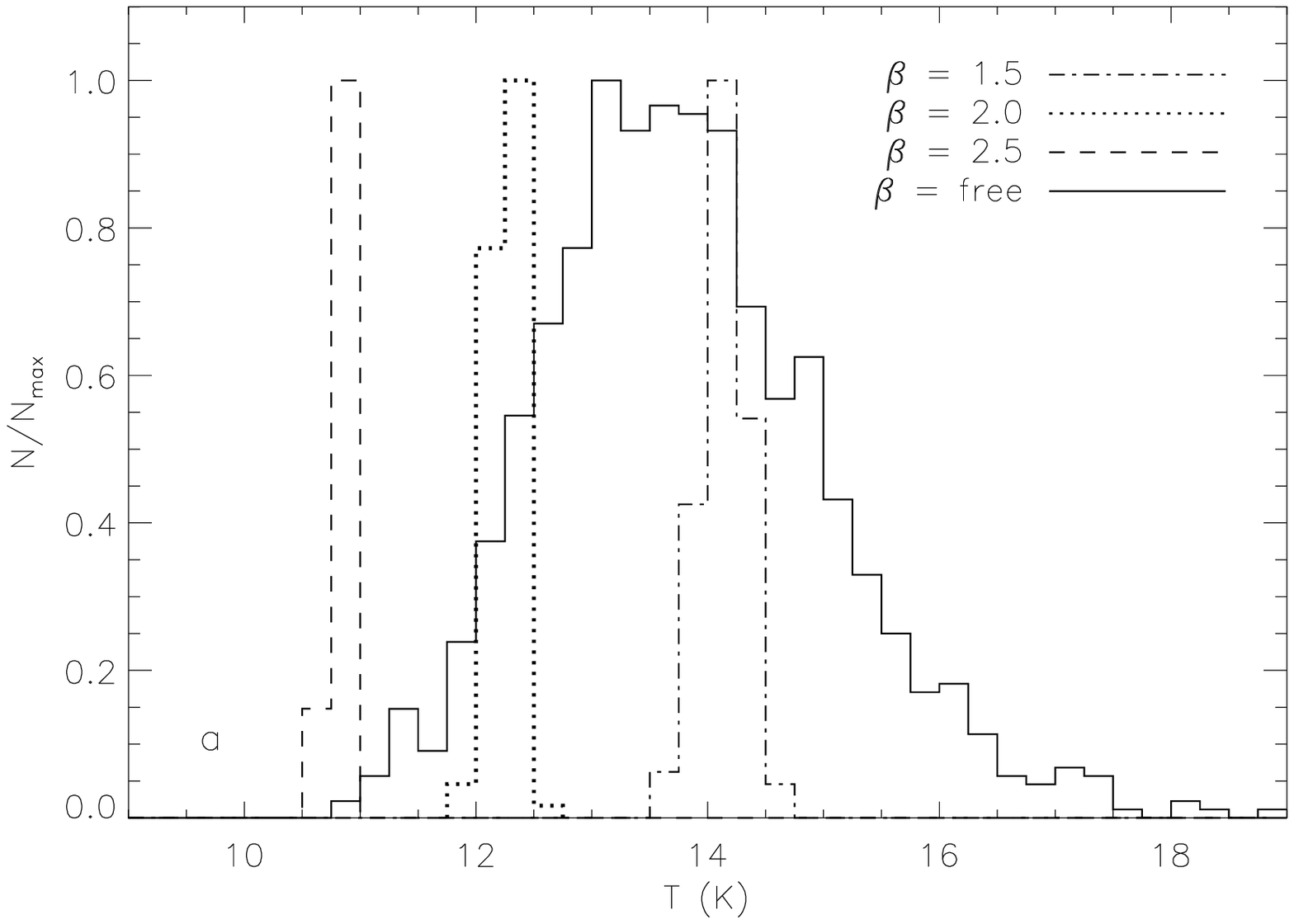}
\includegraphics[scale=0.49]{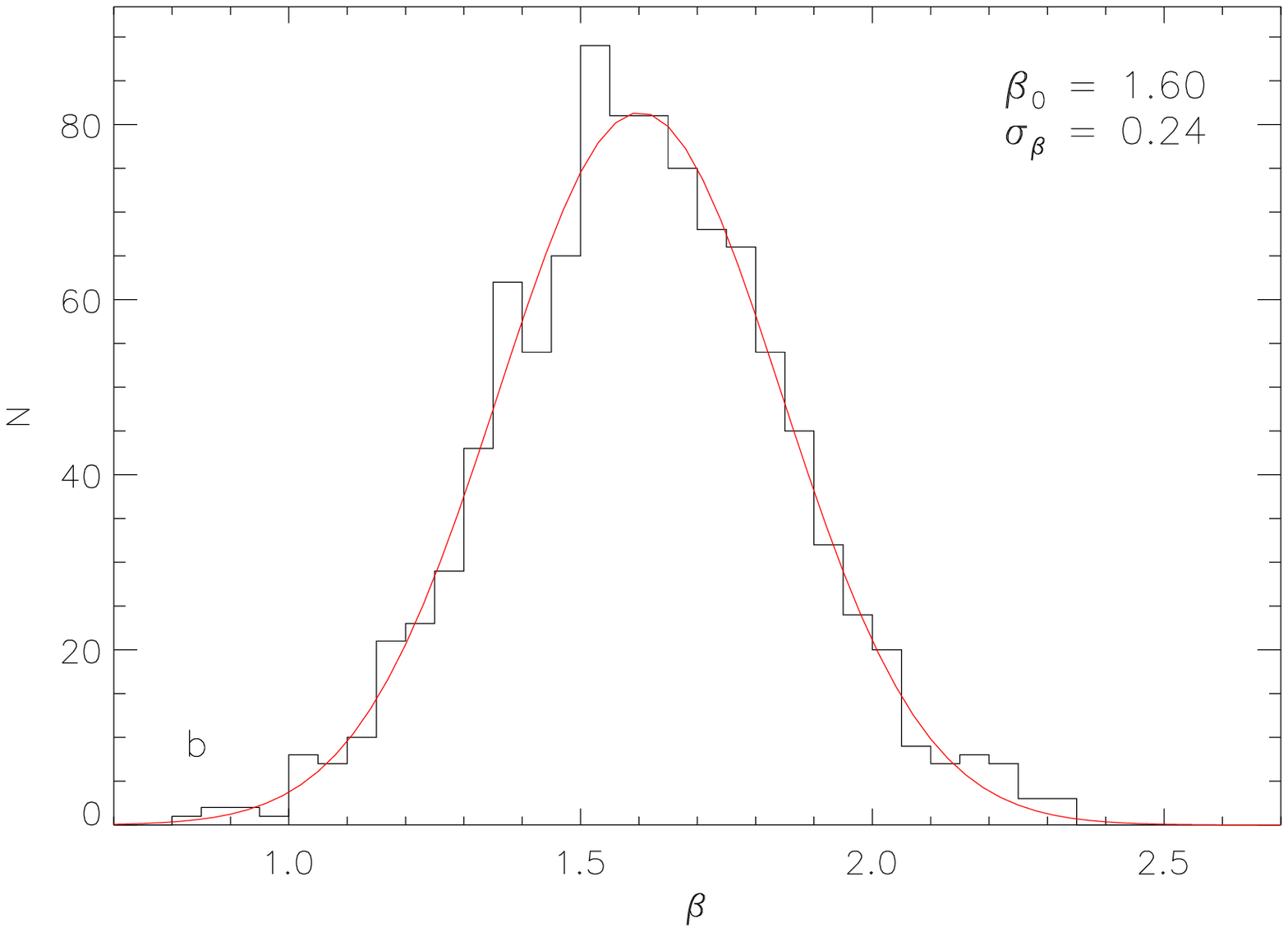}
\caption{Sample results from SED-fitting towards a pixel in the B1-a core. We fit the observed SEDs after applying our colour correction factors and random calibration correction factors (see text). For each calibration correction factor, we fit the modified dust emission with a dusty black body model assuming $\beta = 1.5, 2.0, 2.5$, and free $\beta$. We show the (a) relative temperature distributions for each of these cases of $\beta$ and (b) the $\beta$ distribution when unrestricted. We fit the $\beta$ distribution with a Gaussian (red curve).\label{herschelONLYb2}}
\end{figure}

Most \emph{Herschel} GBS analyses (e.g., \citealt{Andre10}; \citealt{Arzoumanian11}; \citealt{Sadavoy12}) fit SEDs across the $160 - 500$ \um\ \emph{Herschel} bands assuming $\beta = 2$ (\citealt{Hildebrand83}). To test this assumption, we used Equation \ref{modBB} to fit the \emph{Herschel} $160 - 500$ \um\ data assuming  (1) $\beta = 1.5$, (2) $\beta = 2.0$, (3) $\beta = 2.5$, and (4) $\beta$ is a free variable. We applied the colour correction factors (see Table \ref{colourCorr}) and the random distribution of calibration correction factors to the observed emission. For the flux uncertainties, we added in quadrature the colour calibration uncertainties and the map sensitivities (see Table \ref{colourCorr}). Figure \ref{herschelONLYb2} shows the distributions of  SED-derived parameters for a single example pixel towards the B1-a core. For our sample pixel, when $\beta$ is fixed, the temperature distributions are narrow despite the wide range in calibration correction factors. The mean temperature, however, varies from $T \approx 14$ K for $\beta =1.5$ to $T \approx 10.5$ K for $\beta = 2.5$. Allowing $\beta$ to vary produces broad Gaussian-like temperature and $\beta$ distributions with mean values of 13.7 K $\pm\ 1.1$ K and 1.60 $\pm\ 0.24$, respectively. These broad distributions suggest that the \emph{Herschel} calibration uncertainties result in significant uncertainties for the SED parameters. 

\begin{figure}[h!]
\includegraphics[scale=0.8]{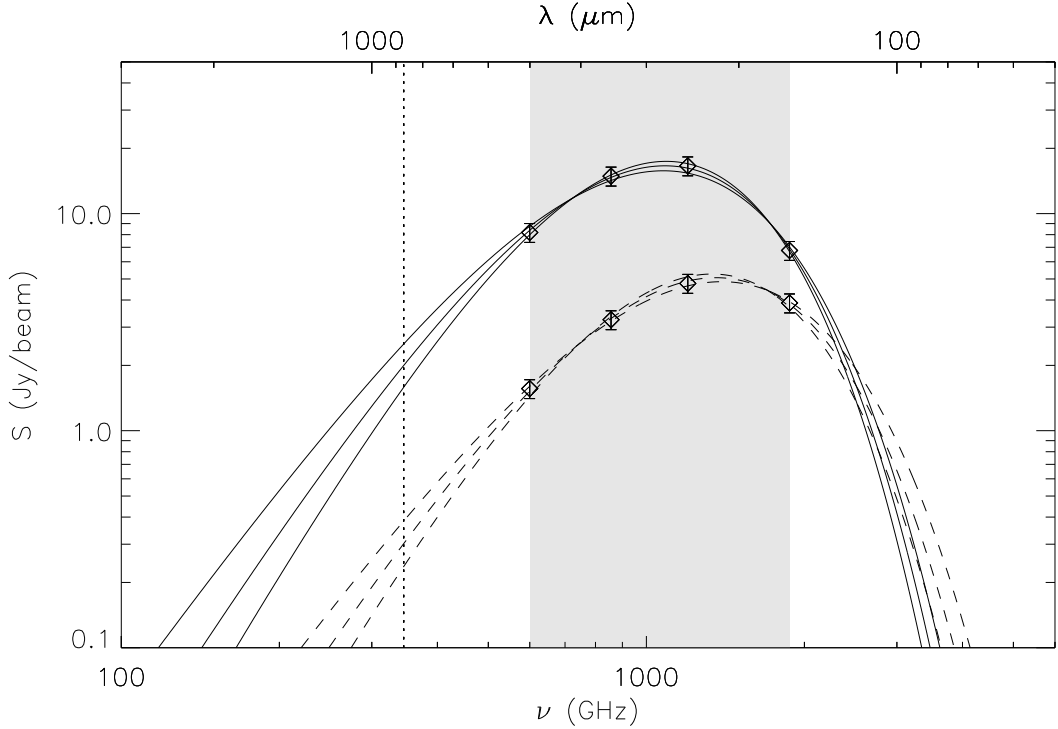}
\caption{Sample modified black body profiles for dust emission corresponding to (solid) $T = 10$ K, $\NHH = 10^{23}$ \cden, and $\beta = 2.25$ and (dashed) $T = 14$ K, $\NHH = 6 \times 10^{21}$ \cden, and $\beta = 1.75$. For both cases, we fit the generated SEDs (diamonds) with (from left to right) $\beta = 1.5, 2.0,$ or 2.5. The \emph{Herschel} $160 - 500$ \um\ bands cover the region highlighted by the solid grey box. The dotted line corresponds to 850 \um. At short wavelengths (i.e., $\lambda \lesssim$ 100 \um), the dust emission may become optically thick, and thus, the SED curves could be independent of $\beta$ at these wavelengths. For simplicity, we do not consider optically thick dust emission at short wavelengths. Thus, $\beta$ cannot be constrained with short wavelength observations and instead, long wavelength observations are necessary. \label{longWave}}
\end{figure}

Furthermore, Figure \ref{longWave} highlights the $\beta$-temperature degeneracy of SED-fitting with the \emph{Herschel} bands only. First, we generated fluxes assuming (1) $T = 10$ K, $\NHH = 10^{23}$ \cden, and $\beta = 2.25$ and (2) $T = 14$ K, $\NHH = 6 \times 10^{21}$ \cden, and $\beta = 1.75$. Second, we fitted the generated \emph{Herschel} band fluxes with our dusty black body models assuming $\beta = 1.5, 2.0,$ and 2.5 and no calibration uncertainties. The solid curves show the best-fit SED models to the dust emission for the first case and the dashed curves show the best-fit SED models for the second case. For both cases, the generated data were moderately well fit by all three values of $\beta$. Note that the \emph{Herschel} bands cover the region illustrated by the grey box where the model curves are largely degenerate. The error bars indicate an uncertainty of 10\%\ to illustrate the range of uncertainty from the calibration. (Recall, instrument calibration is correlated such that the same calibration correction will be required for the three SPIRE bands.)  Conversely, at longer wavelengths, the SED models are more distinct and thus, the SED parameters are much better constrained. Since we have longer wavelength data (i.e., 850 \um), we may circumvent the degeneracy between $\beta$ and temperature (see Section \ref{combo}). 

\section{Determining $\beta$}\label{combo}

As discussed in Section \ref{s2+harp}, the SCUBA-2 observations do not recover large-scale emission and that emission can be significant. For example, the residuals between a prediction of unfiltered 850 \um\ emission (i.e., from SED fitting of the \emph{Herschel} data assuming $\beta=2$) and the observed 850 \um\ dust emission suggest that the observed fluxes at 850 \um\ could be missing $\gtrsim$ 30\%\ of the total emission. Therefore, direct comparisons between PACS+SPIRE and SCUBA-2 observations can be misleading.

In this Section, we explore four different techniques to determine $\beta$ towards B1 using the SCUBA-2 data. In Section \ref{filterH}, we determined $\beta$ using filtered dust emission maps and in Section \ref{recoverS2}, we determined $\beta$ using unfiltered maps.  Since filtered maps do not include any large-scale structure, there is no reason to expect a priori that $\beta$ will be the same for filtered and unfiltered maps. For example, the extended emission may trace more of the warmer, diffuse cloud structure than the cold, dense structures associated with cores, therefore broadening the SED and lowering $\beta$ (\citealt{Martin12}). If true, the absence of the large-scale structure may result in more pronounced variations in $\beta$ towards the small scale features in the filtered maps than in the unfiltered maps.  For all fits to the SEDs, we used the following procedure \emph{unless stated otherwise}:

\begin{description}
\item[1. ] We applied the colour correction factors in Table \ref{colourCorr} to the \emph{Herschel} bands. 
\item[2. ] We adopted flux errors as the quadrature sum of the 1 $\sigma$ rms errors and the uncertainty of the colour corrections (see Table \ref{colourCorr}). 
\item[3. ] We generated different sets of 1000 random calibration correction factors following a Gaussian distribution to represent a 10\%\ or 50\%\ calibration uncertainty. We generated a separate set of calibration correction factors for PACS, SPIRE, SCUBA-2 at 450 \um, and SCUBA-2  at 850 \um.
\item[4. ] We fit SEDs following a minimization of $\chi^2$ routine. We used the modified black body model in Equation \ref{modBB} and the IDL \emph{mpfitfun} routine to minimize the $\chi^2$ fit.
\item[5. ] For the filtered maps (see Section \ref{scuba2}), we only considered regions with $\ge 10\ \sigma$ intensities at 160 - 850 \um\ at 36\arcsec\ resolution (excluding the 450 \um\ band). The filtered maps are only well measured within the masked regions used in the reduction (see Section \ref{scuba2}). Since the 450 \um\ data has a higher 1 $\sigma$ rms noise level ($\sim 100$ \mJybeam), these regions include 450 \um\ emission that is $< 5\ \sigma$.
\end{description}

\subsection{$\beta$ Using Filtered Maps}\label{filterH}

\begin{figure}[h!]
\includegraphics[scale=0.48]{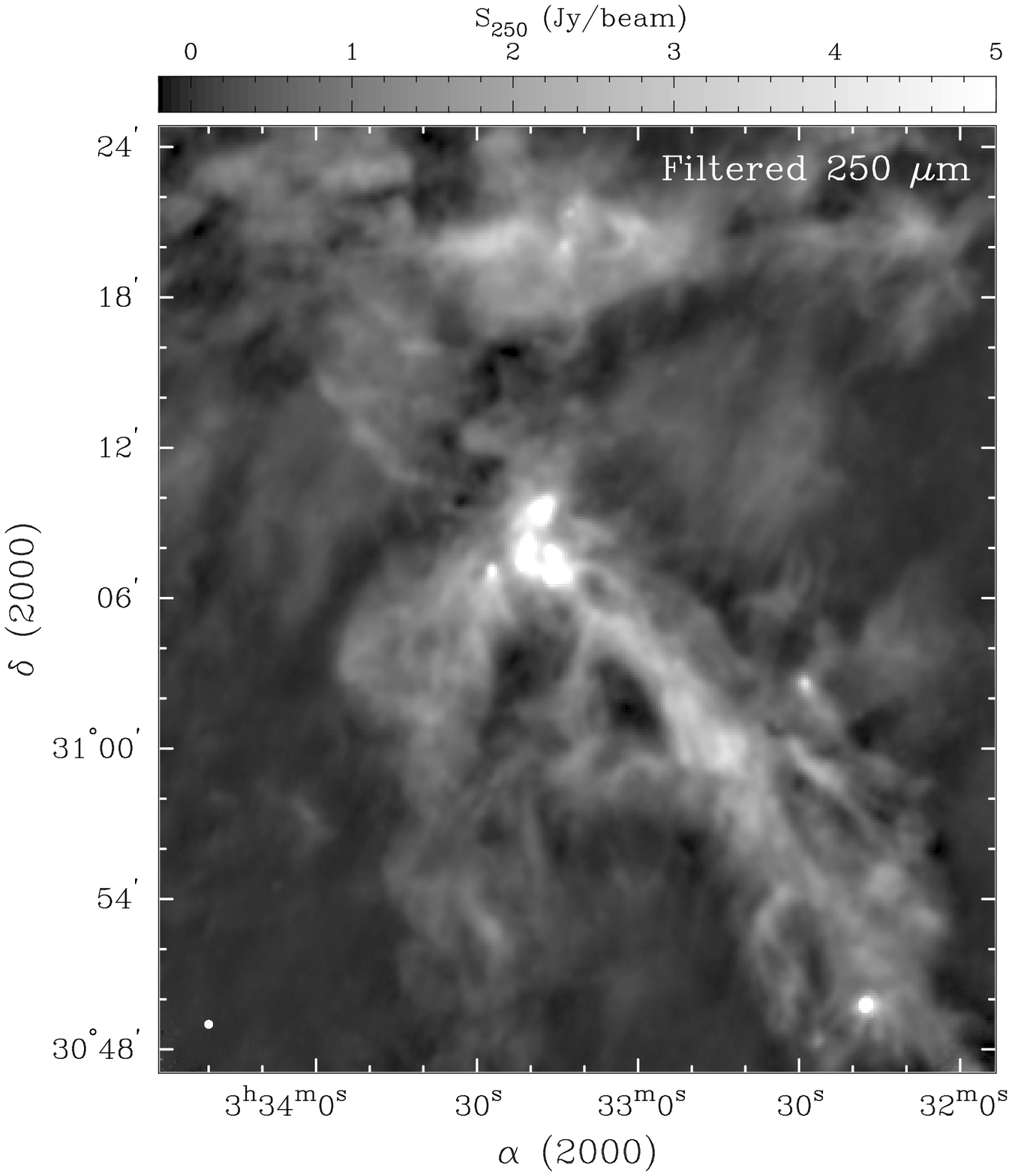}
\includegraphics[scale=0.48]{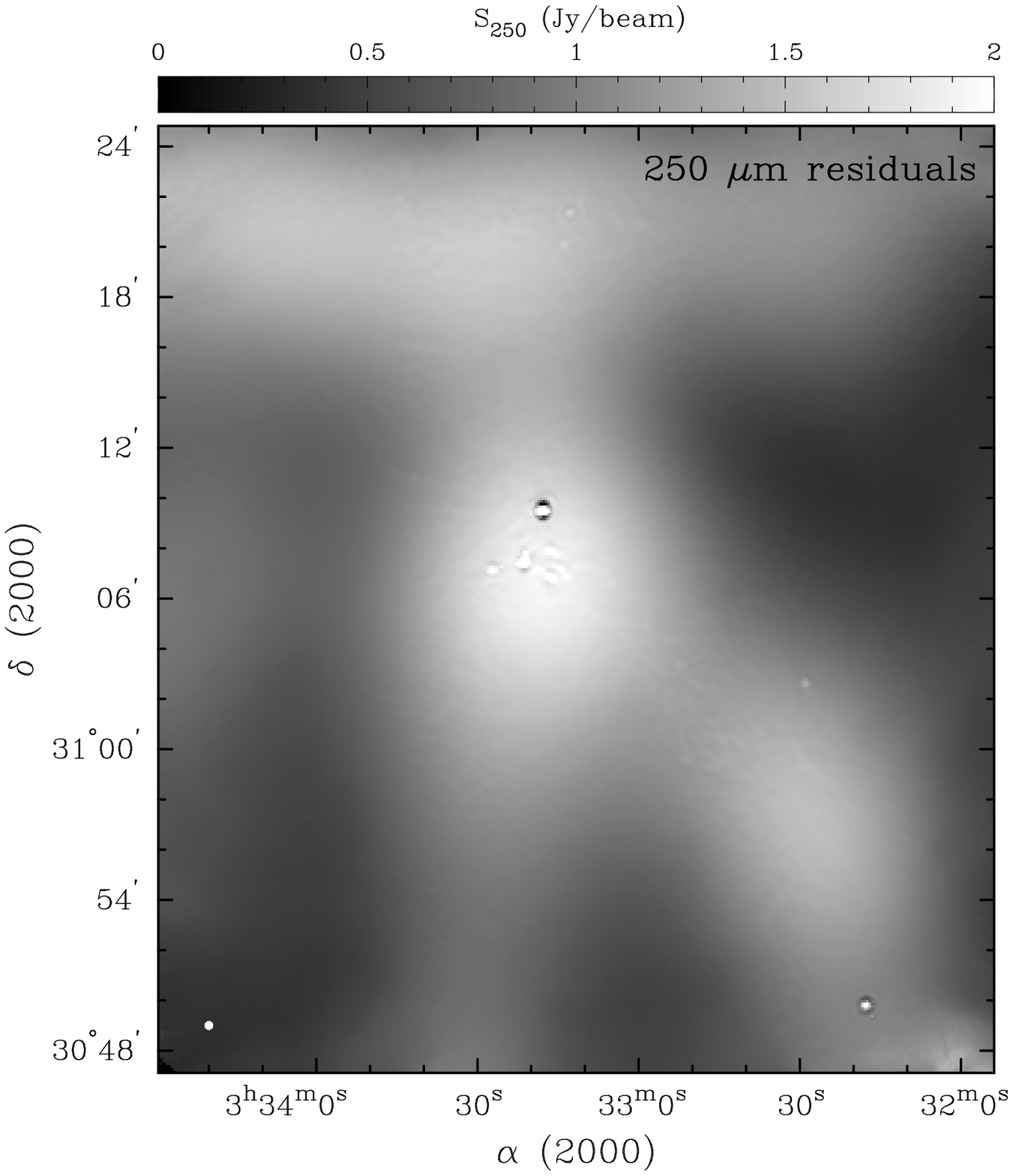}
\caption{Comparison between the filtered \emph{Herschel} map at 250 \um\ and the filtered-out large-scale emission (using the SCUBA-2 mapmaker; see Appendix \ref{filterAppendix}). Map resolutions are 18\arcsec. \label{resH}}
\end{figure}

For our first case (hereafter the Filtered Case), we removed the large-scale emission from the \emph{Herschel} observations by applying the SCUBA-2 reduction pipeline to the observed \emph{Herschel} maps as we did with the CO ($3-2$) line data (see Appendix \ref{filterAppendix} for more information). This technique has the advantage of forcing all the data to include the same range of spatial scales, and thus, $\beta$ and temperature can be determined from directly fitting the SEDs of the $160 - 850$ \um\ emission.  Unlike with our SCUBA-2 data, we can compare the filtered and unfiltered versions of the \emph{Herschel} maps and see how the SCUBA-2 mapmaker removes large-scale emission. Figure \ref{resH} compares the filtered SPIRE 250 \um\ map with the filtered-out large-scale emission. Note that a few artifacts remain at the locations of the brightest cores. These artifacts are very localized and thus, become negligible ($< 5\%$ of the observed emission) when the filtered maps are convolved to 36.3\arcsec\  resolution. As of this writing, the SCUBA-2 reduction pipeline is still evolving. Thus, while some artifacts may be introduced in this version of the SCUBA-2 reduction pipeline, future versions of the pipeline should improve the filtering. 

For our second case (hereafter the SCUBA-2 Ratio Case), we determined $\beta$ using the 450 \um\ and 850 \um\ bands only, i.e., without \emph{Herschel} data (e.g., see \citealt{Hill06}; \citealt{Hatchell12}). Using Equation \ref{modBB}, the ratio of the fluxes at 450 \um\ and 850 \um\ is

\begin{equation}
\frac{I_{450}}{I_{850}} = \left(\frac{\nu_{450}}{\nu_{850}}\right)^{(\beta+3)}\left[\frac{\exp{(h\nu_{850}/kT)}-1.0}{\exp{(h\nu_{450}/kT)}-1.0}\right].\label{S2slope}
\end{equation}

\noindent Therefore, $\beta$ can be determined from the ratio of 450 \um\ to 850 \um\ emission for a given temperature. For simplicity, we find $\beta$ for assumed line-of-sight temperatures. We do not naively expect the line-of-sight temperature to be constant, particularly since the B1 cluster contains known protostellar sources (i.e., B1-c, see Figure \ref{compareB1Cluster}). Additionally, we also assume that the 450 \um\ and 850 \um\ maps include the same range of spatial scales.

\begin{figure}[h!]
\includegraphics[scale=0.9]{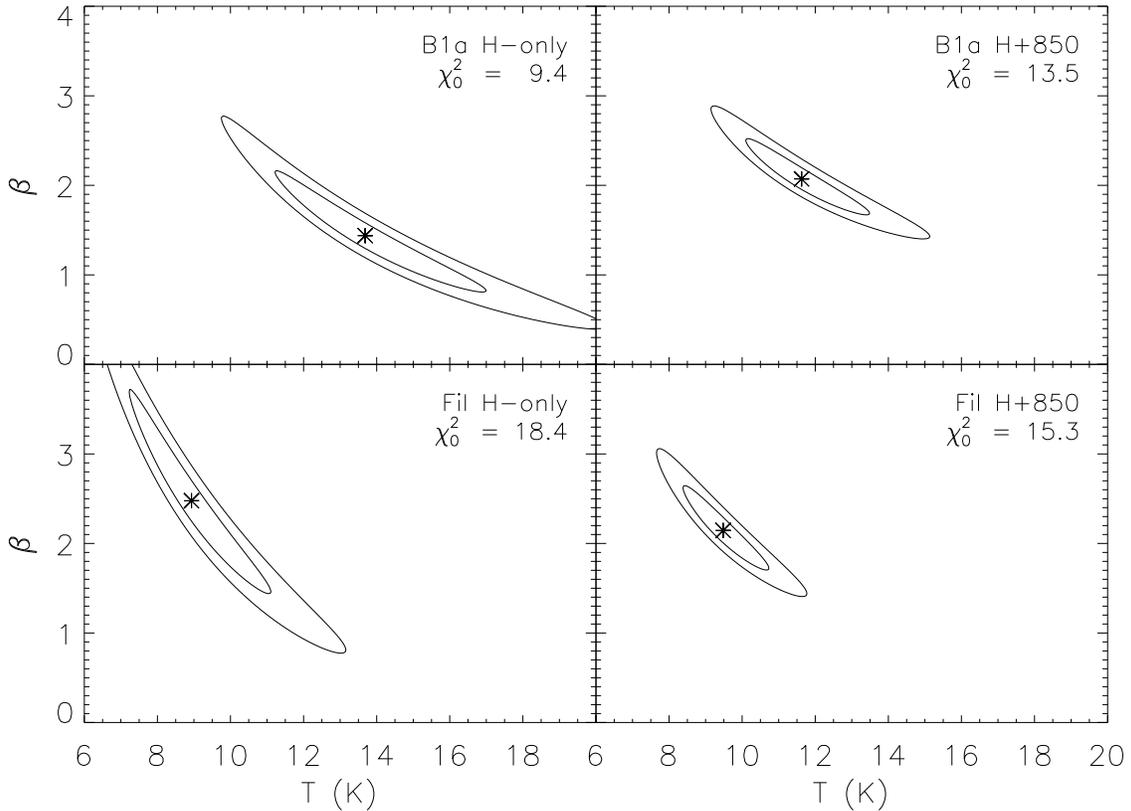}
\caption{Measurements of \emph{relative} reduced $\chi^2$ for assumed values of temperature and $\beta$ for (top) a pixel towards B1-a and (bottom) another pixel towards the brightest part of the filament (see also Figure \ref{filtBetaMaps}). For these plots, we do not assume any calibration uncertainties. The central star corresponds to the best-fit temperature and $\beta$ value from SED-fits to the \emph{Herschel}-only (H-only) data or the \emph{Herschel} with 850 \um\ (H+850) data. The reduced $\chi^2$ measure associated with the best-fit SED, $\chi^2_0$, is shown in the top right corner. The contours correspond to 2$\chi^2_0$ and 4$\chi^2_0$. For the H+850 data, we have four degrees of freedom and for the H-only data, we have three degrees of freedom. \label{contourPlots}}
\end{figure}

For the Filtered Case, Figure \ref{contourPlots} shows contour plots of reduced $\chi^2$ assuming temperatures of 5 K $\le T \le 20$ K and $\beta$ values of $0.5 \le \beta \le 4.0$. These plots highlight the uncertainties from the minimization of $\chi^2$ technique only. For clarity, we do not include the calibration uncertainties for these plots. (We note, however, that the calibration uncertainties will only increase the accepted range of beta and temperature values and would not change the behaviour of the SED fitting itself.)  For simplicity, we show results for a pixel towards B1-a and a pixel towards the brightest part of the filament, though these sample pixels represent the typical behaviour for B1-a and the filament. We compare the results from SED-fitting to the \emph{Herschel} bands alone (H-only) and the \emph{Herschel} bands with the 850 \um\ data (H+850), where the contours correspond to 2$\chi^2_0$ and 4$\chi^2_0$, and $\chi^2_0$ is the reduced $\chi^2$ measure from the best-fit SED to the corresponding data.  Thus, Figure \ref{contourPlots} represents the \emph{relative} uncertainties in $\beta$ and temperature from SED fits using the minimization of $\chi^2$ technique. \emph{By adding the 850 \um\ flux as a new constraint, we see significant improvements in our ability to determine $\beta$ and temperature}. For example, we find improvements in the uncertainty of $\beta$ by a factor of $\sim 2$, and improvements in the uncertainty of temperature by $\sim$ 40\% with the 850 \um\ data.

\begin{figure}[h!]
\includegraphics[scale=0.775,angle=-90]{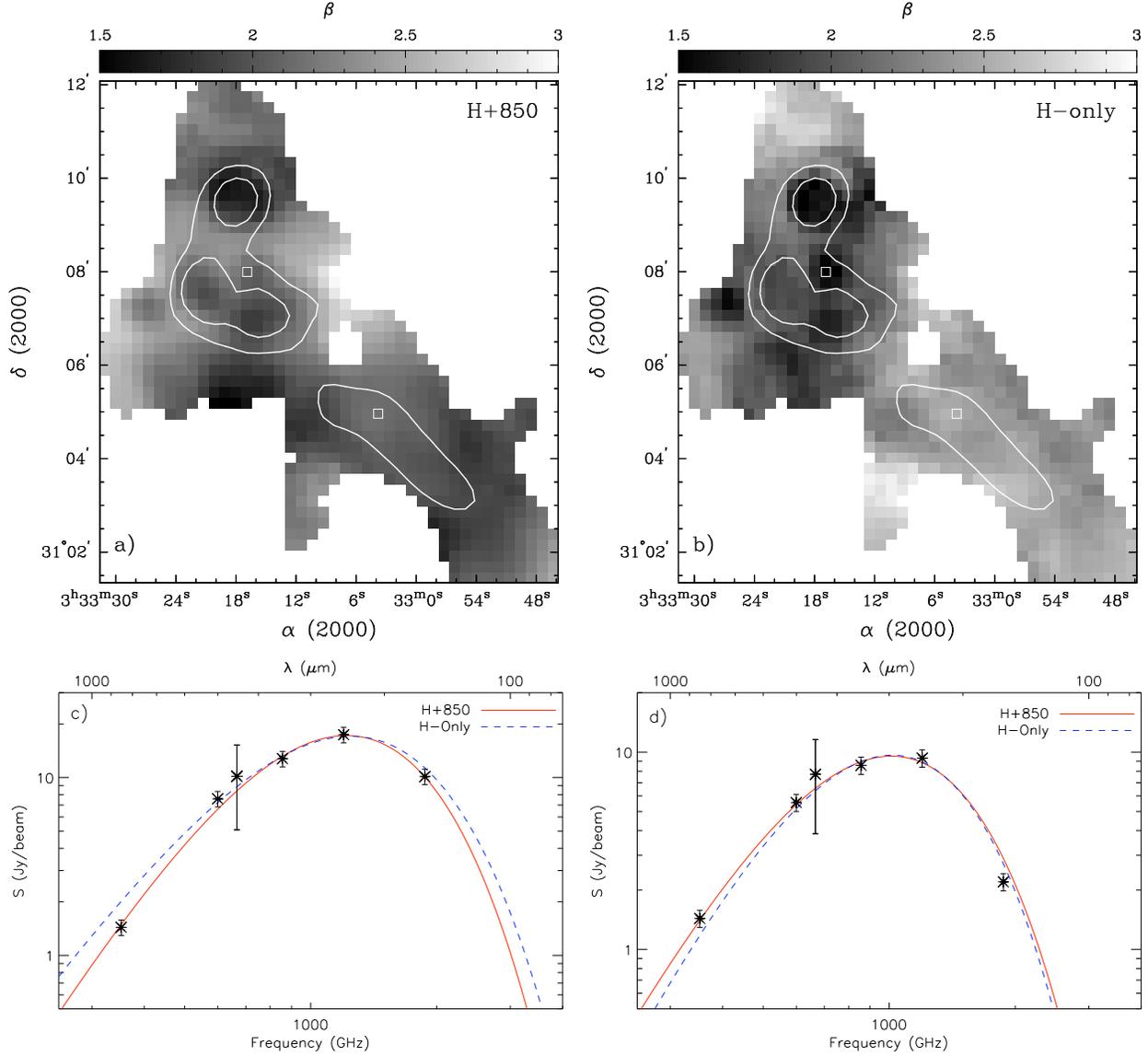}
\caption{Dust emissivity maps towards the B1 main cluster for the Filtered Case. White contours show 850 \um\ flux levels of 1 \Jybeam\ and 2 \Jybeam\ at 36\arcsec\ resolution. \emph{Bottom:} Sample SEDs from (c) B1a and (d) the south-west filament, each marked by squares in the top panel. The curves show the best-fit SEDs. For the solid curves, we used the H+850 bands, whereas for the dashed curves, we used the H-only bands. The error bars on the data points illustrate the range in flux from the calibration uncertainties. The 450 \um\ band is included for comparison; it was not used in the fitting.\label{filtBetaMaps}}
\end{figure}

In addition to temperature and $\beta$ uncertainties from the fitting technique, each instrument has a 10\%\ flux calibration error. As in Section \ref{sedH}, we applied calibration correction factors within a Gaussian distribution to represent the flux calibration errors for each instrument. Figure \ref{filtBetaMaps} shows the mean $\beta$ values and two example SEDs for the H-only and H+850 cases. For the H-only case, we found more extreme values of $\beta$ towards the dense cores and filament. For example, the H-only fits give $\beta \approx$ 1.5 towards B1-a and $\beta \approx 2.5$ towards the filament, whereas the H+850 fits give $\beta \approx 2.0$ towards both. Similarly, the H-only fits overestimated the temperature by $\sim 2$ K towards B1-a and underestimated the temperature by $\lesssim 1$ K towards the filament compared to the H+850 results. In the bottom panels of Figure \ref{filtBetaMaps}, we show the SEDs for the sample pixels towards B1-a and the south-west filament, corresponding to the same example pixels in Figure \ref{contourPlots}. With the 850 \um\ band included, we found a significant difference in the absolute value of $\beta$ (see also Figure \ref{contourPlots}), though the uncertainties related to the calibration correction factors are nearly identical. 
 
\begin{figure}[h!]
\includegraphics[scale=0.47]{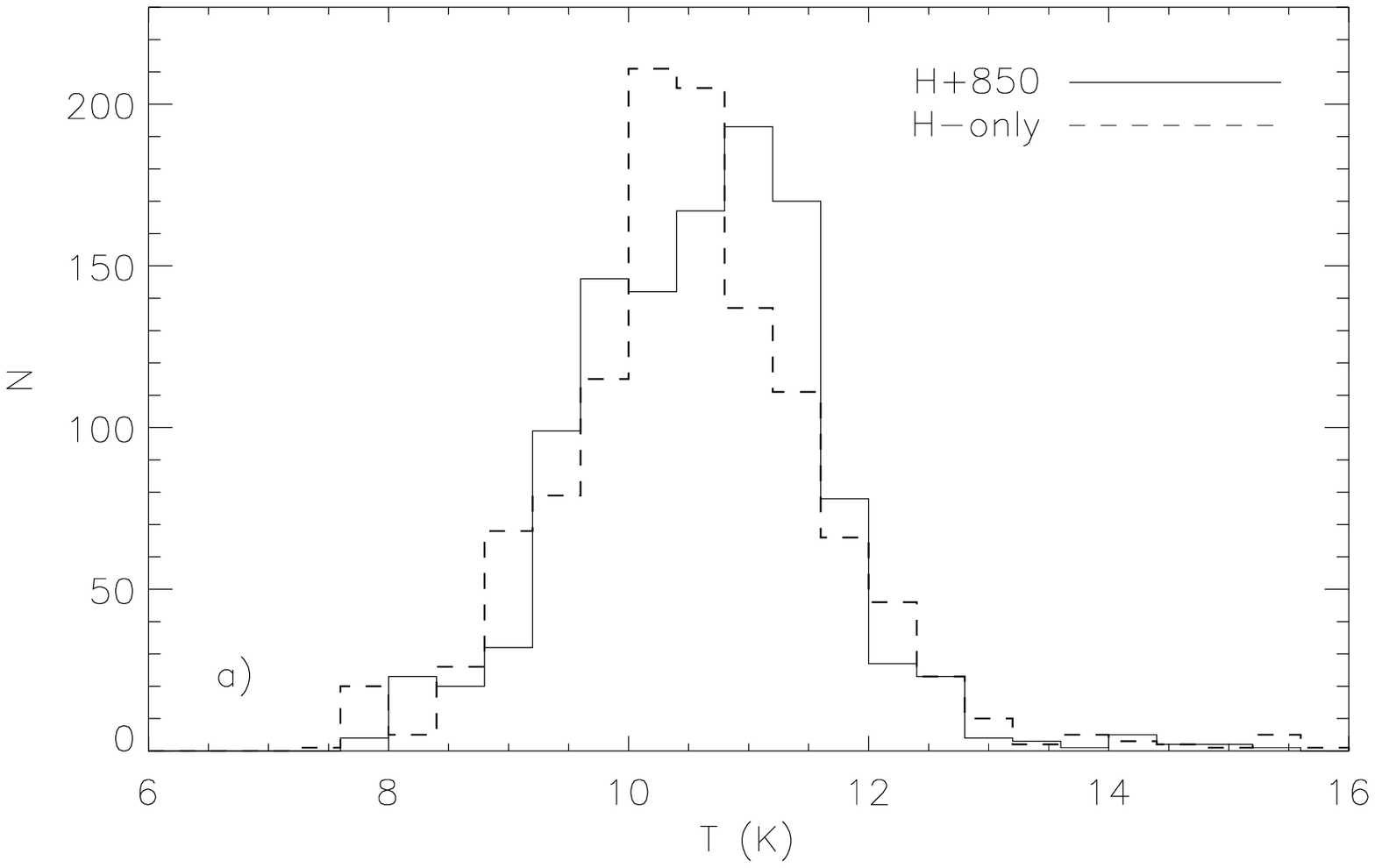}
\includegraphics[scale=0.47]{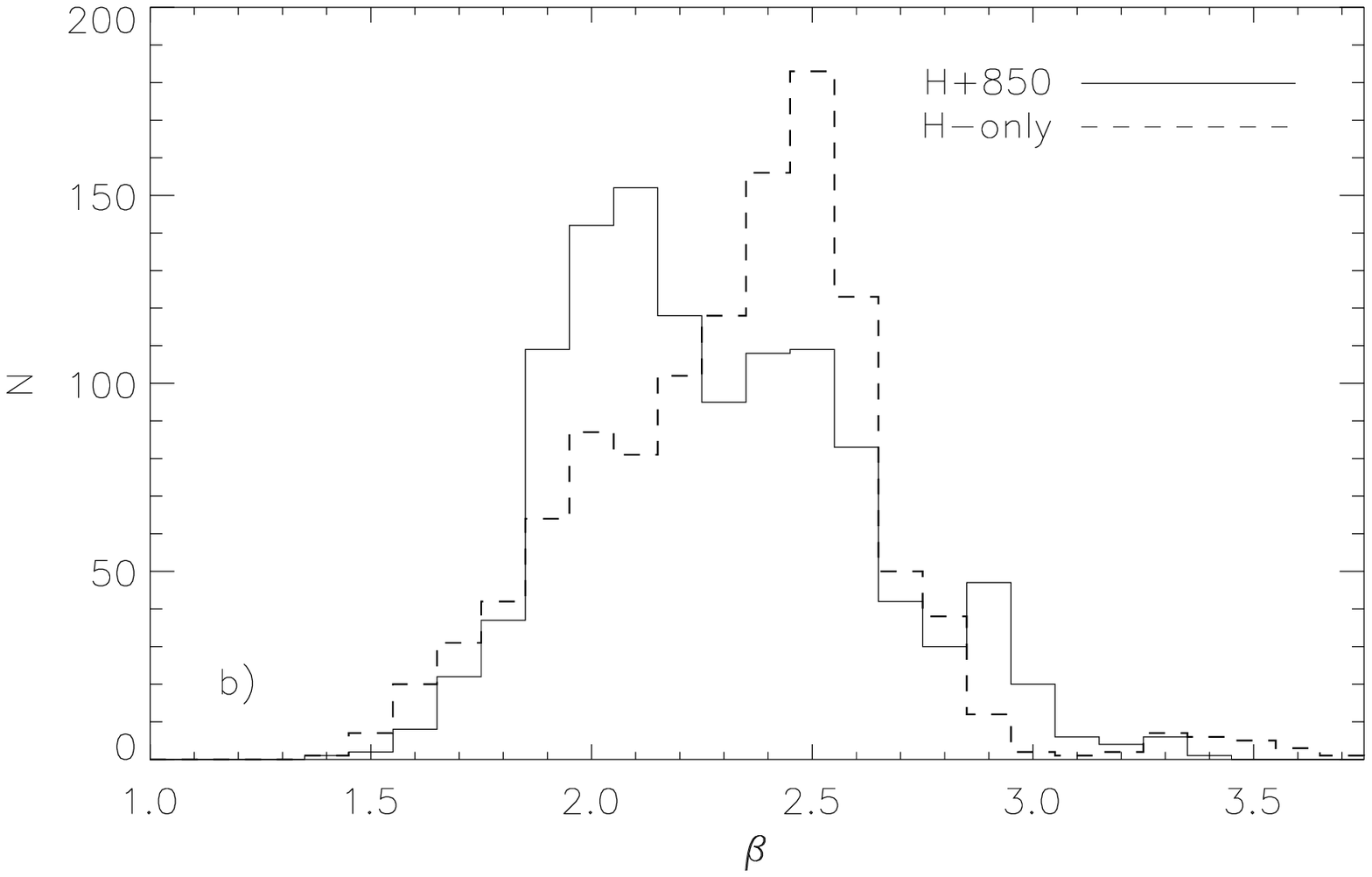}
\caption{Histograms of (a) temperature and (b) $\beta$ for the B1 complex. For comparison, the temperature and $\beta$ values correspond to SED fits when the SCUBA-2 850 \um\ band is included (solid lines) or excluded (dashed lines) with the filtered \emph{Herschel} maps (Filtered Case). In both cases, we only consider pixels with $\ge 10\ \sigma$ fluxes in all five bands.  \label{compareDist}}
\end{figure}

Figure \ref{compareDist} shows the distributions of temperature and $\beta$ for the Filtered Case. These plots highlight the change in absolute value of $\beta$ between the H+850 and H-only fits. For the mean $\beta$ distribution (Figure \ref{compareDist}b), the H-only $\beta$ distribution peaks at $\beta \approx 2.5$ and the H+850 distribution peaks at $\beta \approx 2.0$. For both the H-only and H+850 fits the 1 $\sigma$ standard deviation errors based on the calibration uncertainties alone are $\sigma_T \lesssim 1.5$ K for temperature and $\sigma_{\beta} < 0.3$ for $\beta$. Note that these errors are typically less than the errors from the $\chi^2$ fitting (Figure \ref{contourPlots}), suggesting that the constraints on the SED-fitting are more significant than uncertainties in the flux calibration. For larger calibration uncertainties at 850 \um, however, the uncertainties in temperature and $\beta$ due to the calibration becomes more significant.  For example, if we were to assume 50\% calibration errors at 850 \um, the 1 $\sigma$ standard deviation errors would become $\sigma_T \lesssim 3.7$ K and $\sigma_{\beta} \lesssim 0.8$, and thus, we cannot  constrain well either parameter.  Therefore, long wavelength observations can improve the SED-fitting only if the calibration uncertainties are relatively low, as achieved with SCUBA-2 at 850 \um.

\begin{figure}[h!]
 \centering
 \includegraphics[scale=0.5]{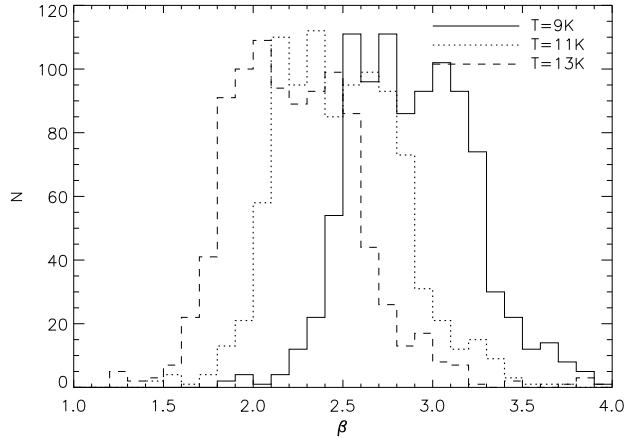}
\caption{Distributions of $\beta$ for fixed temperatures of 9 K, 11 K, and 13 K, and using Equation \ref{S2slope} from the SCUBA-2 Ratio Case. \label{betaComp}}
\end{figure}

Alternatively, with the SCUBA-2 Ratio Case, we can determine $\beta$ if the temperature is known (see Equation \ref{S2slope}). For simplicity, we assumed fixed temperatures across the entire map and for comparison with the Filtered Case, we used the 36.3\arcsec\ convolved maps at 450 \um\ and 850 \um. Additionally, we generated 1000 random calibration correction factors within a Gaussian distribution assuming a calibration uncertainty of 50\%\ at 450 \um\ and 10\%\ at 850 \um. Figure \ref{betaComp} compares the resulting mean $\beta$ distributions from the SCUBA-2 Ratio Case for the assumed temperatures of $T = 9$ K, 11 K, and 13 K.  While the $\beta$ distributions appear to have very similar structures, the distributions themselves are clearly shifted with respect to one another. Using a Kolmogorov-Smirnov test, the $\beta$ distributions at $T = 9$ K and $T = 11$ K agree within 90\%\ for a shift of $\Delta\beta \approx 0.4$, and the distributions at $T = 11$ K and $T = 13$ K agree within 90\%\ for a shift of $\Delta\beta \approx 0.27$.  Furthermore, the high calibration uncertainty at 450 \um\ results in a standard deviation error of $\sigma_{\beta} \approx 0.55$ across the B1 clump regardless of the temperature.

SCUBA-2 observations alone cannot constrain both $\beta$ and temperature. If the temperature or $\beta$ across a cloud is roughly constant, then we can probe relative variations using the SCUBA-2 Ratio Case. For regions with known prestellar and protostellar objects like the B1 complex, it is not ideal to assume constant temperatures or $\beta$ indices, however. For example, Figure \ref{compareDist} reveals a $\sim 6$ K variation in temperature within the central B1 region. A difference of 6 K in temperature results in a significant uncertainty of $\Delta\beta \approx 0.86$. Similarly, when we fixed $\beta$, we found that the temperature distributions shifted by $\sim 2$ K between $\beta = 1.5, 2.0,$ and 2.5 (see Figure \ref{herschelONLYb2}). Thus, Figures \ref{herschelONLYb2} and \ref{betaComp} illustrate the caution needed for analyses which assume either a fixed $\beta$ or a fixed temperature. In either case, a slight increase or decrease in the fixed parameter can result in significant differences in the determined quantity.

\begin{figure}[h!]
\centering
\includegraphics[scale=0.6]{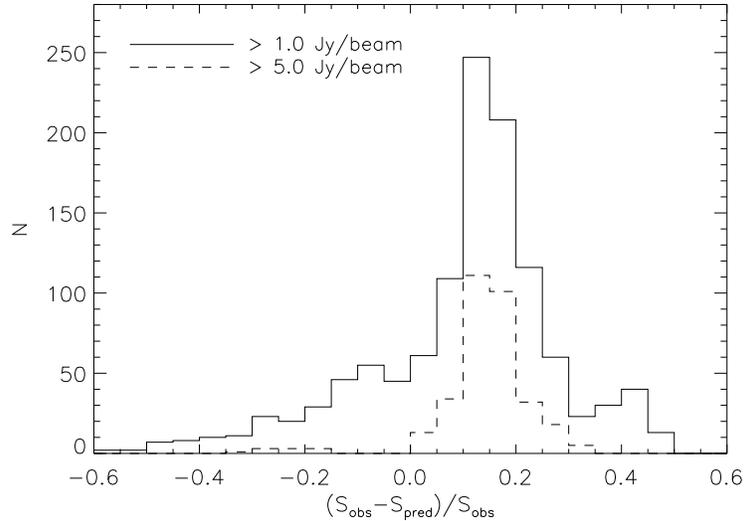}
\caption{Fractional differences between the observed and predicted 450 \um\ emission for pixels with bright ($> 10\ \sigma$) observed 450 \um\ fluxes. These results correspond to the emission predicted from SED fits following the Filtered Case for the H+850 bands. \label{percDiff450}}
\end{figure}

Additionally, in Figure \ref{filtBetaMaps} the SEDs show that the observed 450 \um\ fluxes are roughly 15\%\ greater than those predicted from the SED fits, suggesting that the observed 450 \um\ emission towards these pixels is relatively much brighter than the filtered 500 \um\ emission. Figure \ref{percDiff450} shows the fractional flux increase in the observed 450 \um\ data over the 450 \um\ emission predicted from the best-fit SEDs from the Filtered Case. The ratio of observed 450 \um\ emission to predicted 450 \um\ emission peaks around 10-15\%. (As described in Appendix \ref{COAppendix}, we believe this ``excess'' at 450 \um\ is not due to CO (6-5) contamination.) As a test, we scaled the 450 \um\ emission down by 12.5\%\ and found that $\beta$ decreased by $\Delta\beta \approx 0.18$ based on a Kolmogorov-Smirnov test. These relative fractions may indicate that the 450 \um\ map does not trace the same material as the 850 \um\ map, e.g., due to different spatial scales filtered-out in the reduction or due to temperature variations along the line of sight. Alternatively, these variations could be related to uncertainties in the 450 \um\ fluxes themselves, e.g., from atmospheric variabilities or the beam pattern. Since we cannot assume a single line-of-sight temperature for the B1 clump and the 450 \um\ emission itself is very uncertain, we believe the SCUBA-2 Ratio Case method is less robust than the Filtered Case method.

\subsection{$\beta$ Using Unfiltered Maps}\label{recoverS2}

For the next two techniques, we determined $\beta$ after attempting to recover the missing large-scale emission filtered-out in the SCUBA-2 bands. In Figure \ref{resH}, the median filtered-out large-scale emission at 250 \um\ corresponds to $\sim 1$ \Jybeam\ at 18\arcsec\ resolution (or $\sim 3$ \Jybeam\ at 36.3\arcsec\ resolution). This extended emission peaks towards the central B1 cluster, with additional emission along the south-west filament and to the north. We could naively expect a similar large-scale distribution at 450 \um\ and 850 \um. 

We considered two methods to recover the missing extended emission. The first method, hereafter the Offset Case, seeks to identify the missing large-scale emission at 850 \um\ using a grid of incremental offset values that were added to the observed 850 \um\ map to represent the filtered-out extended emission. We adopted the flux increment that best fit the overall SED to represent the missing emission. For more details, see Appendix \ref{OffsetAppendix}.

The second method, hereafter the Spatial Factor Case, assumes that the observed 450 \um\ and 850 \um\ data correspond to the same spatial scales and that both bands are missing the same fraction of extended emission. For the Spatial Factor Case, we modify the dusty black body function in Equation \ref{modBB} as:
 \begin{equation}
 I_{\nu} = \kappa_{\nu_0}(\nu/\nu_0)^{\beta}B_{\nu}(T)\Sigma F_{SF} \label{modBBv2}
\end{equation}
 where $F_{SF}$ is a scaling factor corresponding to the fraction of recovered emission. For the \emph{Herschel} bands, $F_{SF} = 1$ (i.e., 100\%\ of the emission is recovered), whereas for the SCUBA-2 bands, $F_{SF} \le 1$, and we assume $F_{SF}$ is identical for both the 450 \um\ band and the 850 \um\ band. For more details, see Appendix \ref{SFAppendix}.

Recovering the large-scale emission is very uncertain, and we found large uncertainties in our analyses.  For the Offset Case, we found uncertainties of 20-50\% for the missing extended emission alone, suggesting that we cannot constrain well the diffuse emission.  Thus, the Offset Case cannot be used to constrain $\beta$ from SED fits. For the Spatial Factor Case, the 450 \um\ emission is less certain due to a larger calibration uncertainty of 50\% and as Figure \ref{percDiff450} demonstrates, the observed 450 \um\ emission tends to be 10-15\%\ brighter than the predicted (filtered) emission. Since the scaling factor is very dependent on both the 450 \um\ and 850 \um\ fluxes,  the large flux uncertainties at 450 \um\ will greatly affect the final results.  Finally, based on the filtered \emph{Herschel} maps (see Section \ref{filterH}), the fraction of emission lost in the filtering appears to vary considerably with wavelength. Therefore, $F_{SF}$ may actually vary somewhat between the 450 \um\ and 850 \um\ bands. For the $\beta$ results using these two techniques, see Appendix \ref{OffsetAppendix} and Appendix \ref{SFAppendix}.

\subsection{Comparison of the Techniques}\label{methodComp}

The Filtered Case is the most robust technique to combine SCUBA-2 data with PACS and SPIRE data. For the Filtered Case, we have demonstrated that the 850 \um\ band makes a significant improvement to the uncertainties associated with SED fitting (see Figure \ref{contourPlots}) and impacts the absolute value of $\beta$. For example, we determined that the H-only results generally underestimated $\beta$ towards the dense cores and overestimated $\beta$ along the filament. These observations are in agreement with a recent study by \citet{Kelly12}, which suggests that SED fitting through the minimization of $\chi^2$ may be biased towards lower $\beta$ values for prestellar cores.

Using the Filtered Case technique, we find $\beta \approx 1.6 - 2$ for the dense cores with the lowest values towards the B1-c core. In addition, we find $\beta \approx 2$ along the filament and $\beta \gtrsim 2$ towards the moderately dense clump material. Similarly, \citet{Friesen05} found $\beta \approx 1.3 - 2.1$ using a Fourier Transform Spectrograph at 350 GHz to observe several hot molecular cores. Other studies, however, have identified larger $\beta$ values towards dense cores. For example, \citet{Schnee10} found $\beta \approx 1.7 - 2.7$ towards a starless core using mid-infrared and (sub)millimeter continuum observations, and \citet{Shirley11} found $\beta \approx 2.2 - 2.6$ towards a young protostellar core using radiative transfer models with SCUBA flux ratios and near-infrared colour excesses. Additionally, \citet{Arab12} found $\beta \sim 2$ values towards the cooler ($\sim 40$ K), denser regions of the Orion bar and $\beta \sim 1.1$ towards the warmer ($\sim 70$ K) regions using \emph{Herschel} and Spitzer observations. Thus, a larger study of dense cores and their environments with the same set of observations is necessary to understand these differences in $\beta$.

With the remaining three techniques, $\beta$ is unclear. For the SCUBA-2 Ratio Case (Section \ref{filterH}), a minor variation in temperature greatly affects the measure of $\beta$; for the Offset Case (Section \ref{recoverS2} and Appendix \ref{OffsetAppendix}), the \emph{Herschel} calibration uncertainties result in very uncertain 850 \um\ offsets; and for the Spatial Factor Case (Section \ref{recoverS2} and Appendix \ref{SFAppendix}), the 450 \um\ band is very uncertain, and thus, causes a wide range of plausible scale factors. Additionally, it is not clear that the 450 \um\ and 850 \um\ observations trace the same scales. Both the SCUBA-2 Ratio Case and the Spatial Factor Case require the 450 \um\ band. 

In Section \ref{discussion}, we discuss the results for the Filtered Case only. For comparison, we will consider both the H-only and H+850 results.

%%%%%%%%%%%%%%%%%%%%%%%%%%%%%%%%
%%%%%%%%%%%     Discussion    %%%%%%%%%%%%%
%%%%%%%%%%%%%%%%%%%%%%%%%%%%%%%%
\section{Discussion}\label{discussion}

\subsection{Submillimeter Dust Opacity and Dust Masses}\label{kappaSection}

We are interested in comparing the dust opacity between the dense cores and the ambient clump material. In Equation \ref{modBB}, we assumed that the dust opacity follows a power-law distribution for a reference opacity, $\kappa_{\nu_0}$, at 1000 THz (300 \um). Ideally, we would want to fit $\kappa_{\nu_0}$ as well as $\beta$. According to \citet{Ossenkopf94}, the dust opacity at 300 \um\ increases by a factor of $\lesssim 2$ for coagulated dust grains with icy mantles (this effect is more significant for grains without icy mantles). More recently, \citet{Martin12} and \citet{Roy13} used submillimeter dust emission and near-infrared extinction maps to constrain the submillimeter dust and gas opacity at 250 \um\ for $\beta = 1.8$. Both studies found a steady increase in dust opacity (by factors of $\sim 2-4$) towards colder temperatures, suggesting that the dust opacity should not be fixed at 300 \um. 

Unfortunately, we cannot separate out the degeneracy of $\kappa_{\nu_0}$ with column density in B1 at this time. Although $\kappa_{\nu_0}$ can be constrained by independent measurements of column density, we are interested in changes to the dust over small scales (i.e., between dense cores), and as such we cannot utilize extinction maps (at $\sim 2\arcmin$ resolutions) as done by \citet{Martin12} and \citet{Roy13}. Therefore, while we cannot measure absolute values of $\kappa_{nu}$ with our observations, we can probe the variations in $\kappa_{nu}$ due to our differences in $\beta$.  Thus, we focus on the relative variations in $\kappa_{nu}$ with the H+850 data and the significance of assuming $\beta=2$, as with most \emph{Herschel} analyses of molecular clouds. 

\begin{figure}[h!]
\centering
\includegraphics[scale=0.6]{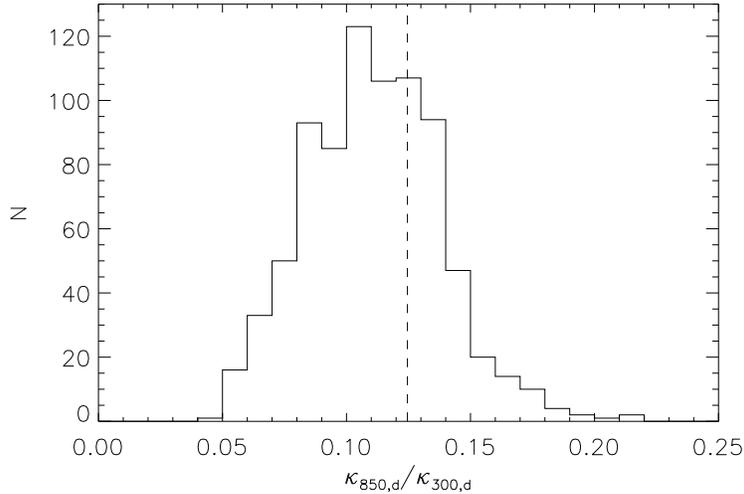}
\caption{Distribution of the dust opacity at 850 \um\ relative to the dust opacity at 300 \um\ using the Filtered Case. The dashed line shows the relative dust opacities for $\beta = 2$.}\label{kappaDist}
\end{figure}

Figure \ref{kappaDist} shows the dust opacity at 850 \um\ relative to the opacity at 300 \um\ for the Filtered Case. This measure does not assume any prior knowledge of the dust opacity at either 300 \um\ or 850 \um.  Thus, Figure \ref{kappaDist} demonstrates that the relative dust opacity at 850 \um\ increases by a factor of $2-3$ between the moderately dense clump material and the dense B1-c core, which suggests the grains are evolving. Dust coagulation alone can increase the submillimeter dust opacity by factors of 4-5 at 850 \um, though this increase is enhanced for bare dust grains and suppressed with surface ices (\citealt{Ossenkopf94}). Alternatively, \citet{Ossenkopf94} noted that moderately elliptical dust grains or higher abundances of metal compounds can increase submillimeter dust opacities at 850 \um\ by factors of a few. Without an independent measure of the dust opacity or column density, however, we cannot directly measure the absolute variations in dust opacity. Thus, further observations of the clump chemistry and dust are necessary to help determine the causes of these opacity variations.

\begin{figure}[h!]
\includegraphics[scale=0.48]{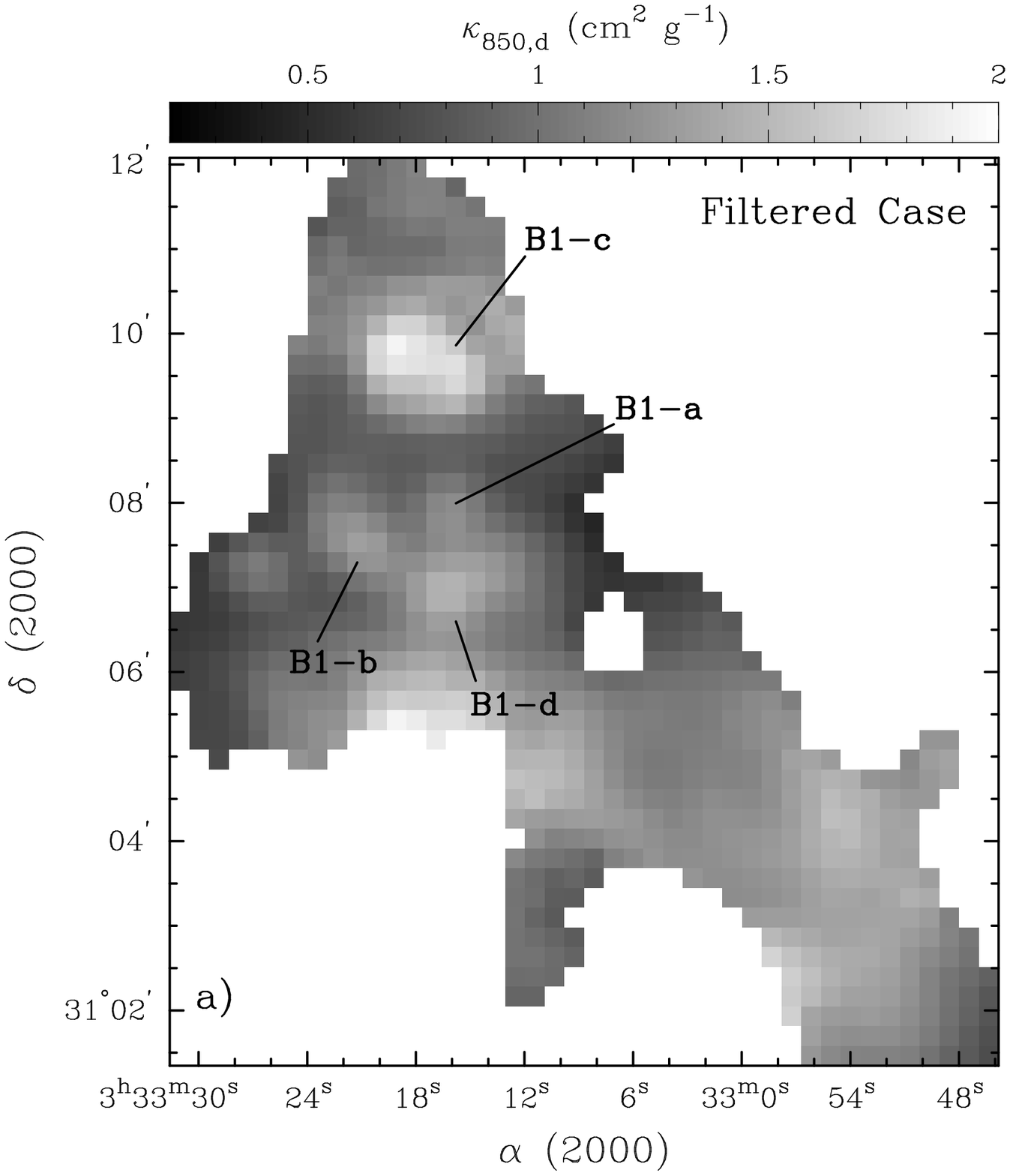}
\includegraphics[scale=0.48]{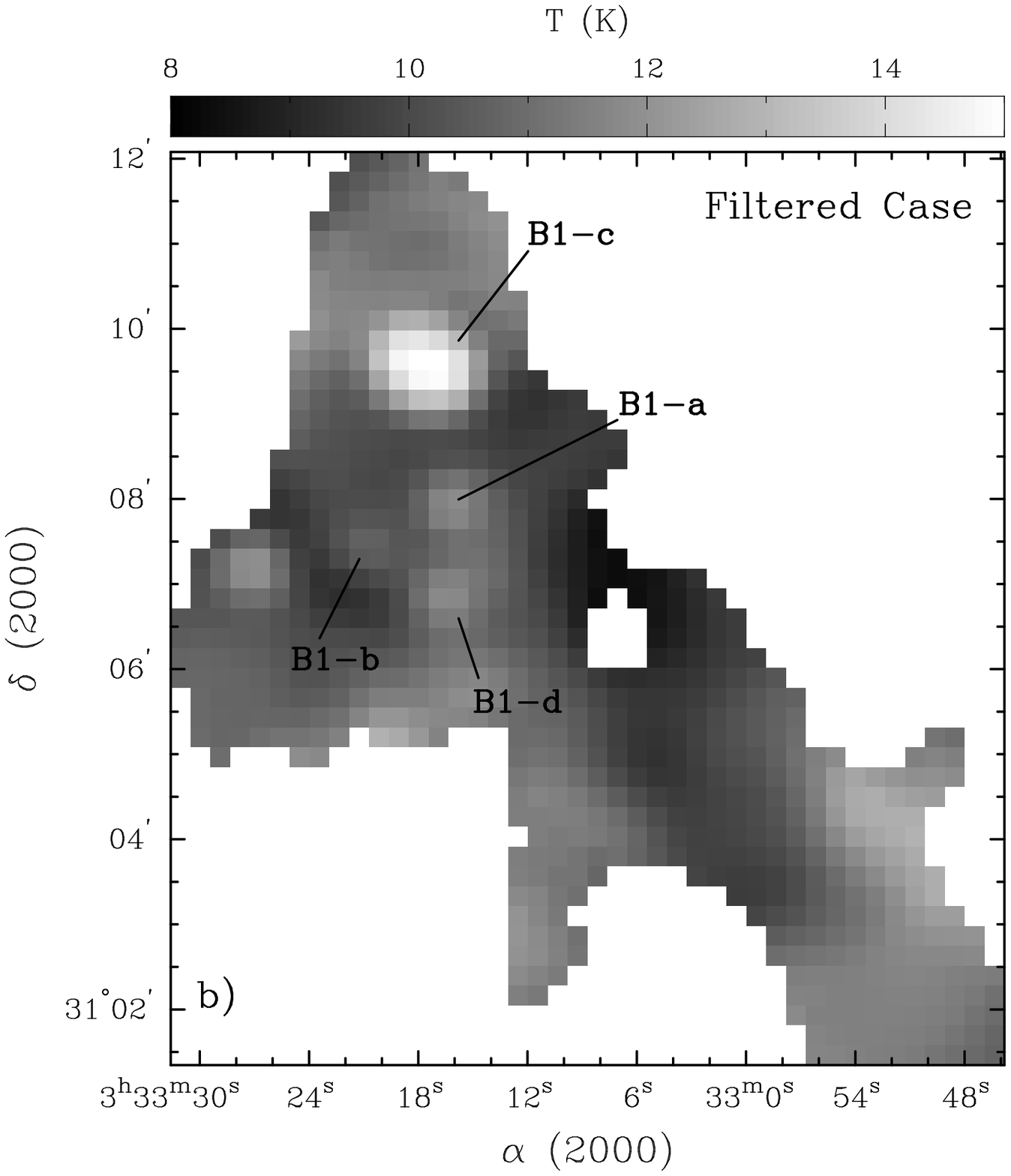}
\caption{Results from the Filtered Case showing (a) the dust opacity at 850 \um\ and (b) the dust temperature. The dust opacity map assumes $\kappanu = 10\ \cmg$ at 300 \um\ and that all variations in $\beta$ manifest themselves as changes in \kappanu\ at 850 \um.}\label{kappaCase1}
\end{figure}

To measure  column densities and masses, however, we must assume a dust opacity law. Thus, we adopted a dust opacity per unit \emph{dust} mass of $\kappanu = 10\ (\nu/\mbox{1 THz})^{\beta}\ \cmg$, following our assumed dust and gas opacity, $\kappa_{\nu_0} = 0.1$ \cmg\ at 1 THz (300 \um), and a dust-to-gas mass ratio of 100.  Figure \ref{kappaCase1} shows the resulting map for the dust opacity at 850 \um and the distribution of best-fit temperatures from the Filtered Case using the H+850 bands.  Based on our $1\ \sigma$ uncertainties for $\beta$, we find that the uncertainties on \kappanu\ are $10-30$\%, where the largest errors are found towards the compact cores. These results are only valid if the dust opacity is kept fixed at 300 \um.  From the temperature map, B1-a, B1-c, and B1-d show internal heating from protostellar sources. B1-b contains two first hydrostatic core candidates (see \citealt{Pezzuto12}), and we see an indication of moderate heating towards these cores (at 36\arcsec\ resolutions, we cannot separate the two components). The uncertainties on the dust temperatures are predominantly $\sigma_{T} < 1.5$ K with the largest errors towards B1-c. The dust temperature is $\sim 10.5$ K and $\sim 12$ K for B1-b and B1-d, respectively, which are similar to the NH$_3$-derived kinetic gas temperatures of $\sim 11.5$ K reported in \citet{Rosolowsky08}. The kinetic gas temperature of B1-c, however, is $\sim 12.4$ K (\citealt{Rosolowsky08}), which is lower than the observed dust temperatures of $\sim 15$ K. This difference in temperature suggests that the NH$_3$ gas is tracing a different region of the core than the dust (e.g., see \citealt{Matthews06}).

\begin{figure}[h!]
\includegraphics[scale=0.48]{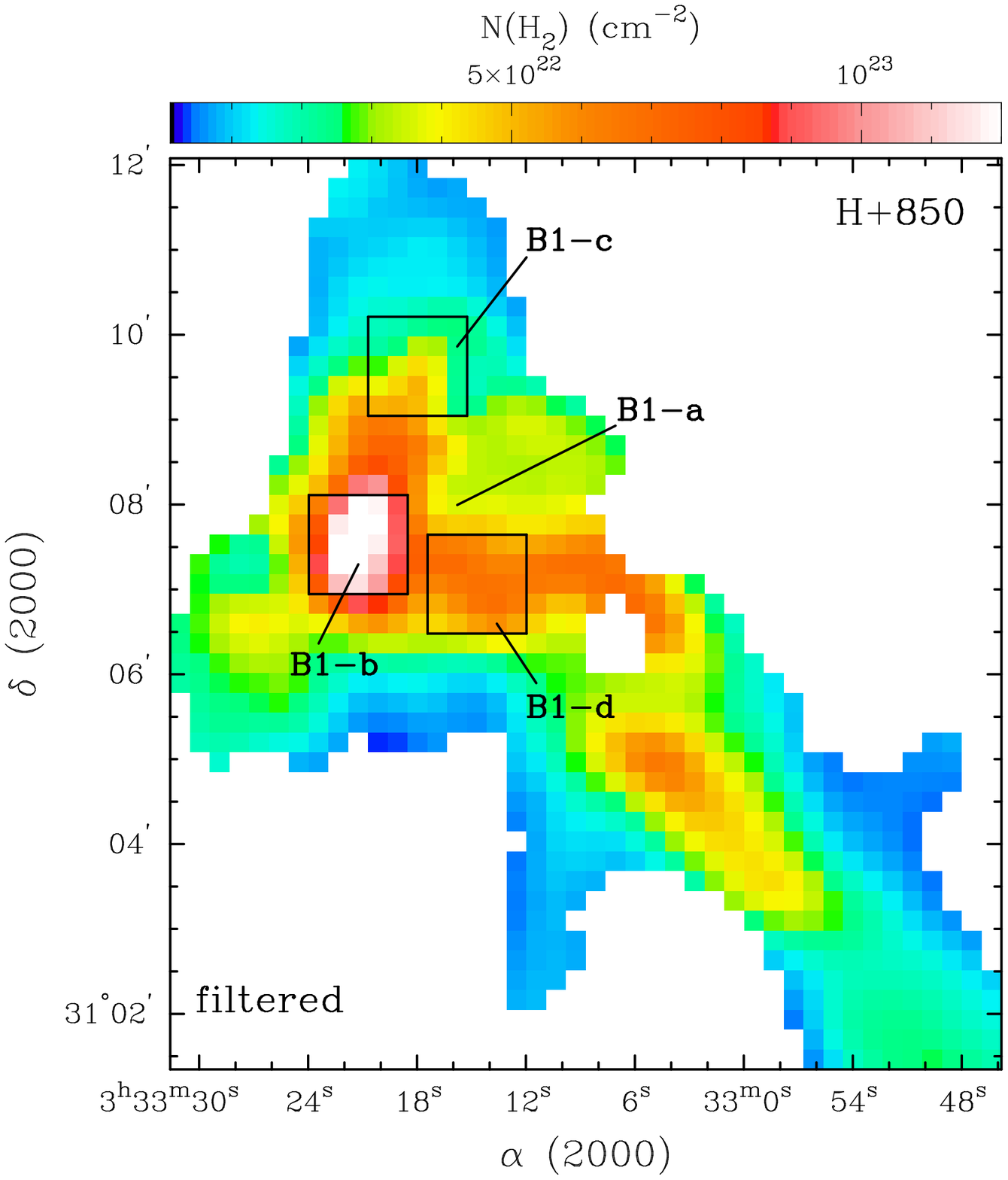}
\includegraphics[scale=0.48]{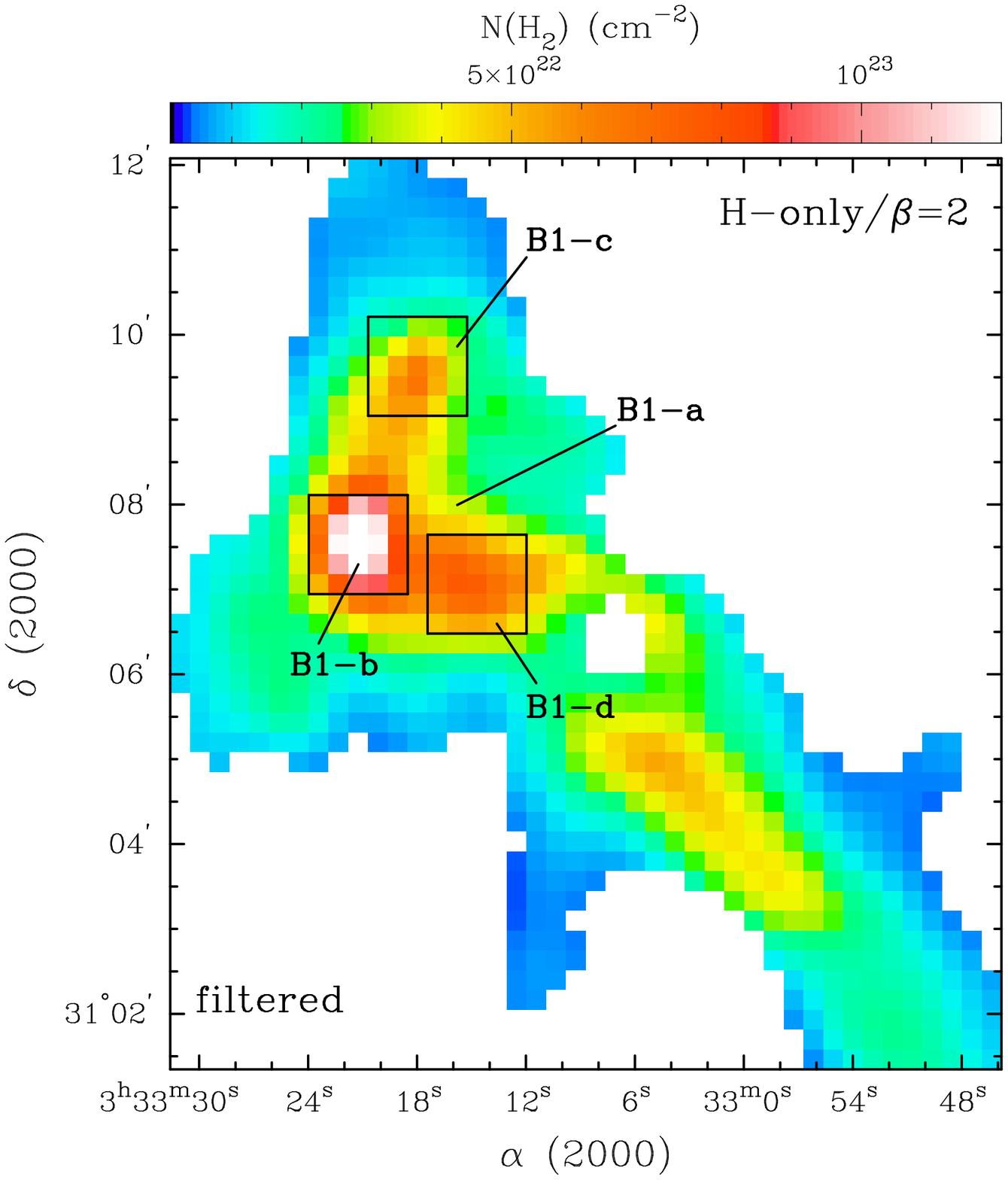}
\caption{Column density maps using the filtered \emph{Herschel} data. For panel (a), we used the H+850 bands and the $\beta$ and temperature results from the Filtered Case, and for panel (b) we used the H-only bands and fixed $\beta = 2$. The black boxes correspond to our mass measurements. \label{columnDen}}
\end{figure}

Figure \ref{columnDen} compares the \NHH\ column densities measured towards the B1 cores for (a) the H+850 bands and (b) the H-only bands with $\beta=2$ (i.e., using the method employed by most \emph{Herschel} GBS papers). The gas column density is determined following Equation \ref{modBB} and assuming $\kappa_{\nu} = 0.1\ \cmg\ $ at 300 \um\ ($\kappanu = 10\ \cmg$ for a gas to dust ratio of 100). Table \ref{massEstimates} lists the total (gas and dust) mass estimates for the B1-b, B1-c, and B1-d cores using these column density maps. For simplicity, we calculated the masses associated with $5\times5$ pixel boxes around the cores. For the \emph{Herschel}-only data with $\beta=2$, we underestimate the columns of mass towards B1-b by $\sim 10$\% and overestimate the columns of mass towards B1-c by $\sim 30$\%\ (the masses towards B1-d agree within 1\%).  Since we find $\beta \approx 2$ for much of the B1 clump (see Figure \ref{compareDist}), assuming $\beta = 2$ with \emph{Herschel}-only data appears to provide a decent first look at the mass distribution of dense cores and the column density of  extended cloud emission. Nevertheless, for regions that deviate from $\beta = 2$, such as B1-c ($\beta \approx 1.6-1.7$), we find that the SED-fitting to the \emph{Herschel}-only bands with $\beta=2$ yields more significant mass uncertainties ($\sim$ 30\%). Therefore, accurate core masses or column densities require direct measurements of $\beta$. For example, assuming $\beta = 2$ could affect our ability to classify cores as dynamically stable or unstable, to properly compare the core mass function to the stellar initial mass function, and to calculate relative (to $H_2$) abundances of molecules.

\begin{table}[h!]
\caption{Core Mass Estimations\tablenotemark{a}}\label{massEstimates}
\begin{tabular}{lccc}
\hline\hline
Core & B1-b & B1-c & B1-d \\
\hline
H+850 & 15 \Msun\ & 5 \Msun\ & 9 \Msun\ \\
H-only\tablenotemark{b} & 13 \Msun\ & 6 \Msun\ & 9 \Msun\ \\
\hline
\end{tabular}
\tablenotetext{a}{\ Masses measured from the column density maps using the Filtered Case and assuming $\kappa_{\nu} = 0.1\ \cmg\ $ at 300 \um. For the H+850 case, the fixed dust opacity at 300 \um\ is the greatest source of uncertainties, so these masses are only accurate within a factor of $\sim 2$.  For the H-only case, forcing $\beta=2$ increases the uncertainties in the masses by, at most, an additional factor of two.}
\tablenotetext{b}{\ Masses measured from filtered H-only maps assuming $\beta=2$.  }
\end{table}

\subsection{Temperature and $\beta$}

In the literature, the dust emissivity index is often assumed to be $\beta = 2$ based on studies of the dielectric functions of graphite and silicate dust grains (e.g., \citealt{DraineLee84}).  Deviations from $\beta = 2$ can arise from processes such as grain growth and the accumulation of icy mantles. For example, $\beta < 2$ has been attributed to the coagulation of bare dust grains (e.g., \citealt{Schwartz82}; \citealt{Ossenkopf94}; \citealt{Lis98}) and \citet{BeckwithSargent91} attributed very low $\beta$ values ($ \sim 1$) towards protostellar disks to grain growth. Large values of $\beta$ (e.g., $ > 2$), however, are harder to interpret.  Thick icy mantles can steepen the dust opacity slope at submillimeter wavelengths, but this appears to be a lesser effect than dust coagulation (e.g., \citealt{Ossenkopf94}). Unfortunately, most dust models do not predict dust opacities with $\beta > 2$ (\citealt{Shirley11}).  \citet{Meny07}) found that $\beta > 2$ may arise due to complex interactions between electromagnetic radiation and the disordered  mass and charge structures of amorphous dust grains.  Thus, $\beta$ highly reflects the dust grain properties, and it is important to understand the relationship between $\beta$ and the surrounding environment. 

Nevertheless, the relationship between $\beta$ and temperature is not well understood.  Many studies that fit submillimeter SEDs to determine $\beta$ use the minimization of $\chi^2$ technique as employed here. Several recent studies, however, have suggested that the minimization of $\chi^2$ produces unclear results due to the degeneracy between $\beta$ and temperature, especially towards the SED peak (e.g., \citealt{Veneziani10}; \citealt{Kelly12}).  Since the \emph{Herschel} bands span the SED peak for cold star forming regions, the method of SED fitting becomes significant.  In particular, \citet{Kelly12} demonstrated that the $\chi^2$ minimization technique gives different temperature and emissivity values than a more detailed Bayesian approach. In addition to the fitting technique, an artificial anti-correlation between $\beta$ and temperature can also be introduced by instrumental errors (e.g., \citealt{Shetty09_noise}) or temperature variations along the line-of-sight (e.g., \citealt{Shetty09}; \citealt{Schnee10}).

Despite these uncertainties, several large studies have found strong inverse relationships between $\beta$ and temperature over large areas and a wide range of temperatures.  These studies include the PRONAOS experiment (e.g., \citealt{Dupac03}), the \emph{Herschel} Hi-Gal project (e.g., \citealt{Paradis10}), and early science results from Planck (e.g., \citealt{Planck_beta2}).  In particular, \citet{Paradis10} employed two techniques to their SED-fitting, a minimization of $\chi^2$ method and a maximum likelihood algorithm.  They found an inverse $\beta$-temperature relationship with both methods.

In Figure \ref{contourPlots}, we find the uncertainties on our values of $\beta$ and temperature are such that small degeneracies may persist between any two values determined, and so any anti-correlation seen between $\beta$ and temperature may be artificial. Thus, despite our improved constraints to determine $\beta$ with the 850 \um\ data, the SED fits may still retain some degeneracies. From our data, we believe the causes of any degeneracy are likely dominated more by our use of the $\chi^2$ technique rather than the noise or single component line-of-sight-temperature assumption. For example, we only selected very bright emission, so issues with noise will be negligible.  Additionally, our analysis used filtered observations and thus, the line-of-sight temperature is less affected by warm diffuse material. The protostellar cores, however, will have warm central objects and cool outer envelopes, and thus assuming a single line-of-sight temperature towards these objects may induce a stronger degeneracy between $\beta$ and temperature. Nevertheless, with only five bands in the submillimeter domain, we cannot apply a multiple temperature component fit.  Thus, with our analyses, we cannot make any robust conclusions regarding any physical $\beta$-temperature relationship.

\subsection{The Addition of SCUBA-2 Data}

Figure \ref{contourPlots} demonstrates that the SED fits are much improved with the inclusion of 850 \um\ observations since the \emph{Herschel} data alone are unable to constrain $\beta$ for cold star-forming regions. Since the 850 \um\ emission is more dependent on $\beta$ (see Figure \ref{longWave}), the SCUBA-2 observations have a significant influence on the best-fit SEDs and the determined values of $\beta$. Therefore, this work demonstrates that the SCUBA-2 850 \um\ data (or equivalent long wavelength observations) are essential for constraining SED fits. 

The SCUBA-2 450 \um\ band has superior resolution ($\sim 9\arcsec$) to the \emph{Herschel} data at $160 - 500$ \um\ presented here (see Figure \ref{obsMaps}), which is necessary for deblending compact sources to obtain more accurate flux measurements. Additionally, with the 450 \um\ band, we can measure dust properties at higher resolutions. For example, in this paper, we convolved our data to a common resolution of 36.3\arcsec\ (corresponding to the 500 \um\ beam), but with the SCUBA-2 450 \um\ band, we could exclude the SPIRE 500 \um\ band and convolve to a common resolution of 25\arcsec\ (the 350 \um\ beam), and find the dust temperature, opacity, and column density over smaller scales, thereby probing variations across cores themselves.

Unfortunately, the 450 \um\ emission has significant uncertainties. For example, in Figures \ref{filtBetaMaps} and \ref{percDiff450}, we see that the observed 450 \um\ emission is roughly $10-15$\%\ brighter than the predicted emission. Additionally, the 450 \um\ band is more susceptible to atmospheric variability. Therefore, we do not use the 450 \um\ band at this time, but note that  with better calibrations and atmospheric modeling, the 450 \um\ band will become more robust and higher resolution analyses will become possible.

\subsection{High Resolution Extinction Maps}

In Section \ref{kappaSection}, we assumed a fixed dust opacity at 300 \um\ to measure column densities and masses.  To circumvent fixing $\kappa_{\nu}$ (300 \um), we need to measure the dust opacity or column density through an independent method, such as from extinction maps.  Since we are interested in measuring changes in dust opacity over small scales, we need high resolution ($< 1\arcmin$) extinction maps.  Such observations would greatly improve the analysis in this paper. 

Typically, extinction maps have resolutions of $\gtrsim 2\arcmin$, which are too coarse to probe changes in dust opacity between dense cores and the ambient clump.  Nevertheless, high resolution ($\sim 20\arcsec$) extinction maps are possible with deep near-infrared observations (e.g., \citealt{RomanZuniga10}). Therefore, future studies of dust opacity using \emph{Herschel} and JCMT observations will greatly benefit from such high resolution extinction mapping.

%%%%%%%%%%%%%%%%%%%%%%%%%%%%%%%%
%%%%%%%%%%%    Conclusions   %%%%%%%%%%%%%
%%%%%%%%%%%%%%%%%%%%%%%%%%%%%%%%
\section{Summary}\label{conc}

We have presented four unique methods to measure the dust emissivity index, $\beta$, from a combination of \emph{Herschel} PACS+SPIRE and SCUBA-2 observations. Of these methods, the Filtered Case, where the \emph{Herschel} maps are filtered using the SCUBA-2 reduction pipeline and $\beta$ is determined from SED-fitting to the filtered \emph{Herschel}+850 bands is the most robust. Our main goal was to examine whether or not the SCUBA-2 850 \um\ band made significant improvements to the SED fits and the \emph{relative} differences in $\beta$ (or \kappanu) that arise from those fits. We summarize our main findings as follows:

\begin{description}
\item[1. ] The \emph{Herschel}-only data do not constrain the SED fits well enough to allow simultaneous fitting of $\beta$ and temperature towards cold clouds. In Figure \ref{contourPlots}, we demonstrated that long wavelength observations are necessary to improve SED fitting to the \emph{Herschel} bands, provided that the long wavelength observations have relatively low flux errors. The 850 \um\ SCUBA-2 band improves our ability to constrain $\beta$ by a factor of $\sim 2$ and our ability to determine temperature by $\sim 40$\%. Additionally, we find significant variations in $\beta$ when the 850 \um\ data are included, i.e., we find $\beta \sim 1.6$ towards the B1-c core and $\beta \sim 2.0$ towards the filament. 
\item[2. ] Based on our SED-fitting, we find the dust opacity at 850 \um\  varies by factors of $2-3$ relative to the dust opacity at 300 \um, with the highest opacities towards the dense cores, particularly B1-c. Therefore, the B1-c core may represent a region where the dust grains have coagulated with some icy mantles, have moderate elliptical shapes, or have higher abundances of metal compounds (\citealt{Ossenkopf94}). Our observations cannot constrain the source of the opacity variations.  Future studies with high resolution extinction mapping are needed to probe the dust opacity in more detail.
\item[3.] With our analysis, we cannot make robust conclusions about the $\beta$-temperature relationship. We found that both $\beta$ and temperature are constrained better with the inclusion of 850 \um\ observations (see Figure \ref{contourPlots}), yet the degeneracy between these parameters is not entirely removed.  While this degeneracy likely arises from using the minimization of $\chi^2$ technique to fit the SEDs, there can be additional degeneracies due to assuming a single component line-of-sight temperature. 
\item[4. ] Assuming a fixed $\beta = 2$ value with the \emph{Herschel} bands alone affects the measured best-fit temperature by $\lesssim 2$ K if $\beta = 1.5$ or $\beta = 2.5$. We found that the core masses as measured with the \emph{Herschel}-only bands and with $\beta=2$ varied by $\lesssim 30$\%\ compared to the results using \emph{Herschel}+850 bands and the same assumptions for the dust opacity ($\kappa_{\nu_0} = 0.1\ \cmg$\ at 300 \um). Thus, the \emph{Herschel}-only bands provide a decent first look at the core masses for an assumed value of $\beta = 2$. Nevertheless, to obtain more accurate core masses, i.e., to measure the dynamical stability of dense cores or the core mass function, long wavelength data and high resolution extinction maps are necessary.

\end{description}

This analysis of the B1 clump with PACS+SPIRE+SCUBA-2 represents the first of many future analyses with these instruments. The SCUBA-2 observations are an excellent complement to the \emph{Herschel} bands, providing a much stronger constraint to SED-fitting and derivations of $\beta$. \emph{Herschel} and long wavelength observations such as those from SCUBA-2 have only recently provided the wavelength and spatial coverage necessary to make these measurements. Future analyses using high resolution extinction maps are also important to probe the changes in $\kappanu$ with environment.

%%%%%%%%%%% ACKNOWLEDGEMENTS %%%%%%%%%%%%%
\vspace{1cm}
\acknowledgments{\noindent \emph{Acknowledgements:} This work was possible with funding from the Natural Sciences and Engineering Research Council of Canada (NSERC) Canadian Graduate Student award. JDF and DIJ acknowledges support by the National Research Council of Canada (NRC) and NSERC (via a Discovery Grant). We thank the anonymous referee for their detailed comments that greatly improved the clarity of this paper.        We thank E. Chapin, J. Dempsey, P. Friberg, A. Gibb, and S. Schnee for their invaluable advice and assistance throughout this project. We thank B. Ali, B. Altieri, and M. Griffin for useful discussions about PACS and SPIRE calibration.         The JCMT is operated by the Joint Astronomy Centre (JAC) on behalf of the Science and Technology Facilities Council (STFC) of the United Kingdom, the National Research Council (NRC) of Canada, and the Netherlands Organisation for ScientiÞc Research. The SCUBA-2 project is funded by the UK Science and Technology Facilities Council (STFC), the JCMT Development Fund and the Canadian Foundation for Innovation (CFI).    \emph{Herschel} is an ESA space observatory with science instruments provided by European-led Principal Investigator consortia and with important participation from NASA.       PACS has been developed by a consortium of institutes led by MPE (Germany) and including UVIE (Austria); KU Leuven, CSL, IMEC (Belgium); CEA, LAM (France); MPIA (Germany); INAF-IFSI/OAA/OAP/OAT, LENS, SISSA (Italy); IAC (Spain). This development has been supported by the funding agencies BMVIT (Austria), ESA-PRODEX (Belgium), CEA/CNES (France), DLR (Germany), ASI/INAF (Italy), and CICYT/MCYT (Spain).        SPIRE has been developed by a consortium of institutes led by Cardiff University (UK) and including Univ. Lethbridge (Canada); NAOC (China); CEA, LAM (France); IFSI, Univ. Padua (Italy); IAC (Spain); Stockholm Observatory (Sweden); Imperial College London, RAL, UCL-MSSL, UKATC, Univ. Sussex (UK); and Caltech, JPL, NHSC, Univ. Colorado (USA). This development has been supported by national funding agencies: CSA (Canada); NAOC (China); CEA, CNES, CNRS (France); ASI (Italy); MCINN (Spain); SNSB (Sweden); STFC (UK); and NASA (USA).}

%%%%%%%%%%%   BIBLIOGRAPHY  %%%%%%%%%%%%%%%%
\bibliographystyle{apj}
\bibliography{references}

\begin{thebibliography}{63}
\expandafter\ifx\csname natexlab\endcsname\relax\def\natexlab#1{#1}\fi

\bibitem[{{Andr{\'e}} {et~al.}(2010){Andr{\'e}}, {Men'shchikov}, {Bontemps},
  {K{\"o}nyves}, {Motte}, {Schneider}, {Didelon}, {Minier}, {Saraceno},
  {Ward-Thompson}, {di Francesco}, {White}, {Molinari}, {Testi}, {Abergel},
  {Griffin}, {Henning}, {Royer}, {Mer{\'{\i}}n}, {Vavrek}, {Attard},
  {Arzoumanian}, {Wilson}, {Ade}, {Aussel}, {Baluteau}, {Benedettini},
  {Bernard}, {Blommaert}, {Cambr{\'e}sy}, {Cox}, {di Giorgio}, {Hargrave},
  {Hennemann}, {Huang}, {Kirk}, {Krause}, {Launhardt}, {Leeks}, {Le Pennec},
  {Li}, {Martin}, {Maury}, {Olofsson}, {Omont}, {Peretto}, {Pezzuto}, {Prusti},
  {Roussel}, {Russeil}, {Sauvage}, {Sibthorpe}, {Sicilia-Aguilar}, {Spinoglio},
  {Waelkens}, {Woodcraft}, \& {Zavagno}}]{Andre10}
{Andr{\'e}}, P., {Men'shchikov}, A., {Bontemps}, S., {K{\"o}nyves}, V.,
  {Motte}, F., {Schneider}, N., {Didelon}, P., {Minier}, V., {Saraceno}, P.,
  {Ward-Thompson}, D., {di Francesco}, J., {White}, G., {Molinari}, S.,
  {Testi}, L., {Abergel}, A., {Griffin}, M., {Henning}, T., {Royer}, P.,
  {Mer{\'{\i}}n}, B., {Vavrek}, R., {Attard}, M., {Arzoumanian}, D., {Wilson},
  C.~D., {Ade}, P., {Aussel}, H., {Baluteau}, J., {Benedettini}, M., {Bernard},
  J., {Blommaert}, J.~A.~D.~L., {Cambr{\'e}sy}, L., {Cox}, P., {di Giorgio},
  A., {Hargrave}, P., {Hennemann}, M., {Huang}, M., {Kirk}, J., {Krause}, O.,
  {Launhardt}, R., {Leeks}, S., {Le Pennec}, J., {Li}, J.~Z., {Martin}, P.~G.,
  {Maury}, A., {Olofsson}, G., {Omont}, A., {Peretto}, N., {Pezzuto}, S.,
  {Prusti}, T., {Roussel}, H., {Russeil}, D., {Sauvage}, M., {Sibthorpe}, B.,
  {Sicilia-Aguilar}, A., {Spinoglio}, L., {Waelkens}, C., {Woodcraft}, A., \&
  {Zavagno}, A. 2010, \aap, 518, L102+

\bibitem[{{Arab} {et~al.}(2012){Arab}, {Abergel}, {Habart}, {Bernard-Salas},
  {Ayasso}, {Dassas}, {Martin}, \& {White}}]{Arab12}
{Arab}, H., {Abergel}, A., {Habart}, E., {Bernard-Salas}, J., {Ayasso}, H.,
  {Dassas}, K., {Martin}, P.~G., \& {White}, G.~J. 2012, \aap, 541, A19

\bibitem[{{Arzoumanian} {et~al.}(2011){Arzoumanian}, {Andr{\'e}}, {Didelon},
  {K{\"o}nyves}, {Schneider}, {Men'shchikov}, {Sousbie}, {Zavagno}, {Bontemps},
  {di Francesco}, {Griffin}, {Hennemann}, {Hill}, {Kirk}, {Martin}, {Minier},
  {Molinari}, {Motte}, {Peretto}, {Pezzuto}, {Spinoglio}, {Ward-Thompson},
  {White}, \& {Wilson}}]{Arzoumanian11}
{Arzoumanian}, D., {Andr{\'e}}, P., {Didelon}, P., {K{\"o}nyves}, V.,
  {Schneider}, N., {Men'shchikov}, A., {Sousbie}, T., {Zavagno}, A.,
  {Bontemps}, S., {di Francesco}, J., {Griffin}, M., {Hennemann}, M., {Hill},
  T., {Kirk}, J., {Martin}, P., {Minier}, V., {Molinari}, S., {Motte}, F.,
  {Peretto}, N., {Pezzuto}, S., {Spinoglio}, L., {Ward-Thompson}, D., {White},
  G., \& {Wilson}, C.~D. 2011, \aap, 529, L6+

\bibitem[{{Bally} {et~al.}(2008){Bally}, {Walawender}, {Johnstone}, {Kirk}, \&
  {Goodman}}]{Bally08}
{Bally}, J., {Walawender}, J., {Johnstone}, D., {Kirk}, H., \& {Goodman}, A.
  2008, {The Perseus Cloud} (Handbook of Star Forming Regions, Volume I, ~ed.,
  by Reipurth, B.), 308--+

\bibitem[{{Beckwith} \& {Sargent}(1991)}]{BeckwithSargent91}
{Beckwith}, S.~V.~W., \& {Sargent}, A.~I. 1991, \apj, 381, 250

\bibitem[{{Bergin} \& {Tafalla}(2007)}]{BerginTafalla07}
{Bergin}, E.~A., \& {Tafalla}, M. 2007, \araa, 45, 339

\bibitem[{{Bernard} {et~al.}(2010){Bernard}, {Paradis}, {Marshall}, {Montier},
  {Lagache}, {Paladini}, {Veneziani}, {Brunt}, {Mottram}, {Martin},
  {Ristorcelli}, {Noriega-Crespo}, {Compi{\`e}gne}, {Flagey}, {Anderson},
  {Popescu}, {Tuffs}, {Reach}, {White}, {Benedetti}, {Calzoletti}, {Digiorgio},
  {Faustini}, {Juvela}, {Joblin}, {Joncas}, {Mivilles-Deschenes}, {Olmi},
  {Traficante}, {Piacentini}, {Zavagno}, \& {Molinari}}]{Bernard10}
{Bernard}, J.-P., {Paradis}, D., {Marshall}, D.~J., {Montier}, L., {Lagache},
  G., {Paladini}, R., {Veneziani}, M., {Brunt}, C.~M., {Mottram}, J.~C.,
  {Martin}, P., {Ristorcelli}, I., {Noriega-Crespo}, A., {Compi{\`e}gne}, M.,
  {Flagey}, N., {Anderson}, L.~D., {Popescu}, C.~C., {Tuffs}, R., {Reach}, W.,
  {White}, G., {Benedetti}, M., {Calzoletti}, L., {Digiorgio}, A.~M.,
  {Faustini}, F., {Juvela}, M., {Joblin}, C., {Joncas}, G.,
  {Mivilles-Deschenes}, M.-A., {Olmi}, L., {Traficante}, A., {Piacentini}, F.,
  {Zavagno}, A., \& {Molinari}, S. 2010, \aap, 518, L88+

\bibitem[{{Buckle} {et~al.}(2009){Buckle}, {Hills}, {Smith}, {Dent}, {Bell},
  {Curtis}, {Dace}, {Gibson}, {Graves}, {Leech}, {Richer}, {Williamson},
  {Withington}, {Yassin}, {Bennett}, {Hastings}, {Laidlaw}, {Lightfoot},
  {Burgess}, {Dewdney}, {Hovey}, {Willis}, {Redman}, {Wooff}, {Berry},
  {Cavanagh}, {Davis}, {Dempsey}, {Friberg}, {Jenness}, {Kackley}, {Rees},
  {Tilanus}, {Walther}, {Zwart}, {Klapwijk}, {Kroug}, \& {Zijlstra}}]{Buckle09}
{Buckle}, J.~V., {Hills}, R.~E., {Smith}, H., {Dent}, W.~R.~F., {Bell}, G.,
  {Curtis}, E.~I., {Dace}, R., {Gibson}, H., {Graves}, S.~F., {Leech}, J.,
  {Richer}, J.~S., {Williamson}, R., {Withington}, S., {Yassin}, G., {Bennett},
  R., {Hastings}, P., {Laidlaw}, I., {Lightfoot}, J.~F., {Burgess}, T.,
  {Dewdney}, P.~E., {Hovey}, G., {Willis}, A.~G., {Redman}, R., {Wooff}, B.,
  {Berry}, D.~S., {Cavanagh}, B., {Davis}, G.~R., {Dempsey}, J., {Friberg}, P.,
  {Jenness}, T., {Kackley}, R., {Rees}, N.~P., {Tilanus}, R., {Walther}, C.,
  {Zwart}, W., {Klapwijk}, T.~M., {Kroug}, M., \& {Zijlstra}, T. 2009, \mnras,
  399, 1026

\bibitem[{{Cavanagh} {et~al.}(2008){Cavanagh}, {Jenness}, {Economou}, \&
  {Currie}}]{Cavanagh08}
{Cavanagh}, B., {Jenness}, T., {Economou}, F., \& {Currie}, M.~J. 2008,
  Astronomische Nachrichten, 329, 295

\bibitem[{{Chapin} {et~al.}(2013){Chapin}, {Berry}, {Gibb}, {Jenness}, {Scott},
  {Tilanus}, {Economou}, \& {Holland}}]{Chapin13}
{Chapin}, E.~L., {Berry}, D.~S., {Gibb}, A.~G., {Jenness}, T., {Scott}, D.,
  {Tilanus}, R.~P.~J., {Economou}, F., \& {Holland}, W.~S. 2013, MNRAS, in
  press, ArXiv e-prints, astro-ph:1301.3652

\bibitem[{{Davis} {et~al.}(2000){Davis}, {Dent}, {Matthews}, {Coulson}, \&
  {McCaughrean}}]{Davis00}
{Davis}, C.~J., {Dent}, W.~R.~F., {Matthews}, H.~E., {Coulson}, I.~M., \&
  {McCaughrean}, M.~J. 2000, \mnras, 318, 952

\bibitem[{{Dempsey} {et~al.}(2013){Dempsey}, {Friberg}, {Jenness}, {Tilanus},
  {Thomas}, {Holland}, {Bintley}, {Berry}, {Chapin}, {Chrysostomou}, {Davis},
  {Gibb}, {Parsons}, \& {Robson}}]{Dempsey13}
{Dempsey}, J.~T., {Friberg}, P., {Jenness}, T., {Tilanus}, R.~P.~J., {Thomas},
  H.~S., {Holland}, W.~S., {Bintley}, D., {Berry}, D.~S., {Chapin}, E.~L.,
  {Chrysostomou}, A., {Davis}, G.~R., {Gibb}, A.~G., {Parsons}, H., \&
  {Robson}, E.~I. 2013, MNRAS, in press, ArXiv e-prints; astro-ph:1301.3773

\bibitem[{{Dempsey} {et~al.}(2012){Dempsey}, {Holland}, {Chrysostomou},
  {Berry}, {Bintley}, {Chapin}, {Craig}, {Coulson}, {Davis}, {Friberg},
  {Jenness}, {Gibb}, {Parsons}, {Scott}, {Thomas}, {Tilanus}, {Robson}, \&
  {Walther}}]{Dempsey12}
{Dempsey}, J.~T., {Holland}, W.~S., {Chrysostomou}, A., {Berry}, D.~S.,
  {Bintley}, D., {Chapin}, E.~L., {Craig}, S.~C., {Coulson}, I.~M., {Davis},
  G.~R., {Friberg}, P., {Jenness}, T., {Gibb}, A.~G., {Parsons}, H.~A.~L.,
  {Scott}, D., {Thomas}, H.~S., {Tilanus}, R.~P.~J., {Robson}, I., \&
  {Walther}, C.~A. 2012, in Society of Photo-Optical Instrumentation Engineers
  (SPIE) Conference Series, Vol. 8452, 845202

\bibitem[{{Di Francesco} {et~al.}(2008){Di Francesco}, {Johnstone}, {Kirk},
  {MacKenzie}, \& {Ledwosinska}}]{difran08}
{Di Francesco}, J., {Johnstone}, D., {Kirk}, H., {MacKenzie}, T., \&
  {Ledwosinska}, E. 2008, \apjs, 175, 277

\bibitem[{{Doty} \& {Leung}(1994)}]{Doty94}
{Doty}, S.~D., \& {Leung}, C.~M. 1994, \apj, 424, 729

\bibitem[{{Drabek} {et~al.}(2012){Drabek}, {Hatchell}, {Friberg}, {Richer},
  {Graves}, {Buckle}, {Nutter}, {Johnstone}, \& {Di Francesco}}]{Drabek12}
{Drabek}, E., {Hatchell}, J., {Friberg}, P., {Richer}, J., {Graves}, S.,
  {Buckle}, J.~V., {Nutter}, D., {Johnstone}, D., \& {Di Francesco}, J. 2012,
  \mnras, 426, 23

\bibitem[{{Draine} \& {Lee}(1984)}]{DraineLee84}
{Draine}, B.~T., \& {Lee}, H.~M. 1984, \apj, 285, 89

\bibitem[{{Dupac} {et~al.}(2003){Dupac}, {Bernard}, {Boudet}, {Giard},
  {Lamarre}, {M{\'e}ny}, {Pajot}, {Ristorcelli}, {Serra}, {Stepnik}, \&
  {Torre}}]{Dupac03}
{Dupac}, X., {Bernard}, J.-P., {Boudet}, N., {Giard}, M., {Lamarre}, J.-M.,
  {M{\'e}ny}, C., {Pajot}, F., {Ristorcelli}, I., {Serra}, G., {Stepnik}, B.,
  \& {Torre}, J.-P. 2003, \aap, 404, L11

\bibitem[{{Enoch} {et~al.}(2009){Enoch}, {Evans}, {Sargent}, \&
  {Glenn}}]{Enoch09}
{Enoch}, M.~L., {Evans}, II, N.~J., {Sargent}, A.~I., \& {Glenn}, J. 2009,
  \apj, 692, 973

\bibitem[{{Friesen} {et~al.}(2005){Friesen}, {Johnstone}, {Naylor}, \&
  {Davis}}]{Friesen05}
{Friesen}, R.~K., {Johnstone}, D., {Naylor}, D.~A., \& {Davis}, G.~R. 2005,
  \mnras, 361, 460

\bibitem[{{Griffin} {et~al.}(2010){Griffin}, {Abergel}, {Abreu}, {Ade},
  {Andr{\'e}}, \& {et al.}}]{Griffin10}
{Griffin}, M.~J., {Abergel}, A., {Abreu}, A., {Ade}, P.~A.~R., {Andr{\'e}}, P.,
  \& {et al.} 2010, \aap, 518, L3+

\bibitem[{{Hatchell} {et~al.}(2007){Hatchell}, {Fuller}, {Richer}, {Harries},
  \& {Ladd}}]{Hatchell07}
{Hatchell}, J., {Fuller}, G.~A., {Richer}, J.~S., {Harries}, T.~J., \& {Ladd},
  E.~F. 2007, \aap, 468, 1009

\bibitem[{{Hatchell} {et~al.}(2012){Hatchell}, {Wilson}, {Drabek}, {Curtis},
  {Richer}, {Nutter}, {Di Francesco}, {Ward-Thompson}, \& {JCMT GBS
  consortium}}]{Hatchell12}
{Hatchell}, J., {Wilson}, T., {Drabek}, E., {Curtis}, E., {Richer}, J.,
  {Nutter}, D., {Di Francesco}, J., {Ward-Thompson}, D., \& {JCMT GBS
  consortium}. 2012, ArXiv:1210.5094

\bibitem[{{Henning} {et~al.}(1995){Henning}, {Michel}, \&
  {Stognienko}}]{Henning95}
{Henning}, T., {Michel}, B., \& {Stognienko}, R. 1995, \planss, 43, 1333

\bibitem[{{Hildebrand}(1983)}]{Hildebrand83}
{Hildebrand}, R.~H. 1983, \qjras, 24, 267

\bibitem[{{Hill} {et~al.}(2006){Hill}, {Thompson}, {Burton}, {Walsh}, {Minier},
  {Cunningham}, \& {Pierce-Price}}]{Hill06}
{Hill}, T., {Thompson}, M.~A., {Burton}, M.~G., {Walsh}, A.~J., {Minier}, V.,
  {Cunningham}, M.~R., \& {Pierce-Price}, D. 2006, \mnras, 368, 1223

\bibitem[{{Hirota} {et~al.}(2008){Hirota}, {Bushimata}, {Choi}, {Honma},
  {Imai}, {Iwadate}, {Jike}, {Kameya}, {Kamohara}, {Kan-Ya}, {Kawaguchi},
  {Kijima}, {Kobayashi}, {Kuji}, {Kurayama}, {Manabe}, {Miyaji}, {Nagayama},
  {Nakagawa}, {Oh}, {Omodaka}, {Oyama}, {Sakai}, {Sasao}, {Sato}, {Shibata},
  {Tamura}, \& {Yamashita}}]{Hirota08}
{Hirota}, T., {Bushimata}, T., {Choi}, Y.~K., {Honma}, M., {Imai}, H.,
  {Iwadate}, K., {Jike}, T., {Kameya}, O., {Kamohara}, R., {Kan-Ya}, Y.,
  {Kawaguchi}, N., {Kijima}, M., {Kobayashi}, H., {Kuji}, S., {Kurayama}, T.,
  {Manabe}, S., {Miyaji}, T., {Nagayama}, T., {Nakagawa}, A., {Oh}, C.~S.,
  {Omodaka}, T., {Oyama}, T., {Sakai}, S., {Sasao}, T., {Sato}, K., {Shibata},
  K.~M., {Tamura}, Y., \& {Yamashita}, K. 2008, \pasj, 60, 37

\bibitem[{{Holland} {et~al.}(2013){Holland}, {Bintley}, {Chapin},
  {Chrysostomou}, {Davis}, {Dempsey}, {Duncan}, {Fich}, {Friberg}, {Halpern},
  {Irwin}, {Jenness}, {Kelly}, {MacIntosh}, {Robson}, {Scott}, {Ade},
  {Atad-Ettedgui}, {Berry}, {Craig}, {Gao}, {Gibb}, {Hilton}, {Hollister},
  {Kycia}, {Lunney}, {McGregor}, {Montgomery}, {Parkes}, {Tilanus}, {Ullom},
  {Walther}, {Walton}, {Woodcraft}, {Amiri}, {Atkinson}, {Burger}, {Chuter},
  {Coulson}, {Doriese}, {Dunare}, {Economou}, {Niemack}, {Parsons},
  {Reintsema}, {Sibthorpe}, {Smail}, {Sudiwala}, \& {Thomas}}]{Holland13}
{Holland}, W.~S., {Bintley}, D., {Chapin}, E.~L., {Chrysostomou}, A., {Davis},
  G.~R., {Dempsey}, J.~T., {Duncan}, W.~D., {Fich}, M., {Friberg}, P.,
  {Halpern}, M., {Irwin}, K.~D., {Jenness}, T., {Kelly}, B.~D., {MacIntosh},
  M.~J., {Robson}, E.~I., {Scott}, D., {Ade}, P.~A.~R., {Atad-Ettedgui}, E.,
  {Berry}, D.~S., {Craig}, S.~C., {Gao}, X., {Gibb}, A.~G., {Hilton}, G.~C.,
  {Hollister}, M.~I., {Kycia}, J.~B., {Lunney}, D.~W., {McGregor}, H.,
  {Montgomery}, D., {Parkes}, W., {Tilanus}, R.~P.~J., {Ullom}, J.~N.,
  {Walther}, C.~A., {Walton}, A.~J., {Woodcraft}, A.~L., {Amiri}, M.,
  {Atkinson}, D., {Burger}, B., {Chuter}, T., {Coulson}, I.~M., {Doriese},
  W.~B., {Dunare}, C., {Economou}, F., {Niemack}, M.~D., {Parsons}, H.~A.~L.,
  {Reintsema}, C.~D., {Sibthorpe}, B., {Smail}, I., {Sudiwala}, R., \&
  {Thomas}, H.~S. 2013, MNRAS, in press, ArXiv e-prints, astro-ph:1301.3650

\bibitem[{{Holland} {et~al.}(1999){Holland}, {Robson}, {Gear}, {Cunningham},
  {Lightfoot}, {Jenness}, {Ivison}, {Stevens}, {Ade}, {Griffin}, {Duncan},
  {Murphy}, \& {Naylor}}]{Holland99}
{Holland}, W.~S., {Robson}, E.~I., {Gear}, W.~K., {Cunningham}, C.~R.,
  {Lightfoot}, J.~F., {Jenness}, T., {Ivison}, R.~J., {Stevens}, J.~A., {Ade},
  P.~A.~R., {Griffin}, M.~J., {Duncan}, W.~D., {Murphy}, J.~A., \& {Naylor},
  D.~A. 1999, \mnras, 303, 659

\bibitem[{{Jenness} {et~al.}(2011){Jenness}, {Berry}, {Chapin}, {Economou},
  {Gibb}, \& {Scott}}]{Jenness11}
{Jenness}, T., {Berry}, D., {Chapin}, E., {Economou}, F., {Gibb}, A., \&
  {Scott}, D. 2011, in Astronomical Society of the Pacific Conference Series,
  Vol. 442, Astronomical Data Analysis Software and Systems XX, ed. I.~N.
  {Evans}, A.~{Accomazzi}, D.~J. {Mink}, \& A.~H. {Rots}, 281

\bibitem[{{Jenness} {et~al.}(2008){Jenness}, {Cavanagh}, {Economou}, \&
  {Berry}}]{Jenness08}
{Jenness}, T., {Cavanagh}, B., {Economou}, F., \& {Berry}, D.~S. 2008, in
  Astronomical Society of the Pacific Conference Series, Vol. 394, Astronomical
  Data Analysis Software and Systems XVII, ed. R.~W. {Argyle}, P.~S.
  {Bunclark}, \& J.~R. {Lewis}, 565

\bibitem[{{Johnstone} {et~al.}(2003){Johnstone}, {Boonman}, \& {van
  Dishoeck}}]{Johnstone03}
{Johnstone}, D., {Boonman}, A.~M.~S., \& {van Dishoeck}, E.~F. 2003, \aap, 412,
  157

\bibitem[{{J{\o}rgensen} {et~al.}(2007){J{\o}rgensen}, {Johnstone}, {Kirk}, \&
  {Myers}}]{Jorgensen07}
{J{\o}rgensen}, J.~K., {Johnstone}, D., {Kirk}, H., \& {Myers}, P.~C. 2007,
  \apj, 656, 293

\bibitem[{{J{\o}rgensen} {et~al.}(2008){J{\o}rgensen}, {Johnstone}, {Kirk},
  {Myers}, {Allen}, \& {Shirley}}]{Jorgensen08}
{J{\o}rgensen}, J.~K., {Johnstone}, D., {Kirk}, H., {Myers}, P.~C., {Allen},
  L.~E., \& {Shirley}, Y.~L. 2008, \apj, 683, 822

\bibitem[{{Kauffmann} {et~al.}(2008){Kauffmann}, {Bertoldi}, {Bourke}, {Evans},
  \& {Lee}}]{Kauffmann08}
{Kauffmann}, J., {Bertoldi}, F., {Bourke}, T.~L., {Evans}, II, N.~J., \& {Lee},
  C.~W. 2008, \aap, 487, 993

\bibitem[{{Kelly} {et~al.}(2012){Kelly}, {Shetty}, {Stutz}, {Kauffmann},
  {Goodman}, \& {Launhardt}}]{Kelly12}
{Kelly}, B.~C., {Shetty}, R., {Stutz}, A.~M., {Kauffmann}, J., {Goodman},
  A.~A., \& {Launhardt}, R. 2012, \apj, 752, 55

\bibitem[{{Kirk} {et~al.}(2006){Kirk}, {Johnstone}, \& {Di Francesco}}]{Kirk06}
{Kirk}, H., {Johnstone}, D., \& {Di Francesco}, J. 2006, \apj, 646, 1009

\bibitem[{{Lis} {et~al.}(1998){Lis}, {Serabyn}, {Keene}, {Dowell}, {Benford},
  {Phillips}, {Hunter}, \& {Wang}}]{Lis98}
{Lis}, D.~C., {Serabyn}, E., {Keene}, J., {Dowell}, C.~D., {Benford}, D.~J.,
  {Phillips}, T.~G., {Hunter}, T.~R., \& {Wang}, N. 1998, \apj, 509, 299

\bibitem[{{Markwardt}(2009)}]{Markwardt09}
{Markwardt}, C.~B. 2009, in Astronomical Society of the Pacific Conference
  Series, Vol. 411, Astronomical Data Analysis Software and Systems XVIII, ed.
  D.~A. {Bohlender}, D.~{Durand}, \& P.~{Dowler}, 251

\bibitem[{{Martin} {et~al.}(2012){Martin}, {Roy}, {Bontemps},
  {Miville-Desch{\^e}nes}, {Ade}, {Bock}, {Chapin}, {Devlin}, {Dicker},
  {Griffin}, {Gundersen}, {Halpern}, {Hargrave}, {Hughes}, {Klein}, {Marsden},
  {Mauskopf}, {Netterfield}, {Olmi}, {Patanchon}, {Rex}, {Scott}, {Semisch},
  {Truch}, {Tucker}, {Tucker}, {Viero}, \& {Wiebe}}]{Martin12}
{Martin}, P.~G., {Roy}, A., {Bontemps}, S., {Miville-Desch{\^e}nes}, M.-A.,
  {Ade}, P.~A.~R., {Bock}, J.~J., {Chapin}, E.~L., {Devlin}, M.~J., {Dicker},
  S.~R., {Griffin}, M., {Gundersen}, J.~O., {Halpern}, M., {Hargrave}, P.~C.,
  {Hughes}, D.~H., {Klein}, J., {Marsden}, G., {Mauskopf}, P., {Netterfield},
  C.~B., {Olmi}, L., {Patanchon}, G., {Rex}, M., {Scott}, D., {Semisch}, C.,
  {Truch}, M.~D.~P., {Tucker}, C., {Tucker}, G.~S., {Viero}, M.~P., \& {Wiebe},
  D.~V. 2012, \apj, 751, 28

\bibitem[{{Matthews} {et~al.}(2006){Matthews}, {Hogerheijde}, {J{\o}rgensen},
  \& {Bergin}}]{Matthews06}
{Matthews}, B.~C., {Hogerheijde}, M.~R., {J{\o}rgensen}, J.~K., \& {Bergin},
  E.~A. 2006, \apj, 652, 1374

\bibitem[{{Meny} {et~al.}(2007){Meny}, {Gromov}, {Boudet}, {Bernard},
  {Paradis}, \& {Nayral}}]{Meny07}
{Meny}, C., {Gromov}, V., {Boudet}, N., {Bernard}, J.-P., {Paradis}, D., \&
  {Nayral}, C. 2007, \aap, 468, 171

\bibitem[{{Ossenkopf} \& {Henning}(1994)}]{Ossenkopf94}
{Ossenkopf}, V., \& {Henning}, T. 1994, \aap, 291, 943

\bibitem[{{Paradis} {et~al.}(2010){Paradis}, {Veneziani}, {Noriega-Crespo},
  {Paladini}, {Piacentini}, {Bernard}, {de Bernardis}, {Calzoletti},
  {Faustini}, {Martin}, {Masi}, {Montier}, {Natoli}, {Ristorcelli}, {Thompson},
  {Traficante}, \& {Molinari}}]{Paradis10}
{Paradis}, D., {Veneziani}, M., {Noriega-Crespo}, A., {Paladini}, R.,
  {Piacentini}, F., {Bernard}, J.~P., {de Bernardis}, P., {Calzoletti}, L.,
  {Faustini}, F., {Martin}, P., {Masi}, S., {Montier}, L., {Natoli}, P.,
  {Ristorcelli}, I., {Thompson}, M.~A., {Traficante}, A., \& {Molinari}, S.
  2010, \aap, 520, L8

\bibitem[{{Pezzuto} {et~al.}(2012){Pezzuto}, {Elia}, {Schisano}, {Strafella},
  {Di Francesco}, {Sadavoy}, {Andr{\'e}}, {Benedettini}, {Bernard}, {di
  Giorgio}, {Facchini}, {Hennemann}, {Hill}, {K{\"o}nyves}, {Molinari},
  {Motte}, {Nguyen-Luong}, {Peretto}, {Pestalozzi}, {Polychroni}, {Rygl},
  {Saraceno}, {Schneider}, {Spinoglio}, {Testi}, {Ward-Thompson}, \&
  {White}}]{Pezzuto12}
{Pezzuto}, S., {Elia}, D., {Schisano}, E., {Strafella}, F., {Di Francesco}, J.,
  {Sadavoy}, S., {Andr{\'e}}, P., {Benedettini}, M., {Bernard}, J.~P., {di
  Giorgio}, A.~M., {Facchini}, A., {Hennemann}, M., {Hill}, T., {K{\"o}nyves},
  V., {Molinari}, S., {Motte}, F., {Nguyen-Luong}, Q., {Peretto}, N.,
  {Pestalozzi}, M., {Polychroni}, D., {Rygl}, K.~L.~J., {Saraceno}, P.,
  {Schneider}, N., {Spinoglio}, L., {Testi}, L., {Ward-Thompson}, D., \&
  {White}, G.~J. 2012, \aap, 547, A54

\bibitem[{{Pilbratt} {et~al.}(2010){Pilbratt}, {Riedinger}, {Passvogel},
  {Crone}, {Doyle}, {Gageur}, {Heras}, {Jewell}, {Metcalfe}, {Ott}, \&
  {Schmidt}}]{Pilbratt10}
{Pilbratt}, G.~L., {Riedinger}, J.~R., {Passvogel}, T., {Crone}, G., {Doyle},
  D., {Gageur}, U., {Heras}, A.~M., {Jewell}, C., {Metcalfe}, L., {Ott}, S., \&
  {Schmidt}, M. 2010, \aap, 518, L1

\bibitem[{{Planck Collaboration} {et~al.}(2011){Planck Collaboration},
  {Abergel}, {Ade}, {Aghanim}, {Arnaud}, {Ashdown}, {Aumont}, {Baccigalupi},
  {Balbi}, {Banday}, \& et~al.}]{Planck_beta2}
{Planck Collaboration}, {Abergel}, A., {Ade}, P.~A.~R., {Aghanim}, N.,
  {Arnaud}, M., {Ashdown}, M., {Aumont}, J., {Baccigalupi}, C., {Balbi}, A.,
  {Banday}, A.~J., \& et~al. 2011, \aap, 536, A25

\bibitem[{{Poglitsch} {et~al.}(2010){Poglitsch}, {Waelkens}, {Geis},
  {Feuchtgruber}, {Vandenbussche}, {Rodriguez}, {Krause}, {Renotte}, {van
  Hoof}, {Saraceno}, {Cepa}, {Kerschbaum}, \& {et al.}}]{Poglitsch10}
{Poglitsch}, A., {Waelkens}, C., {Geis}, N., {Feuchtgruber}, H.,
  {Vandenbussche}, B., {Rodriguez}, L., {Krause}, O., {Renotte}, E., {van
  Hoof}, C., {Saraceno}, P., {Cepa}, J., {Kerschbaum}, F., \& {et al.} 2010,
  \aap, 518, L2+

\bibitem[{{Rom{\'a}n-Z{\'u}{\~n}iga} {et~al.}(2010){Rom{\'a}n-Z{\'u}{\~n}iga},
  {Alves}, {Lada}, \& {Lombardi}}]{RomanZuniga10}
{Rom{\'a}n-Z{\'u}{\~n}iga}, C.~G., {Alves}, J.~F., {Lada}, C.~J., \&
  {Lombardi}, M. 2010, \apj, 725, 2232

\bibitem[{{Rosolowsky} {et~al.}(2008){Rosolowsky}, {Pineda}, {Foster},
  {Borkin}, {Kauffmann}, {Caselli}, {Myers}, \& {Goodman}}]{Rosolowsky08}
{Rosolowsky}, E.~W., {Pineda}, J.~E., {Foster}, J.~B., {Borkin}, M.~A.,
  {Kauffmann}, J., {Caselli}, P., {Myers}, P.~C., \& {Goodman}, A.~A. 2008,
  \apjs, 175, 509

\bibitem[{{Roussel}(2012)}]{Roussel12}
{Roussel}, H. 2012, ArXiv:1205.2576

\bibitem[{{Roy} {et~al.}(2013){Roy}, {Martin}, {Polychroni}, {Bontemps},
  {Abergel}, {Andr{\'e}}, {Arzoumanian}, {Di Francesco}, {Hill}, {Konyves},
  {Nguyen-Luong}, {Pezzuto}, {Schneider}, {Testi}, \& {White}}]{Roy13}
{Roy}, A., {Martin}, P.~G., {Polychroni}, D., {Bontemps}, S., {Abergel}, A.,
  {Andr{\'e}}, P., {Arzoumanian}, D., {Di Francesco}, J., {Hill}, T.,
  {Konyves}, V., {Nguyen-Luong}, Q., {Pezzuto}, S., {Schneider}, N., {Testi},
  L., \& {White}, G. 2013, \apj, 763, 55

\bibitem[{{Sadavoy} {et~al.}(2012){Sadavoy}, {Di Francesco}, {Andr{\'e}},
  {Pezzuto}, {Bernard}, {Bontemps}, {Bressert}, {Chitsazzadeh}, {Fallscheer},
  {Hennemann}, {Hill}, {Martin}, {Motte}, {Nguyn Luong}, {Peretto}, {Reid},
  {Schneider}, {Testi}, {White}, \& {Wilson}}]{Sadavoy12}
{Sadavoy}, S.~I., {Di Francesco}, J., {Andr{\'e}}, P., {Pezzuto}, S.,
  {Bernard}, J.-P., {Bontemps}, S., {Bressert}, E., {Chitsazzadeh}, S.,
  {Fallscheer}, C., {Hennemann}, M., {Hill}, T., {Martin}, P., {Motte}, F.,
  {Nguyn Luong}, Q., {Peretto}, N., {Reid}, M., {Schneider}, N., {Testi}, L.,
  {White}, G.~J., \& {Wilson}, C. 2012, \aap, 540, A10

\bibitem[{{Schnee} {et~al.}(2010){Schnee}, {Enoch}, {Noriega-Crespo}, {Sayers},
  {Terebey}, {Caselli}, {Foster}, {Goodman}, {Kauffmann}, {Padgett}, {Rebull},
  {Sargent}, \& {Shetty}}]{Schnee10}
{Schnee}, S., {Enoch}, M., {Noriega-Crespo}, A., {Sayers}, J., {Terebey}, S.,
  {Caselli}, P., {Foster}, J., {Goodman}, A., {Kauffmann}, J., {Padgett}, D.,
  {Rebull}, L., {Sargent}, A., \& {Shetty}, R. 2010, \apj, 708, 127

\bibitem[{{Schwartz}(1982)}]{Schwartz82}
{Schwartz}, P.~R. 1982, \apj, 252, 589

\bibitem[{{Shetty} {et~al.}(2009{\natexlab{a}}){Shetty}, {Kauffmann}, {Schnee},
  \& {Goodman}}]{Shetty09_noise}
{Shetty}, R., {Kauffmann}, J., {Schnee}, S., \& {Goodman}, A.~A.
  2009{\natexlab{a}}, \apj, 696, 676

\bibitem[{{Shetty} {et~al.}(2009{\natexlab{b}}){Shetty}, {Kauffmann}, {Schnee},
  {Goodman}, \& {Ercolano}}]{Shetty09}
{Shetty}, R., {Kauffmann}, J., {Schnee}, S., {Goodman}, A.~A., \& {Ercolano},
  B. 2009{\natexlab{b}}, \apj, 696, 2234

\bibitem[{{Shirley} {et~al.}(2011){Shirley}, {Huard}, {Pontoppidan}, {Wilner},
  {Stutz}, {Bieging}, \& {Evans}}]{Shirley11}
{Shirley}, Y.~L., {Huard}, T.~L., {Pontoppidan}, K.~M., {Wilner}, D.~J.,
  {Stutz}, A.~M., {Bieging}, J.~H., \& {Evans}, II, N.~J. 2011, \apj, 728, 143

\bibitem[{{Sun} {et~al.}(2006){Sun}, {Kramer}, {Ossenkopf}, {Bensch},
  {Stutzki}, \& {Miller}}]{Sun06}
{Sun}, K., {Kramer}, C., {Ossenkopf}, V., {Bensch}, F., {Stutzki}, J., \&
  {Miller}, M. 2006, \aap, 451, 539

\bibitem[{{Tothill} {et~al.}(2002){Tothill}, {White}, {Matthews}, {McCutcheon},
  {McCaughrean}, \& {Kenworthy}}]{Tothill02}
{Tothill}, N.~F.~H., {White}, G.~J., {Matthews}, H.~E., {McCutcheon}, W.~H.,
  {McCaughrean}, M.~J., \& {Kenworthy}, M.~A. 2002, \apj, 580, 285

\bibitem[{{Veneziani} {et~al.}(2010){Veneziani}, {Ade}, {Bock}, {Boscaleri},
  {Crill}, {de Bernardis}, {De Gasperis}, {de Oliveira-Costa}, {De Troia}, {Di
  Stefano}, {Ganga}, {Jones}, {Kisner}, {Lange}, {MacTavish}, {Masi},
  {Mauskopf}, {Montroy}, {Natoli}, {Netterfield}, {Pascale}, {Piacentini},
  {Pietrobon}, {Polenta}, {Ricciardi}, {Romeo}, \& {Ruhl}}]{Veneziani10}
{Veneziani}, M., {Ade}, P.~A.~R., {Bock}, J.~J., {Boscaleri}, A., {Crill},
  B.~P., {de Bernardis}, P., {De Gasperis}, G., {de Oliveira-Costa}, A., {De
  Troia}, G., {Di Stefano}, G., {Ganga}, K.~M., {Jones}, W.~C., {Kisner},
  T.~S., {Lange}, A.~E., {MacTavish}, C.~J., {Masi}, S., {Mauskopf}, P.~D.,
  {Montroy}, T.~E., {Natoli}, P., {Netterfield}, C.~B., {Pascale}, E.,
  {Piacentini}, F., {Pietrobon}, D., {Polenta}, G., {Ricciardi}, S., {Romeo},
  G., \& {Ruhl}, J.~E. 2010, \apj, 713, 959

\bibitem[{{Ward-Thompson} {et~al.}(2007){Ward-Thompson}, {Di Francesco},
  {Hatchell}, {Hogerheijde}, {Nutter}, {Bastien}, {Basu}, {Bonnell}, {Bowey},
  {Brunt}, {Buckle}, {Butner}, {Cavanagh}, {Chrysostomou}, {Curtis}, {Davis},
  {Dent}, {van Dishoeck}, {Edmunds}, {Fich}, {Fiege}, {Fissel}, {Friberg},
  {Friesen}, {Frieswijk}, {Fuller}, {Gosling}, {Graves}, {Greaves}, {Helmich},
  {Hills}, {Holland}, {Houde}, {Jayawardhana}, {Johnstone}, {Joncas}, {Kirk},
  {Kirk}, {Knee}, {Matthews}, {Matthews}, {Matzner}, {Moriarty-Schieven},
  {Naylor}, {Padman}, {Plume}, {Rawlings}, {Redman}, {Reid}, {Richer},
  {Shipman}, {Simpson}, {Spaans}, {Stamatellos}, {Tsamis}, {Viti}, {Weferling},
  {White}, {Whitworth}, {Wouterloot}, {Yates}, \& {Zhu}}]{Ward-T07GBS}
{Ward-Thompson}, D., {Di Francesco}, J., {Hatchell}, J., {Hogerheijde}, M.~R.,
  {Nutter}, D., {Bastien}, P., {Basu}, S., {Bonnell}, I., {Bowey}, J., {Brunt},
  C., {Buckle}, J., {Butner}, H., {Cavanagh}, B., {Chrysostomou}, A., {Curtis},
  E., {Davis}, C.~J., {Dent}, W.~R.~F., {van Dishoeck}, E., {Edmunds}, M.~G.,
  {Fich}, M., {Fiege}, J., {Fissel}, L., {Friberg}, P., {Friesen}, R.,
  {Frieswijk}, W., {Fuller}, G.~A., {Gosling}, A., {Graves}, S., {Greaves},
  J.~S., {Helmich}, F., {Hills}, R.~E., {Holland}, W.~S., {Houde}, M.,
  {Jayawardhana}, R., {Johnstone}, D., {Joncas}, G., {Kirk}, H., {Kirk}, J.~M.,
  {Knee}, L.~B.~G., {Matthews}, B., {Matthews}, H., {Matzner}, C.,
  {Moriarty-Schieven}, G.~H., {Naylor}, D., {Padman}, R., {Plume}, R.,
  {Rawlings}, J.~M.~C., {Redman}, R.~O., {Reid}, M., {Richer}, J.~S.,
  {Shipman}, R., {Simpson}, R.~J., {Spaans}, M., {Stamatellos}, D., {Tsamis},
  Y.~G., {Viti}, S., {Weferling}, B., {White}, G.~J., {Whitworth}, A.~P.,
  {Wouterloot}, J., {Yates}, J., \& {Zhu}, M. 2007, \pasp, 119, 855

\bibitem[{{White} {et~al.}(2003){White}, {Araki}, {Greaves}, {Ohishi}, \&
  {Higginbottom}}]{White03}
{White}, G.~J., {Araki}, M., {Greaves}, J.~S., {Ohishi}, M., \& {Higginbottom},
  N.~S. 2003, \aap, 407, 589

\end{thebibliography}

\appendix

%%%%%%%%%%%  APPENDICES %%%%%%%%%%%%%%

\section{SCUBA-2 Filtering}\label{filterAppendix}

Since SCUBA-2 is a ground-based detector, we filter out large-scale astronomical emission in the process of removing the emission from a bright and variable atmosphere. Thus, SCUBA-2 is sensitive only to spatial scales where emission is fully recovered. To determine the largest such spatial scale, we introduced artificial Gaussian sources of various sizes ranging from 1\arcmin\ to 7.5\arcmin\ FWHM to the original SCUBA-2 data, and then ran these modified maps through the pipeline. Since the SCUBA-2 pipeline determines which map pixel is covered by which bolometer at any given time, the extra signal corresponding to the Gaussian sources is simply added to the bolometer timeseries. Following our reduction process from Section \ref{scuba2}, the smaller Gaussians were fully recovered, e.g., Gaussians of 1\arcmin\ or 2.5\arcmin\ sizes. Larger Gaussians, however, were only partially recovered or fully filtered out. For example, peak fluxes for $\sim$ 5\arcmin\ Gaussians were only 70\%\ recovered. Based on these tests, the SCUBA-2 observations appear to fully recover objects on size scales $\lesssim$ 2.5\arcmin.

Conversely, \emph{Herschel} is a space-based observatory, and thus, the PACS+SPIRE data  are unaffected by atmospheric emission. For our HARP CO ($3-2$) line observations, we used an emission-free off position to remove the atmospheric contributions. Thus, both PACS+SPIRE and HARP observations are unfiltered, and such data cannot be directly compared to the SCUBA-2 maps. One simple solution is to filter out these large-scale structures, such that the \emph{Herschel} continuum and HARP line observations are sensitive to the same spatial scales as the SCUBA-2 data.

We filtered out the large-scale emission from the CO ($3-2$) and \emph{Herschel} maps using the SCUBA-2 reduction pipeline in the same manner as the artificial Gaussian sources.  We created ``artificial maps'' by adding the unfiltered CO ($3-2$) integrated intensity map (in \Jybeam) or the \emph{Herschel} data to the SCUBA-2 bolometer timeseries by adjusting the original signal observed at every bolometer to include the emission observed by HARP, PACS, or SPIRE.  Using these artificial signals, we repeated the SCUBA-2 reduction as given in Section \ref{scuba2} four times, once for each of the four Grade 1 observations. Finally, we combined the products into a single mosaic. Since this mosaic also includes the original 850 \um\ data, we subtracted out the reduced 850 \um\ map to leave behind the filtered version of the other data (see Figure \ref{resH}).

\section{CO Line Contamination}\label{COAppendix}

The CO (3-2) line transition coincides with the 850 \um\ SCUBA-2 band, and thus, will contaminate the continuum emission. Significant CO ($3-2$) contamination is mostly identified towards outflows and highly energetic regions such as HII regions, where the contamination levels can dominate ($> 50$\%) the total continuum emission (e.g., \citealt{Davis00}; \citealt{Tothill02}; \citealt{Drabek12}). Although B1 is a low-mass star-forming complex, and such regions generally contain weaker CO ($3-2$) line emission (\citealt{Drabek12}), this clump contains several protostellar sources driving outflows  (\citealt{Bally08}). 

To evaluate the CO ($3-2$) contamination, we determined the contribution from the line flux over the entire SCUBA-2 850 \um\ passband.  Following \citet{Drabek12}, the line flux at frequency $\nu_0$ across the continuum band is,
\begin{equation}
\frac{F_{\nu_0}}{\mJybeam} = \frac{2k\nu_0^3}{c^3}\frac{g_{\nu_0,line}}{\int{g_{\nu,cont} d\nu}}\ \Omega\int{T_{MB}dv},
\end{equation}
\noindent where $k$ is the Boltzmann constant, $c$ is the speed of light, $\Omega$ is the telescope beam, $T_{MB}$ is the main beam temperature of the line, $g_{\nu_0,line}$ is the transmission of the submillimeter filter at the frequency of the molecular line ($\nu_0$), and $g_{\nu,cont}$ is the transmission profile of the entire submillimeter filter. \citet{Drabek12} adopted a ``conversion factor'' as,
\begin{equation}
\frac{C}{\mJybeam\ (\Kkms)^{-1}} = \frac{2k\nu_0^3}{c^3}\frac{g_{\nu_0,line}}{\int{g_{\nu,cont} d\nu}}\ \Omega.
\end{equation}
The SCUBA-2 filter profile, $g_{\nu,cont}$, varies significantly with atmospheric conditions. For example, the filter response is best for dry conditions ($\tau_{225} < 0.05$) and can drop by more than 50\%\ in wet conditions ($\tau_{225} > 0.12$). We used the S2SV observations made in very dry conditions, $\tau_{225} < 0.05$, resulting in $C \approx 0.63\ \mJybeam\ (\Kkms)^{-1}$ (\citealt{Drabek12}). This factor, however, assumes that the beam of SCUBA-2 at 850 \um\ is 13.8\arcsec. We have adopted here an effective beam of 14.2\arcsec\ (see Section \ref{scuba2}), which considers the larger error beam. Thus, our conversion factor should be modified by $\Omega_{eff}/\Omega = 1.06$. Therefore, we use
\begin{equation}
C = 0.67\ \mJybeam\ (\Kkms)^{-1}. \label{factor}
\end{equation}
In addition to CO $(3-2)$ line contamination, other (weaker) molecular line emission will contaminate the continuum bands  (\citealt{Drabek12}). For example, CO ($6-5$) at 691.473 GHz coincides with the 450 \um\ SCUBA-2 band and CO ($7-6$) at 806.652 GHz coincides with the SPIRE 350 \um\ band. Nevertheless, Drabek et al. demonstrated that such high excitation line transitions are generally negligible compared to the dust emission. For example, the dust emission rises towards shorter wavelengths as $\nu^{\beta+2}$ whereas the emission from higher transitions of CO does not rise as steeply. For example, we do not believe the excess brightness at 450 \um\ relative to the filtered 500 \um\ emission (see Figures \ref{filtBetaMaps} and Figure \ref{percDiff450}) is due to CO (6-5) contamination at 450 \um.  Since the CO ($3-2$) line contamination is relatively minor, we can also assume that the contamination from less abundant molecules (i.e., $^{13}$CO or C$^{18}$O) should be negligible.

\section{Herschel Colour Corrections}\label{ColourCorrectionAppendix}

The PACS and SPIRE fluxes from HIPE\footnote{The \emph{Herschel} Interactive Processing Environment (HIPE) is a joint development software by the \emph{Herschel} Science Ground Segment Consortium, consisting of ESA, the NASA \emph{Herschel} Science Center, and the HIFI, PACS, and SPIRE consortia.} contain several sources of uncertainty. First, both instruments have a 10\%\ flux error based on the uncertainties in the calibration. For PACS, flux calibrations are mainly based on stellar sources and bright asteroids (\citealt{Poglitsch10}), whereas for SPIRE, flux calibrations primarily use Neptune (\citealt{Griffin10}). Second, both PACS and SPIRE calibrations assume that the observed spectral profile follows a $\nu^{-1}$ power-law such that $\nu{S_{\nu}}$ is flat (\citealt{Poglitsch10}; \citealt{Griffin10}). For cold molecular clouds ($T \sim 10$ K), emission in the PACS and SPIRE bands will not follow a $\nu^{-1}$ power-law, and thus, colour corrections are necessary (e.g., see \citealt{Pezzuto12}). Furthermore, the spectral colour corrections are not systematic, and they depend on the SED shape and thus, will vary with frequency, temperature, and opacity. 

Calculating the spectral colour corrections is nontrivial. First, these correction factors require foreknowledge of the SED profiles. Second, the colour corrections compare the flux from the true SED profile to the assumed SED profile, both weighted by the filter response function. For SPIRE, the beam solid angle varies with wavelength, resulting in more power at longer wavelengths. While this effect is insignificant for point sources, extended sources (i.e., objects that uniformly fill the entire beam) will be biased towards longer wavelengths. Thus, colour corrections vary between point sources and extended objects\footnote{For additional information, see the SPIRE Observers' Manual (2011), HERSCHEL-DOC-0798, v2.4, http://herschel.esac.esa.int/Docs/SPIRE/pdf/spire\_om.pdf}. PACS calibration does not differentiate between point sources and extended sources. Since we are interested in the extended \emph{Herschel} map structures, we will adopt the extended colour corrections for SPIRE. We caution that these structures likely do not fill the filter beam uniformly at all wavelengths, and that these colour corrections represent our best estimate. For SPIRE, these extended source corrections are generally minor (few percent) and should not greatly affect our results (\citealt{Griffin10}). For PACS, however, the colour corrections of cold objects can be more significant, particularly at 70 \um\ (\citealt{Poglitsch10}).  We do not include 70 \um\ emission in the SED fits in this paper, however.

Since the SED profiles are unknown a priori, we found the colour corrections associated with cold dust emission at temperatures between 10 K and 15 K and dust emissivity indices between 1.5 and 2.5. For PACS, the colour corrections for cold and dusty environments are given in M\"{u}ller et al. (2011)\footnote{PACS Photometer Passbands and Colour Correction Factors for Various Source SEDs, PICC-ME-TN-038, v1.0, http://herschel.esac.esa.int/twiki/pub/Public/PacsCalibrationWeb/cc\_report\_v1.pdf}, and we extrapolated those tabulated values to identify the colour corrections associated with our adopted ranges of temperature and beta.  For SPIRE, we calculated the colour corrections by integrating model SEDs weighted by the relative spectral response function (RSRF)\footnote{For extended emission, the RSRF profile is multiplied by $\lambda^2$ to account for the differences in the beam size between the short wavelength end and the long wavelength end of the filter.  For more details, see the SPIRE Observers' Manual.} of each band (e.g., see \citealt{Pezzuto12}). We scaled the RSRF profiles such that an SED with a $\nu^{-1}$ profile would yield colour corrections of 1.0 (i.e., what was initially assumed in the calibration). Assuming various power-law profiles, our colour corrections agree with the corrections in the SPIRE Observers' Manual to within a few percent. For both PACS and SPIRE, we adopted the midpoint between the maximum and minimum colour correction, taking the difference of the average value with respect to the range of accepted values as our uncertainty. Table \ref{colourCorr} lists our adopted colour corrections for each wavelength. We caution that these values are only valid for $T \approx 10 - 15$ K and  $\beta \approx 1.5 - 2.5$.

\section{Measuring $\beta$ from Unfiltered Maps}

In this Appendix, we outline the techniques for measuring $\beta$ while recovering the missing large-scale structure in the SCUBA-2 maps. These techniques are the Offset Case (Appendix \ref{OffsetAppendix}) and the Spatial Factor Case (Appendix \ref{SFAppendix}). We highlight the method and discuss the uncertainties in the determined values of $\beta$.

\subsection{Offset Case}\label{OffsetAppendix}

\begin{figure}[h!]
\includegraphics[angle=-90,scale=0.7]{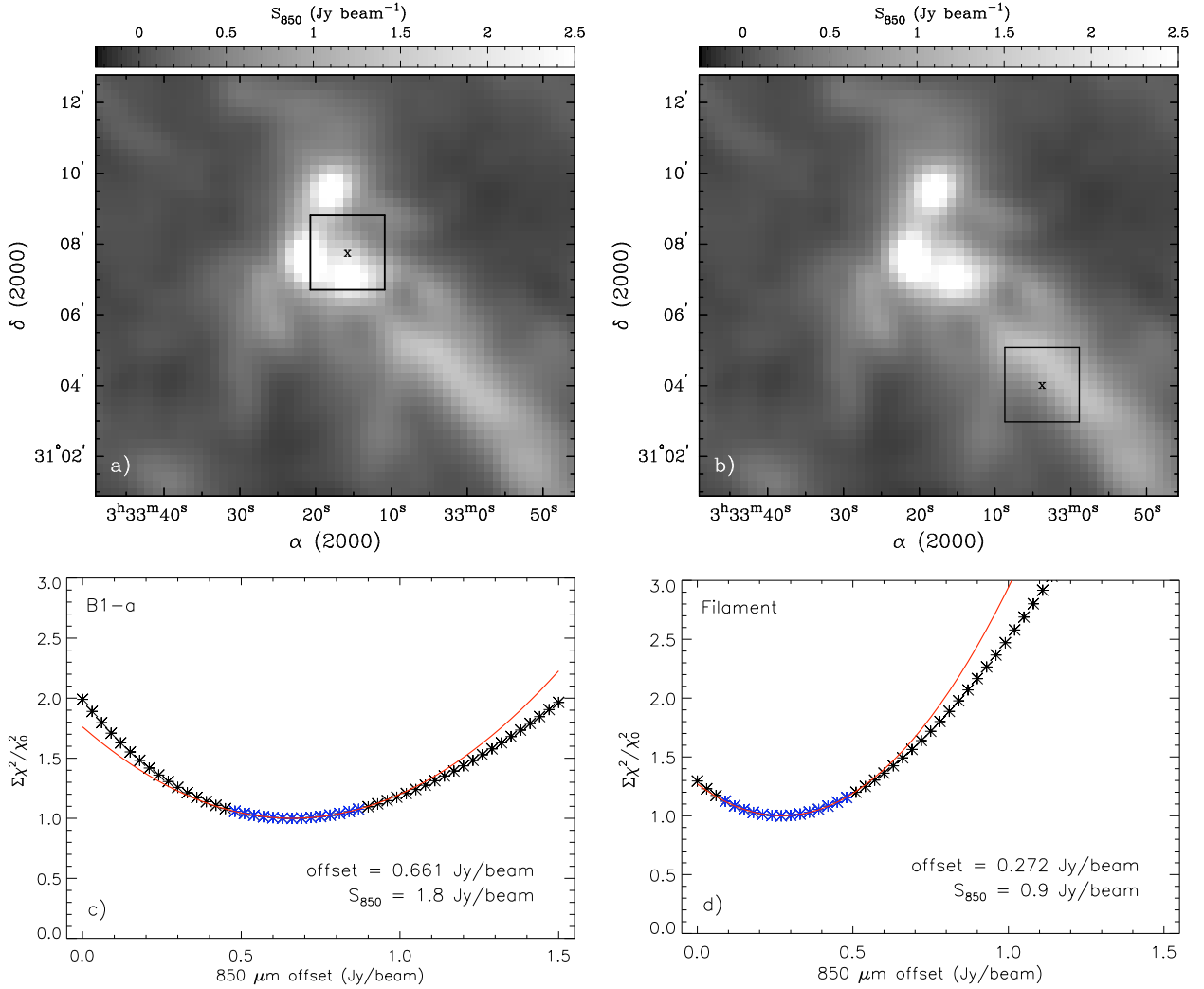}
\caption{Technique to measure the missing 850 \um\ emission. \emph{Top:} Two examples of $9\times9$ pixel subregions centered on (a) B1-a and (b) the filament. For a given subregion, the SEDs of each pixel are fit for a range of 850 \um\ offset values from $0-1.5$ \Jybeam. Then, for each offset estimate, we sum the individual $\chi^2$ measurements from every best-fit SED in the subregion. \emph{Bottom:} The \emph{relative} total $\chi^2$ measure at each offset for the two example subregions. The lowest $\chi^2$ value ($\chi_0$) is determined from a parabolic fit to a set of values (shown in blue) near the distribution minimum. The 850 \um\ offsets corresponding with these minima are given in the bottom right corners along with the observed emission at 850 \um\ (S$_{850}$). \label{offsetFigs}}
\end{figure}

We use the \emph{Herschel} data to recover the large-scale structure at 850 \um. We assume that the missing large-scale emission corresponds to scales $> 2.5$\arcmin, and thus, we do not expect large variations in the extended structure for scales $< 2.5\arcmin$. The method is outlined below and in Figure \ref{offsetFigs}:

\begin{description}
\item[1. ] For each pixel, we select a $9 \times 9$ pixel subregion (corresponding to $\sim 2.1\arcmin$ for 14\arcsec\ pixels). The extended emission should not vary over this subregion. See the top panels of Figure \ref{offsetFigs}.
\item[2. ] We create a grid of 850 \um\ flux offsets ranging from 0 \Jybeam\ to 1.5 \Jybeam\ in increments of 0.03 \Jybeam\ to represent the filtered-out emission at 850 \um.
\item[3. ] For each 850 \um\ offset, we fit the $160 - 850$ \um\ emission from all pixels within the $9\times9$ pixel subregion.  
\item[4. ] We total the individual $\chi^2$ measures across the entire subregion for each estimation of the 850 \um\ offset. If the 850 \um\ offset is close to the filtered-out extended emission, the total $\chi^2$ will be low, and if the 850 \um\ offset differs from the true extended emission, then the total $\chi^2$ will be high.
\item[5. ] For the central pixel in the subregion, we adopt the 850 \um\ offset value that corresponds to the minimum total $\chi^2$. For simplicity, we fit a parabolic function to the distribution of total $\chi^2$ with offset. See the bottom panels of Figure \ref{offsetFigs}.
\end{description}

Figure \ref{offsetFigs} illustrates our technique. The top panels show two example subregions with $9\times9$ pixel boxes used to measure the offset at the marked positions towards a pixel in B1-a and the south-west filament. For each estimate of the 850 \um\ offset ranging from 0 \Jybeam\ to 1.5 \Jybeam, we found the best-fit SEDs across these subregions, and then summed the $\chi^2$ values from all 81 pixels. The bottom panels show the distributions of total $\chi^2$ with 850 \um\ offsets for our two example subregions. As expected, when the 850 \um\ offset is greatly underestimated or overestimated, the total $\chi^2$ is large. We used parabolic fits to identify the offset corresponding to the minima. For simplicity, Figure \ref{offsetFigs} does not include the 10\%\ calibration uncertainties for each instrument.

\begin{figure}[h!]
\includegraphics[scale=0.475]{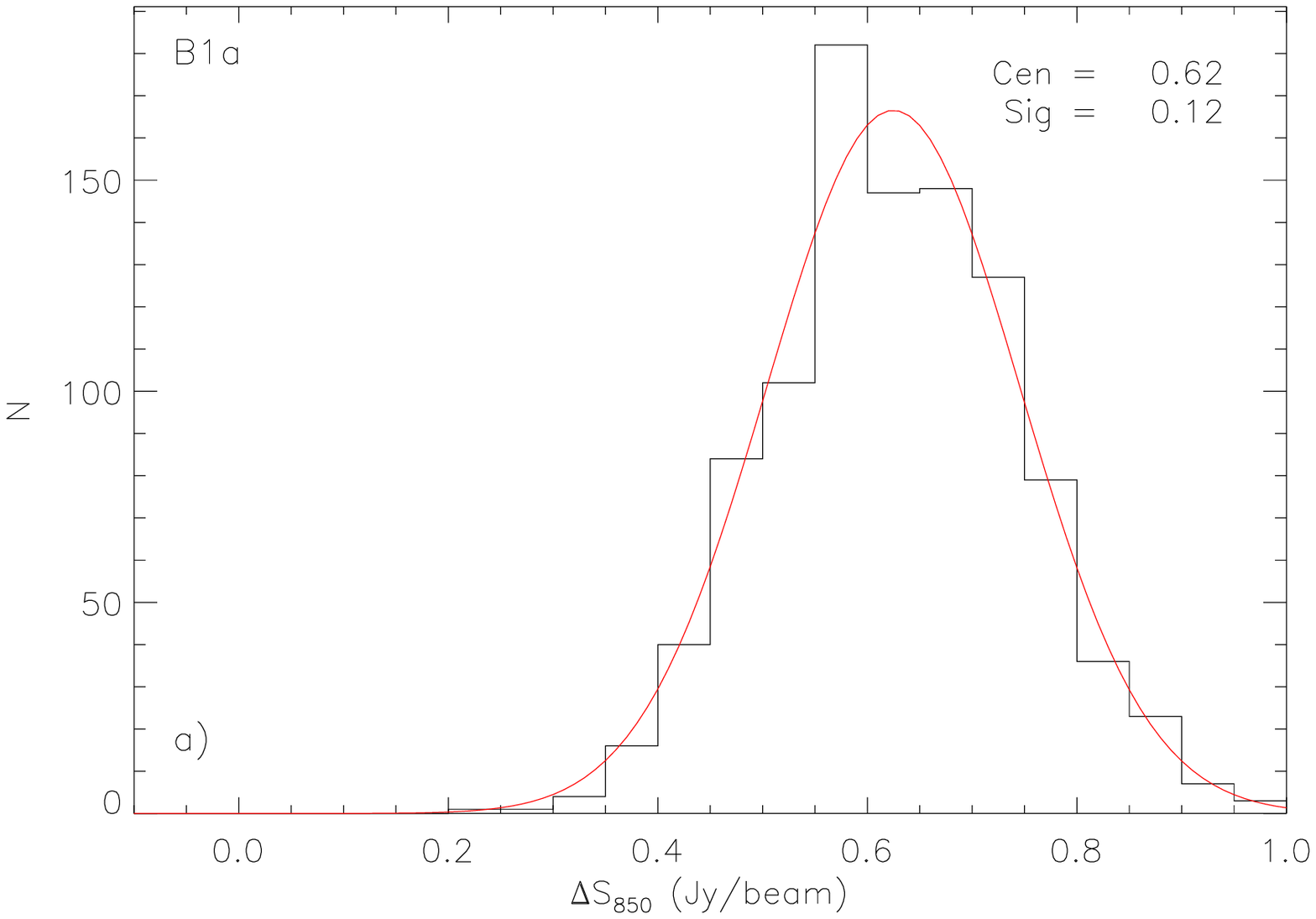}
\includegraphics[scale=0.475]{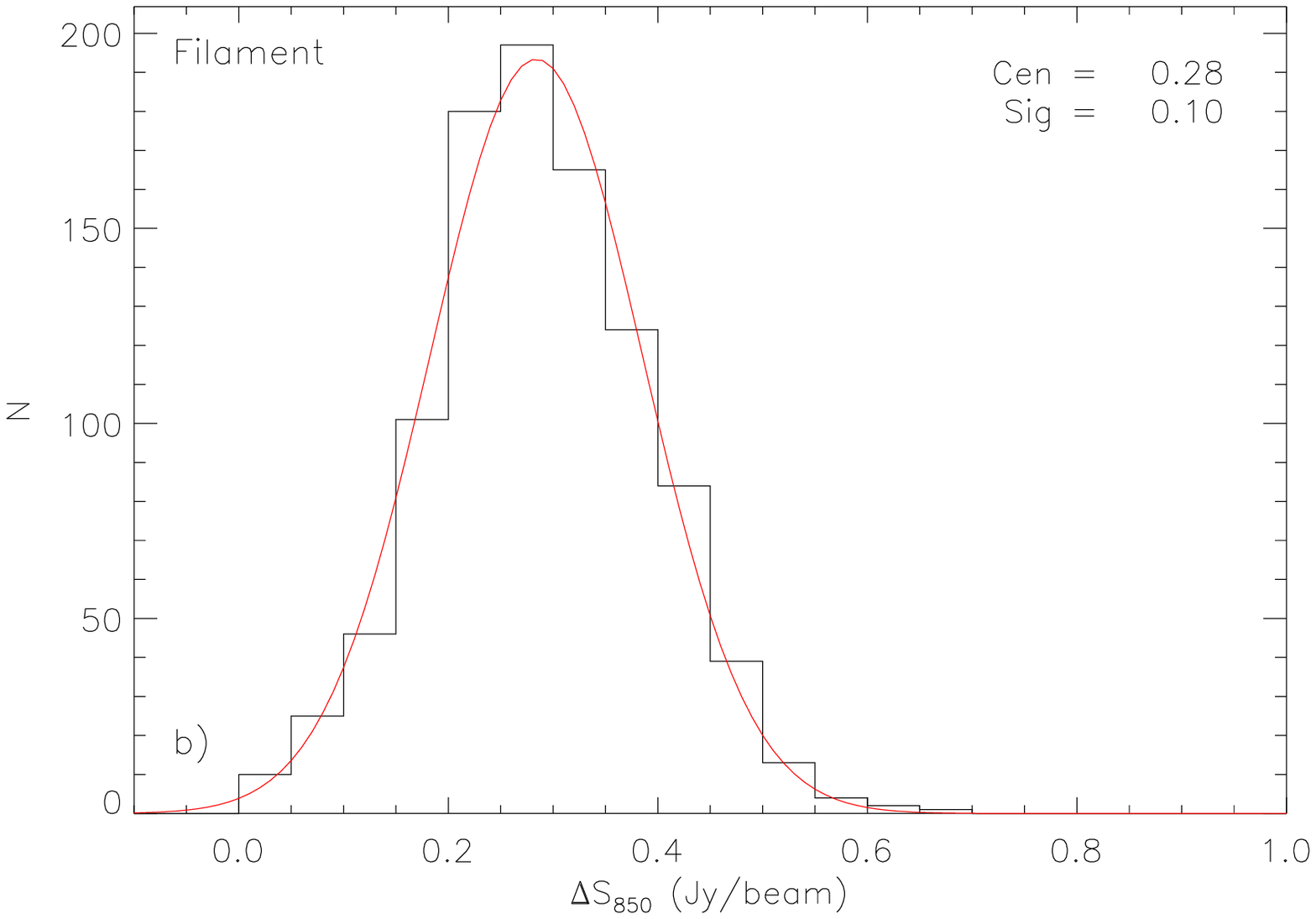}
\caption{Range of best-fit 850 \um\ offsets. We generated 1000 random calibration correction factors as described in Section \ref{combo}, and then found the best-fit 850 \um\ offset for each set of correction factors according to our technique (see text and Figure \ref{offsetFigs}). The red curves show a Gaussian fit to the histograms. \label{offsetErr}}
\end{figure}

Figure \ref{offsetErr} shows the range of best-fit 850 \um\ offsets within the 10\%\ calibration uncertainties for the two pixels highlighted in Figure \ref{offsetFigs}.  The range of acceptable offsets appears fairly broad, suggesting that this technique does not constrain well the missing 850 \um\ large-scale emission and these uncertainties overwhelm the uncertainties of the observed 850 \um\ data (10\%). Applying 1000 random calibration correction factors to each subregion (effectively 81 SEDs per increment) and each estimated 850 \um\ offset (50 possible increments) is computationally expensive. Therefore, we sampled a subset of 10 pixels and found that the $1\ \sigma$ standard deviation errors in the mean offsets are $20 - 50\%$ of the respective offset measures.

\begin{figure}[h!]
\includegraphics[scale=0.475]{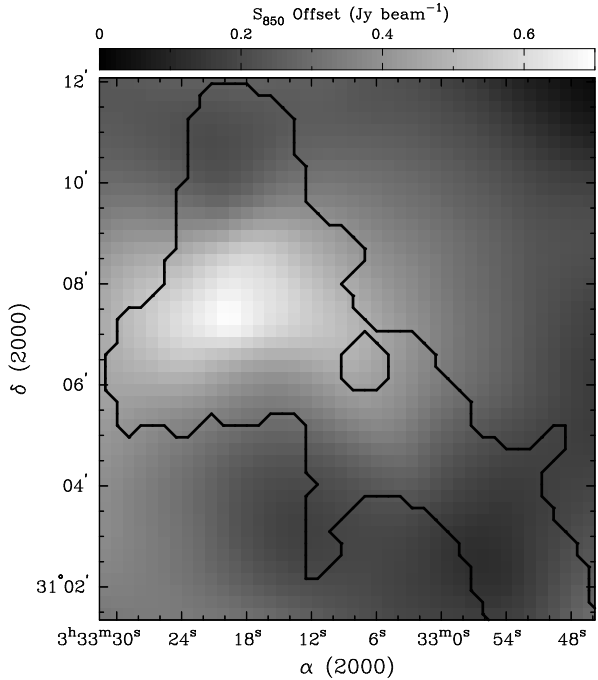}
\caption{The recovered diffuse emission at 850 \um\ obtained with our technique. The black contours correspond to the $10\ \sigma$ region in Section \ref{filterH}.  \label{offsetMap}}
\end{figure}

Figure \ref{offsetMap} shows the ``recovered'' 850 \um\ large-scale emission. The black contours show the $10\ \sigma$ flux level from Section \ref{filterH}. The recovered 850 \um\ extended emission shows some similarities and yet some variations with the filtered-out large-scale emission at 250 \um\ shown in Figure \ref{resH}. For example, we see the recovered 850 \um\ large-scale emission peak towards the central B1 region at both wavelengths. Conversely, the recovered 850 \um\ large-scale emission map appears more structured, with less extended emission towards the main filament and B1-c. Since the 850 \um\ emission traces very cold dust, these differences could reflect differences in dust temperature. Similar differences, however, are not seen between the \emph{Herschel} bands at 250 \um\ and 500 \um. 
 
\begin{figure}[h!]
\includegraphics[scale=0.75,angle=-90]{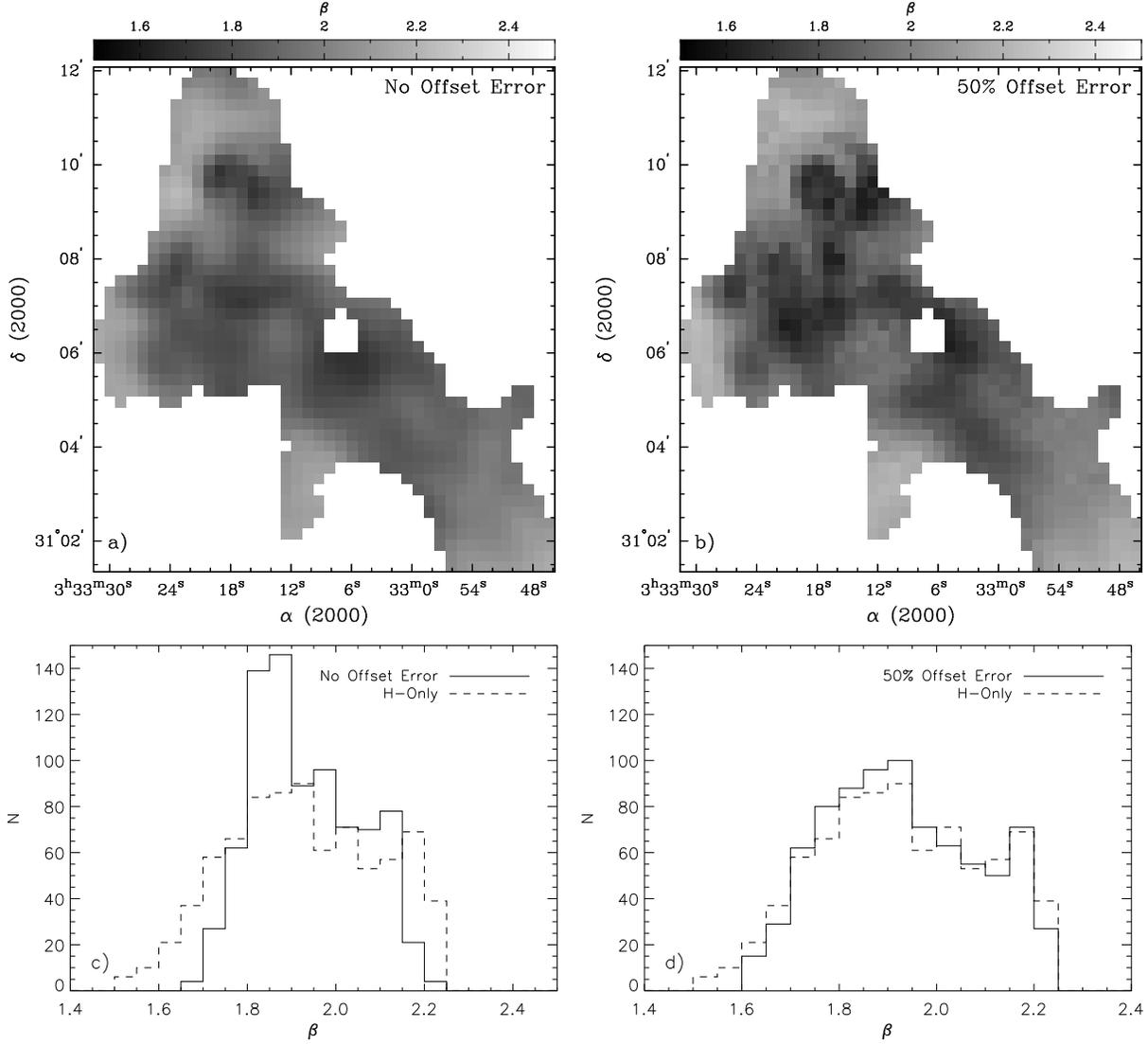}
\caption{\emph{Top:} Dust emissivity maps assuming (a) no errors for the 850 \um\ offsets or (b) 50\% errors for the 850 \um\ offsets (see Appendix \ref{OffsetAppendix}). The black contours indicate the region correspond to $> 10\ \sigma$ from Section \ref{filterH}. \emph{Bottom:} Histograms comparing $\beta$ distributions from the 850 \um\ offset analyses and the distribution obtained from H-only data. We assume no 850 \um\ offset uncertainties in (c) and 50\%\ offset uncertainties in (d). \label{offsetBetaMaps}}
\end{figure}

Figure \ref{offsetBetaMaps} shows the $\beta$ maps determined from SED fitting with the \emph{Herschel}+850 bands, where the 850 \um\ emission includes the recovered extended structure from Figure \ref{offsetMap}. We measured $\beta$ assuming two extremes, i.e., (a) no uncertainties in the extended emission and (b) 50\%\ uncertainties in the extended emission. In the bottom panels, we compare the derived $\beta$ distributions with the equivalent H-only distribution. Most noticeably, there is a spike at $\beta \approx 1.8 - 1.9$ in Figure \ref{offsetBetaMaps}c, which is similar to the $\beta \approx 2$ peak seen in Figure \ref{compareDist}. Nevertheless, as shown in Figure \ref{offsetBetaMaps}d, this spike disappears if we assume 50\%\ errors for the 850 \um\ extended emission, likely due to the 850 \um\ data being too uncertain to constrain well the SEDs. Indeed, the $\beta$ distribution with 50\%\ flux errors in the 850 \um\ extended structure appears more similar to the H-only distribution, with a slight offset towards lower $\beta$ values. Therefore, the Offset Case approach to recover the filtered-out, large-scale 850 \um\ emission makes the 850 \um\ data too uncertain to be an effective constraint.

\subsection{Spatial Factor Case}\label{SFAppendix}

For the Spatial Factor Case, we attempt to recover the missing large-scale emission by scaling the 450 \um\ and 850 \um\ emission by a spatial filtering factor to account for the missing extended flux. We assume that the 450 \um\ and 850 \um\ observations are missing the same fraction of extended emission, and then fit the SEDs according to Equation \ref{modBBv2}, solving for the SED parameters and the spatial filtering factor, $F_{SF}$. Since Figure \ref{percDiff450} demonstrates that the observed 450 \um\ emission tends to be 10-15\%\ brighter than the predicted emission, we adjusted the 450 \um\ emission downward by 12.5\%. This adjustment appears most reasonable for the brightest parts of the B1 clump (i.e., $I_{450} > 5$ \Jybeam), however, and may cause erroneous results outside of these locations.
 
\begin{figure}[h!]
\includegraphics[scale=0.75,angle=-90]{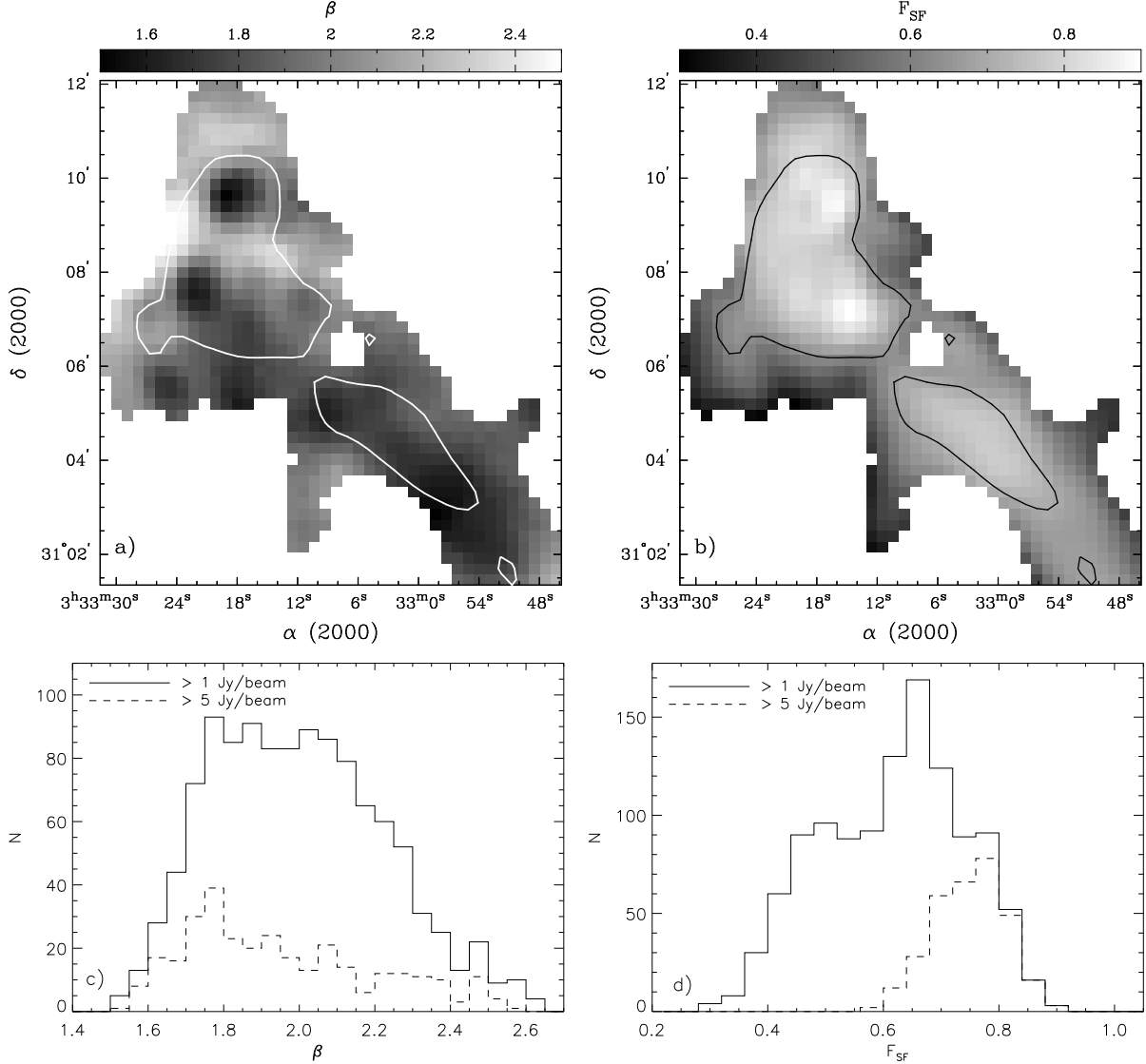}
\caption{\emph{Top:} Results from the Spatial Factor Case highlighting (a) the dust emissivity and (b) spatial filtering factor ($F_{SF}$). The dust emission was fit using Equation \ref{modBBv2}, assuming an equal scaling factor, $F_{SF}$ for the 450 \um\ band and the 850 \um\ band. The contours indicate the region with $I_{450} > 5$ \Jybeam. \emph{Bottom:} Histograms for $\beta$ and spatial filtering factor. The solid lines show the histograms for the entire region in the upper panels, and the dashed lines show the histograms for the $I_{450} > 5$ \Jybeam\ regions, only. For these analyses, the 450 \um\ data were scaled by 12.5\% to account for the fractional increase in brightness in the observed maps over the predicted maps (see Figure \ref{percDiff450}). \label{spatBetaMaps}}
\end{figure}

The top panels of Figure \ref{spatBetaMaps} shows the $\beta$ values and spatial filtering factors derived using Equation \ref{modBBv2}. The contours correspond to the regions with $I_{450} > 5$ \Jybeam, and where a scaling of $-12.5$\%\ is more reasonable (see Figure \ref{percDiff450}). Similar to the previous measures of $\beta$, we find that $\beta$ is lowest towards B1-c. On the other hand, we find $\beta \sim 2.3$ towards B1-a and $\beta \sim 1.6$ towards B1-b.  Additionally, we obtain $\beta \sim 1.7$ along the filament. For the spatial filtering factor, we find $F_{SF} \sim 0.8 - 0.9$ towards the dense, compact cores and $F_{SF} \sim 0.7$ towards the filament. Towards the edges of the 10 $\sigma$ region, however, the spatial filtering factor drops to $\lesssim 0.4$, suggesting that most of that emission was removed by the SCUBA-2 reduction pipeline. Figure \ref{spatBetaMaps} also shows the distributions of $\beta$ and $F_{SF}$ for the $> 1$ \Jybeam\ and $> 5$ \Jybeam\ regions. Unsurprisingly, the brightest regions (i.e., the locations of the compact objects) correspond with the highest fraction of recovered emission. Nevertheless, we find a wide range of $\beta$ values even for the $> 5$ \Jybeam\ region. Within the calibration uncertainties, we find $\sigma_{\beta} \lesssim 1.0$. Therefore, the Spatial Filtering Case is not a robust method to determine $\beta$.

\end{document}